\documentclass[aps,prb,twocolumn,preprintnumbers,amsmath,amssymb,superscriptaddress]{revtex4-2}

\usepackage{tabularx}
\usepackage[utf8]{inputenc} 
\usepackage[T1]{fontenc}    
\usepackage{hyperref}       
\usepackage{url}            
\usepackage{amsfonts}       
\usepackage{amsmath,amsthm,amssymb} 
\usepackage{bm}
\usepackage{bbm}
\usepackage{dcolumn}
\usepackage{nicefrac}       
\usepackage{gensymb}
\usepackage{graphicx}
\usepackage{xcolor, etoolbox}
\usepackage{mathtools}
\usepackage{mathrsfs}
\usepackage{ragged2e}
\usepackage{microtype}      
\usepackage{braket}      
\usepackage{enumitem}
\usepackage{xspace}
\usepackage{ulem} 
\usepackage{dsfont}
\usepackage{float}
\usepackage{soul}
\usepackage{slashed}
\usepackage{esvect}

\usepackage[nameinlink]{cleveref}

\makeatletter
\AddToHook{cmd/appendix/before}{\def\cref@section@alias{appendix}}
\AddToHook{cmd/appendix/before}{\def\cref@subsection@alias{appendix}}
\AddToHook{cmd/appendix/before}{\def\cref@subsubsection@alias{appendix}}
\AddToHook{cmd/appendix/before}{\def\cref@paragraph@alias{appendix}}
\makeatother
\crefname{appendix}{Appendix}{Appendices}
\hypersetup{
	colorlinks,
	linkcolor={red!75!black},
	citecolor={blue!75!black},
	urlcolor={blue!75!black}
}

\usepackage{physics}
\usepackage{ulem}
\newcommand{\Eqref}[1]{Eq.~\eqref{#1}}

\newcommand\varpm{\mathbin{\vcenter{\hbox{%
  \oalign{\hfil$\scriptstyle+$\hfil\cr
          \noalign{\kern-.3ex}
          $\scriptscriptstyle({-})$\cr}%
}}}}
\newcommand\varmp{\mathbin{\vcenter{\hbox{%
  \oalign{$\scriptstyle({+})$\cr
          \noalign{\kern-.3ex}
          \hfil$\scriptscriptstyle-$\hfil\cr}%
}}}}

\begin{document}

\title{Quantum critical fan and emergent relativistic symmetry\\  
of two-dimensional Dirac semimetals}

\author{Friederike Ihssen}
\email{friederike.ihssen@rub.de}
\affiliation{Theoretische Physik III, Ruhr-Universit\"at Bochum, D-44801 Bochum, Germany}

\author{Bilal Hawashin}
\email{bilal.hawashin@rub.de}
\affiliation{Theoretische Physik III, Ruhr-Universit\"at Bochum, D-44801 Bochum, Germany}

\author{Mireia Tolosa-Sime\'on}
\email{tolosa@tp3.rub.de}
\affiliation{Theoretische Physik III, Ruhr-Universit\"at Bochum, D-44801 Bochum, Germany}

\author{Michael M. Scherer}
\email{scherer@tp3.rub.de}
\affiliation{Theoretische Physik III, Ruhr-Universit\"at Bochum, D-44801 Bochum, Germany}

\begin{abstract}
Two-dimensional Dirac semimetals near a quantum critical point can be described by Gross--Neveu--Yukawa models.
In view of recent experimental advances exhibiting a transition from Dirac semimetal to insulator in highly-tunable van-der-Waals heterostructures, a better understanding of finite-temperature effects is mandatory.
Here, we study the Gross--Neveu--Yukawa phase diagram of the chiral Ising model with a non-perturbative field-theory approach at zero and finite temperature, both in the semimetallic phase and in the insulating phase with spontaneously broken $\mathbb{Z}_2$ symmetry. At zero temperature, we find a quantum critical point with critical exponents that are close to the ones of the chiral Ising universality class, and show that relativistic symmetry is emergent close to the quantum critical point. At finite temperature, the ordered phase survives up to a finite critical temperature, at which we observe a classical phase transition into the disordered phase. We confirm that this transition lies in the two-dimensional Ising universality class. Finally, we determine the extent and scaling properties of the quantum critical fan, and the behavior of the quasiparticle weight, therein.
In summary, we present a unified field-theoretical framework for the phase diagram of the chiral Ising model in the surroundings of its quantum critical point.
\end{abstract}

\maketitle

\section{Introduction}

Quantum critical points (QCP) play an important role in strongly-correlated materials as they can leave an imprint on thermodynamics and transport in an extended range of their phase diagram, even at elevated temperatures~\cite{Vojta_2003,Sachdev2011}.
Yet, QCPs in metals with an extended Fermi surface are notoriously hard to grasp for theory~\cite{doi:10.1080/001075199181602,RevModPhys.73.797,RevModPhys.79.1015,doi:10.1063/1.3554314,RevModPhys.88.025006,doi:10.1146/annurev-conmatphys-031016-025531,doi:10.1146/annurev-conmatphys-031218-013339,Xu_2019,RevModPhys.94.035004}.
A more accessible but still  challenging version of QCPs occurs in two-dimensional Dirac semimetals~\cite{Wehling_2014,Vafek_2014}.
Their low-energy description can be cast into a quasi-relativistic form where the electronic excitations obey the massless Dirac equation.
Then, the Fermi-surface is point-like and the effective description exhibits a high degree of symmetry, which facilitates systematic theory development.

Dirac semimetals in the presence of sizable interactions undergo continuous quantum phase transitions into spontaneously symmetry-broken ordered states~\cite{Boyack:2020xpe}.
Near such a transition, the effective theory can then be written in terms of relativistic Gross--Neveu--Yukawa-type models and can be considered as an analogue to dynamical mass generation in elementary particle physics.
The specific types of symmetry-breaking order, and hence the Yukawa couplings and bosonic sectors, depend on microscopic details of the system and include, e.g., charge order, magnetism, and topological phases~\cite{sorella1992semi,PhysRevB.62.2806,Herbut2006,PhysRevLett.98.186809,PhysRevLett.100.146404,Raghu2008,Weeks2010,Grushin2013,PhysRevX.8.031089,PhysRevX.10.031034,PhysRevX.11.011014,PhysRevX.12.011061,Classen_2022}.
In fact, the Gross--Neveu--Yukawa (GNY) model with a single real scalar field which breaks a (discrete) chiral $\mathbbm{Z}_2$~symmetry across the quantum phase transition, i.e., the chiral Ising model, is arguably one of the simplest models for fermions in two spatial dimensions with a quantum critical point featuring three-dimensional universality~\cite{Rosenstein:1993zf}.

Over the last years, theoretical efforts have focused on the quantitative determination of three-dimensional GNY universality~\cite{Boyack:2020xpe,Herbut:2023xgz}.
As a result, estimates for critical exponents of the chiral Ising model and the closely related (yet distinct~\cite{Erramilli2023}) parity-breaking GNY model have converged and now agree well between perturbative field theory~\cite{Zerf2017,Ihrig2018}, functional renormalization~\cite{PhysRevB.89.205403,Knorr2016,Tolosa-Simeon:2025fot}, conformal bootstrap~\cite{Iliesiu:2015qra,Iliesiu:2017nrv,Erramilli2023}, and quantum Monte Carlo simulations~\cite{PhysRevD.88.021701,PhysRevB.101.064308,PhysRevB.108.L121112}.
For other GNY models, e.g., for the physically relevant chiral Heisenberg model~\cite{Herbut2006,lr2x-nnks}, there is no satisfying agreement, yet~\cite{Herbut:2023xgz}.

Interestingly, GNY-type universality is expected to be available for scrutiny via the highly-tunable platform of two-dimensional van-der-Waals materials~\cite{lu2019superconductors, Parthenios2023,huang2024angletunedgrossneveuquantumcriticality,biedermann2024twisttunedquantumcriticalitymoire,PhysRevResearch.5.043173,tchf-w8h7,lr2x-nnks}.
In particular, the recent observation of a semimetal-to-insulator transition in the Dirac material tetralayer tungsten diselenide (WSe$_2$)~\cite{ma2024relativisticmotttransitionstrongly} has provided evidence for such an analogue of chiral symmetry breaking in a real solid upon twist-angle tuning.
The nature of the observed insulating state has not yet been determined experimentally, but has been theoretically argued to be of antiferromagnetic nature~\cite{ma2024relativisticmotttransitionstrongly,PhysRevResearch.5.043173,tchf-w8h7,lr2x-nnks}. This would correspond to a GNY model with a three-component vector order parameter field, known as the chiral Heisenberg model. 

This development suggests that GNY-type models perspectively offer a controlled and systematic pathway to connect quantum many-body theory for interacting Dirac materials with experimental data.
Establishing this connection between Lorentz symmetric GNY-type models and an actual material that does not feature relativistic symmetry comes with certain challenges.
In fact, an appropriate description takes into account that Dirac excitations and order-parameter fluctuations have different bare velocities.
In Dirac materials, relativistic symmetry with equal velocities is then expected to emerge near the QCP~\cite{Roy:2015zna}.
Yet, experiments are, naturally, carried out at finite -- albeit small -- temperature, introducing an additional scale.

\begin{figure}[t!]
    \includegraphics[width=0.95\linewidth]{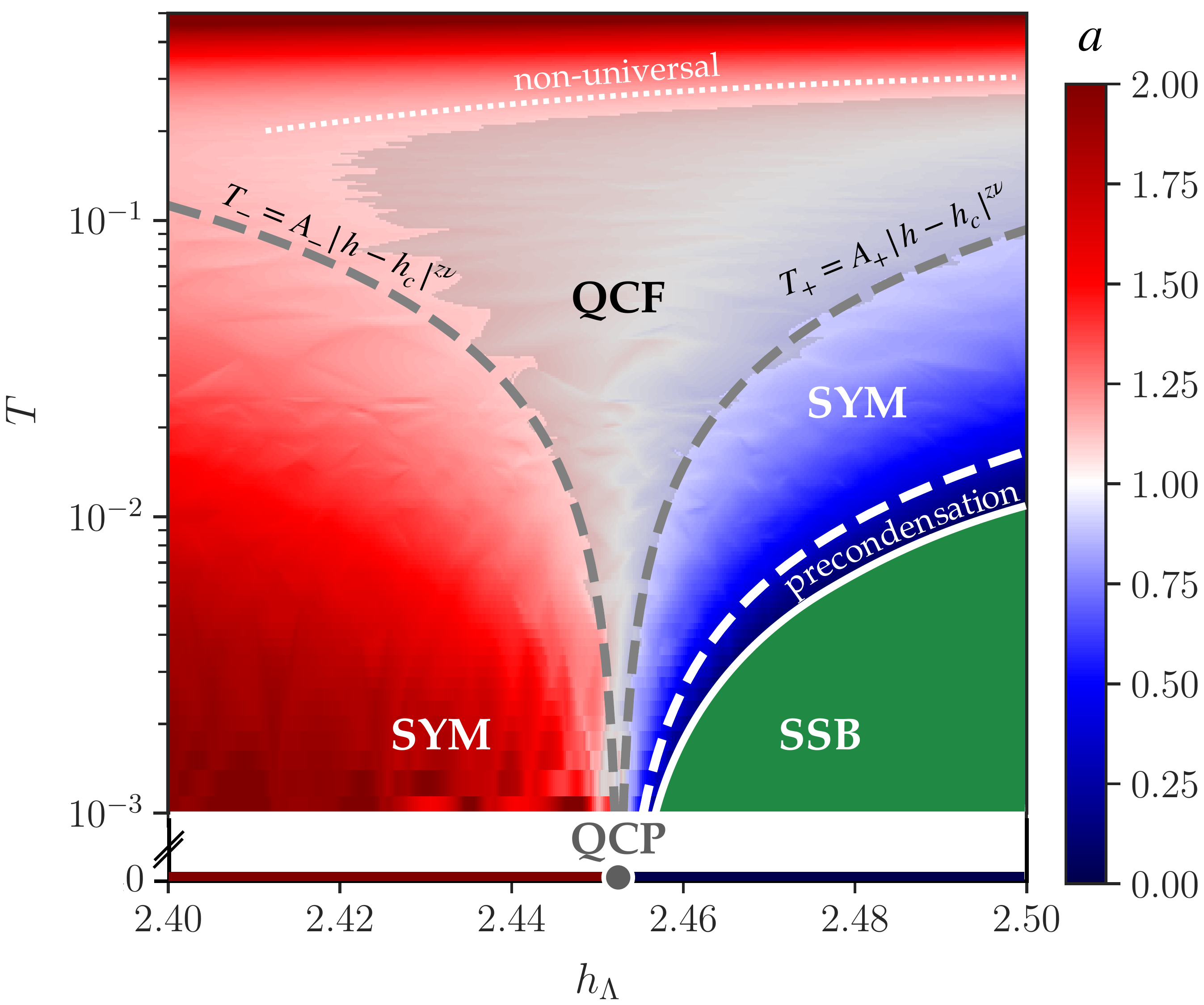}
    \caption{\textbf{GNY phase diagram and quantum critical fan.} The color scheme shows the calculated scaling exponent of the thermal correlation length $a$, cf.~\Eqref{eq:ExponentA}, as a function of the coupling $h_\Lambda$ and temperature $T$,
    which can be used to identify the quantum critical fan~(QCF). 
    The crossovers delimiting the expected extent of the QCF are shown as dashed gray lines,  cf.~\Eqref{eq:QCF}.
    Within this region, the gray-shaded area indicates the region where ${|z - a|<0.1}$, i.e., where the temperature dependence of the thermal correlation length follows the scaling imprinted by the QCP. 
    The other features shown in the phase diagram are discussed in the main text, e.g., the QCP, the precondensation regime, the phase transition line, and the spontaneously symmetry-broken (SSB) regime. 
    }
    \label{fig:Fan_nonUniversal}
\end{figure}

A faithful theoretical description needs to systematically take such effects into account, i.e., the breaking of relativistic symmetry and finite temperature.
Such a theory is also required to identify quantitative deviations from the expected critical behavior away from the QCP.
It is the purpose of the present contribution to set up a systematic non-perturbative field-theoretical approach that faithfully captures all of the aspects mentioned above.
Here, we first develop this approach for the paradigmatic case of the Gross--Neveu--Yukawa model breaking discrete chiral $\mathbbm{Z}_2$ symmetry, i.e., the chiral Ising model, because it most directly allows us to benchmark our results with established results from other methods~\footnote{We note that the Gross--Neveu--Yukawa archipelago studied with the conformal bootstrap approach considers that case of parity breaking, instead. Numerical evidence suggests, however, that the scaling dimensions or critical exponents at the quantum critical point are extremely close to the ones from the chiral Ising model.}.

To that end, we employ a functional renormalization group~(fRG) approach, which allows us to map out the phase diagram of the GNY model, calculate quantum  critical exponents, determine the extent of the quantum critical fan (QCF), estimate the size of ordered domains, and extract the size of fermionic mass gaps in the symmetry-broken phase.
Our present approach expands over previous fRG work~\cite{Tolosa-Simeon:2025fot} in two aspects: (1)~We explicitly track the velocity renormalization of the Dirac fermions and the order-parameter field, and (2)~we resolve the fully field-dependent effective potential using state-of-the-art fluid dynamical methods to solve the highly non-linear partial differential fRG flow equations.

In particular, the second point now allows us to also quantitatively access the classical critical region near the finite-temperature transition into the phase with broken $\mathbbm{Z}_2$~symmetry.
While our theory development is rooted in and benchmarked against the established results for Gross--Neveu--Yukawa universality, it systematically also explores other universal and non-universal aspects of the phase diagram within a unified field-theoretical framework.

The paper is organized as follows: In \Cref{sec:model}, we specify the GNY model and draw connections to symmetries and scales of real Dirac materials.
\Cref{sec:method} discusses the fRG method, the employed truncation, regulators, and initial conditions.
In \Cref{sec:results}, we discuss emergent relativistic symmetry, quantum criticality, the finite-temperature transition, the phenomenon of precondensation, the quantum critical fan, and the fermionic quasiparticle weight.
We draw conclusions and provide an outlook in \Cref{sec:outlook}.

\subsection*{Key results}\label{sec:key}

In most previous work, relativistic symmetry has been assumed for the extraction of quantum critical exponents of Dirac semimetals with high precision~\cite{Erramilli:2022kgp}.
Evidence for its emergence has only been collected using perturbative methods, so far~\cite{Roy:2015zna,lr2x-nnks}.
Here, we employ a non-perturbative fRG approach to add evidence for this phenomenon at zero temperature and specifically exhibit that relativistic symmetry also emerges near the QCP on the side of the quantum phase transition, where a dynamical mass gap has been generated, see \Cref{sec:emLS}, i.e., where the discrete chiral $\mathbbm{Z}_2$ symmetry is spontaneously broken.
We further introduce a spatial regulator scheme with convenient analytical properties and benchmark this regulator scheme to show that it faithfully reproduces the established  critical exponents and the emergent relativistic symmetry at the QCP, see \Cref{sec:QCbehavior}.

Next, we track the finite-temperature transition of the chiral Ising model into the $\mathbbm{Z}_2$ broken phase and show that its critical exponents are consistent with the classical two-dimensional Ising transition~\Cref{sec:classicalIsing}.
This is for example possible through the use of the sophisticated numerical framework provided by \texttt{DiFfRG}~\cite{Sattler:2024ozv}, which allows us to resolve the fully field-dependent effective potential.
Finally, we also calculate the scaling behavior and extent of the quantum critical fan in the presence of different bare boson and fermion velocities (\Cref{sec:QCF}) and the fermionic quasiparticle weight (\Cref{sec:quasiparticle}).
In total, we provide a complete unified framework that quantitatively describes both, the universal quantum and finite-temperature critical behavior of the chiral Ising model as well as the non-universal regions in the phase diagram near the QCP and at elevated temperatures, see \Cref{fig:Fan_nonUniversal}.
Our developments prepare the calculation of thermodynamic observables and spectral functions, which we aim for in future work.

\section{Model}\label{sec:model}

We consider spin-1/2 electrons on the (effective) honeycomb lattice as relevant for graphene~\cite{Herbut2006} and twisted tetralayer WSe$_2$~\cite{ma2024relativisticmotttransitionstrongly,lr2x-nnks,tchf-w8h7}.
The effective low-energy description  features eight-component Dirac fermions ${\psi=(\psi_\uparrow,\psi_\downarrow)^T}$, where $\uparrow, \downarrow$ denote the spin projection of the four-component spinors $\psi_\uparrow, \psi_\downarrow$.
We aim to describe the phase diagram of such a system near the transition to a state with broken $\mathbbm{Z}_2$ sublattice-exchange symmetry of the underlying honeycomb lattice, e.g., a charge-density-wave phase~\cite{Herbut2006,Herbut:2009qb,lr2x-nnks}.
Such a $\mathbbm{Z}_2$-symmetry broken state corresponds to a finite vacuum expectation value $\langle \bar\psi\psi\rangle\neq 0$, which we represent by a real-valued scalar order-parameter field~$\phi$ such that $\langle\phi\rangle\propto \langle \bar\psi\psi\rangle$.

The resulting model is a $2+1$-dimensional Euclidean field theory, known as the chiral Ising model, defined by the action $S = \int_0^{1/T} d\tau \int_{\mathbb{R}^2}  d^2x \, \mathcal{L}$ with Lagrangian 
\begin{align}
    \mathcal{L} &= \bar\psi\left(\gamma_0\partial_\tau +v_\psi\vec{\gamma}\cdot\vec{\partial}\right)\psi
    + \frac{1}{2}\phi\left(-\partial_\tau^2-v_\phi^2\vec{\partial}^{\,2} +r\right)\phi\nonumber\\
    &\quad+h\phi\bar\psi\psi+\lambda\phi^4+\mathcal{O}(\phi^6, \psi^4,\psi^2\phi^2)\,.
    \label{eq:lagrangian}
\end{align}
Here, $\bar\psi=\psi^\dagger(\mathds{1}_2\otimes\gamma_0)$ denotes the Dirac conjugate, where $\gamma_0$ and the two components $\gamma_i$ of $\vec{\gamma}=(\gamma_1,\gamma_2)$ are ${4 \times 4}$ matrices that satisfy the Clifford algebra $\{\gamma_\mu,\gamma_\nu\}=2\delta_{\mu\nu}\mathds{1}_{4}$ with $\mu,\nu\in\{0,1,2\}$, and $T$ is the temperature. 
Due to the reducibility of the representation, there are two additional matrices that commute with all $\gamma_\mu$, which we denote as $\gamma_3$ and $\gamma_5$. 

For twisted tetralayer tungsten diselenide, one possible representation is $(\gamma_0,\gamma_1,\gamma_2)=(\tau_0\sigma_3,\tau_3\sigma_2,\tau_0\sigma_1)$, where $\tau_i$ and $\sigma_i$ with $i\in \{1,2,3\}$ denote Pauli matrices acting in the minivalley and sublattice space of the emergent moiré honeycomb lattice, respectively, while $\tau_0$ and $\sigma_0$ are identity matrices, see~Ref.~\cite{lr2x-nnks} for more details.
For the case of graphene, we refer to Ref.~\cite{PhysRevB.89.205403}.
The Fermi velocity~$v_\psi$ in twisted tetralayer tungsten diselenide is of order $\mathcal{O}(10^4\,\mathrm{m/s})$ for twist angles in the range of a semimetal-to-insulator transition, i.e., $\theta\in [2^\circ,3^\circ]$~\cite{ma2024relativisticmotttransitionstrongly}.

The scalar order-parameter field $\phi$, with mass tuning parameter~$r$ and self-interaction~$\lambda$, couples to the Dirac fermions via a Yukawa interaction~$h$. 
Symmetry-compatible higher-order interactions are represented by~$\mathcal{O}(\phi^6, \psi^4,\psi^2\phi^2)$.
In the $\mathbbm{Z}_2$-broken phase, the order
parameter acquires a finite expectation value $\langle\phi\rangle$, which dynamically generates a mass gap for the Dirac fermions corresponding to a phase with staggered charge order. 
At the microscopic scale, the order-parameter velocity $v_\phi$ is generally different from the Fermi velocity, i.e., $v_\phi\neq v_\psi$. 
This implies the explicit breaking of relativistic symmetry at zero temperature, see below.
Previous works, however, have shown the emergence of relativistic symmetry towards the infrared near the quantum critical point in closely related models~\cite{Roy:2015zna,lr2x-nnks}.

The Lagrangian admits various global symmetries:\medskip

\noindent (a)~\textit{Rotational symmetry:} The action is invariant under
\begin{align}
        &\psi(x) \to D(\Lambda_s) \psi(\Lambda_s^{-1}x)\,, \\
        &\bar{\psi}(x) \to  \bar{\psi}(\Lambda_s^{-1}x) D(\Lambda_s)^{-1}\,, \\
        &\phi(x) \to \phi(\Lambda_s^{-1}x)\,,
\end{align}
where $\Lambda_s \in \mathrm{O}(2)$ acts only non-trivially on $\vec{x}$, and the Dirac representation $D(\Lambda_s) = (\mathds{1}_2 \otimes e^{-i \sigma_{12} \omega_{12}/2})$ is generated by ${\sigma_{12} = \frac{i}{2}[\gamma_1, \gamma_2]}$ with ${\omega_{12} \in \mathbb{R}}$.

At $T=0$ and for vanishing~$h$ \textit{or/and} ${v_\psi = v_\phi}$, this symmetry is enlarged to full relativistic symmetry. 
Then, the action is invariant under the transformation
\begin{align}
        &\psi(x) \to D(\Lambda) \psi(g_\psi^{-1}\Lambda^{-1}g_\psi x)\,, \\
        &\bar{\psi}(x) \to  \bar{\psi}(g_\psi^{-1}\Lambda^{-1}g_\psi x) D(\Lambda)^{-1}\,, \\
        &\phi(x) \to \phi(g_\phi^{-1}\Lambda^{-1}g_\phi x)\,,
\end{align}
with $\Lambda \in \mathrm{O}(3)$ and metric ${g_i = \mathrm{diag}(1, v_i, v_i)}$ with $ {i \in \{\phi, \psi\}}$. 
The Dirac representation is then given by ${D(\Lambda) = (\mathds{1}_2 \otimes e^{-i \sigma_{\mu \nu} \omega_{\mu \nu} / 2})}$ and is generated by ${\sigma_{\mu \nu} = \frac{i}{2}[\gamma_\mu, \gamma_\nu]}$ with $\omega_{\mu \nu} \in \mathbb{R}$.\medskip
    
\noindent (b)~\textit{$\mathrm{U}(1)$ charge symmetry:} The transformation
\begin{align}
        \psi \to e^{i \alpha} \psi, \quad \bar{\psi} \to \bar{\psi} e^{-i \alpha}, \quad \phi \to \phi\quad \text{for}\quad \alpha \in \mathbb{R}\,,
\end{align}
leaves the action invariant. 
This implies that $\psi$ and $\bar{\psi}$ can only appear in pairs ensuring charge neutrality.\medskip
    
\noindent (c)~\textit{$\mathrm{SU}(2)$ spin symmetry:} Under spin rotation, the Dirac fields transform as
\begin{align}
        \psi \to (U \otimes \mathds{1}_{4}) \psi, \quad \bar{\psi} \to \bar{\psi} (U^\dagger \otimes \mathds{1}_{4})\,,
\end{align}
with $U \in \mathrm{SU}(2)$.\medskip
    
\noindent (d)~\textit{Discrete chiral $\mathbb{Z}_2$ symmetry:}
The transformation
\begin{align}
        \psi \to \gamma_5 \psi, \quad \bar{\psi} \to -\bar{\psi} \gamma_5\,,
\end{align}
corresponds to sublattice exchange.\medskip

Since we explore the phase diagram of the chiral Ising model also away from universality, we need to specify some energy and length scales.
To that end, as an orientation, we take into account input from the experiments on twisted tetralayer WSe$_2$~\cite{ma2024relativisticmotttransitionstrongly} -- at least in terms of orders of magnitude. 
The ultraviolet (UV) cutoff scale is chosen to correspond to the inverse of the moir\'e lattice scale~$a_\mathrm{M}$, which can be estimated to be of order $1/\Lambda\sim a_\mathrm{M}\sim\mathcal{O}(1-10\,\mathrm{nm})$.
The typical extent of the uniform moir\'e region in samples is of order $L\sim \mathcal{O}(1-10\,\mu\mathrm{m})$~\cite{xia2025superconductivity}. 
We can use $1/L$ to define a natural infrared cutoff for the fluctuations in our system.
The lowest temperatures studied in experimental setups are of order $T\sim\mathcal{O}(20\,\mathrm{mK})$~\cite{ma2024relativisticmotttransitionstrongly}.
The self-interaction of the order parameter field is expected to be small at the moir\'e lattice scale, because the relevant order-parameter fluctuations are only built up near the transition to the ordered state.
The interesting range of values of the Yukawa coupling will be determined by tuning through the quantum phase transition, see below, when we discuss the initial conditions of our functional RG approach.

\section{Functional Renormalization group}\label{sec:method}

\subsection{Method}

A suitable method to simultaneously resolve the models' universal critical behavior and non-universal phase diagrams near the quantum phase transition is the functional renormalization group~\cite{Wetterich:1992yh,Berges:2002,Dupuis:2020fhh}.
In the context of two-dimensional Dirac materials, it was employed to identify the leading ordering tendencies in the presence of short-range interactions~\cite{PhysRevLett.100.146404,Raghu2008,PhysRevB.87.094521,PhysRevB.92.155137}, to
directly access their quantum critical behavior in $2+1$ spacetime dimensions~\cite{PhysRevLett.86.958,PhysRevB.66.205111,PhysRevB.89.205403,PhysRevB.93.125119,PhysRevB.94.245102,PhysRevB.96.115132,Knorr2018,PhysRevB.97.125137,PhysRevB.97.041117,PhysRevResearch.2.013034,PhysRevResearch.2.022005,PhysRevB.103.155160}, and 
to describe finite temperature effects in pertinent GNY-like models~\cite{Braun_2012,Scherer:2013many,PhysRevD.90.076002,Tolosa-Simeon:2025fot, 
Stoll:2021ori}.

In a recent work~\cite{Tolosa-Simeon:2025fot}, a unified fRG approach was put forward to simultaneously access the zero- and  finite-temperature behavior of three relevant GNY-type models, exhibiting, e.g., the scaling behavior in the quantum critical fan, the mass gap formation for the Dirac fermions, precondensation, and condensate melting in models with continuous symmetries in agreement with the Mermin--Wagner theorem.
To that end, a simplified version of the model in \Eqref{eq:lagrangian} was considered with ${v_\psi=v_\phi}$ set to unity.
Here, we extend the analysis as detailed below.

The method is based on introducing an infrared~(IR) cutoff with scale $k$ into the partition function, ${\mathcal{Z}=\int_\Lambda \mathcal{D}\Phi \, e^{-S\left[\Phi\right]}\to \mathcal Z_k}$.
This is achieved by adding a regulator term bilinear in the fields ${\Phi(q)=\begin{pmatrix}
    \phi(q), 
    \psi(q), 
    \bar{\psi}^T(q)
\end{pmatrix}^T}$ to the microscopic action
\begin{align}\label{eq:deltaS}
    S&\rightarrow S+ \Delta S_k \notag\\[1ex]
    &= S+\int_p\int_q \Big[\frac{1}{2}\phi( -p) R_k^{\phi}\phi(q)+\bar\psi(-p) R_k^{\psi}\psi(q)\Big]\,,
\end{align}
with $R_k^{\phi}=R_k^{\phi}(p,q)$ and $R_k^{\psi}=R_k^{\psi}(p,q)$ being the bosonic and fermionic infrared regulators, respectively, and ${\int_q=\int\frac{d^{d}q}{(2\pi)^{d}}}$.
The scale-dependent flowing action ${\Gamma_k + \Delta S_k}$ is defined as the Legendre transform of the scale-dependent Schwinger functional ${W_k=\ln \mathcal Z_k}$~\cite{Dupuis:2020fhh,Wipf2021}.

The fRG evolution of $\Gamma_k$ interpolates between the microscopic action at large scale, $k\rightarrow \Lambda$, where the classical action is recovered $\Gamma_{k\rightarrow\Lambda} \simeq S $, and the full effective action at $k\rightarrow 0$, i.e., $\Gamma_{k\rightarrow0} = \Gamma $.
The evolution of the flowing action $\Gamma_k$ is governed by the Wetterich equation~\cite{Wetterich:1992yh}
\begin{equation}
    \partial_t \Gamma_k [\Phi] = \frac{1}{2} \text{STr} \left\{ \left[\Gamma_k^{(2)} [\Phi] + R_k\right]^{-1} \left(\partial_t R_k\right) \right\},
    \label{eq:WetterichEq}
\end{equation}
where the RG time is $t=\ln(k/\Lambda)$ with $\partial_t = k \frac{\partial}{\partial k}$. 
The second-order functional derivative of the effective action with respect to the fluctuating fields is given by
$
{\big(\Gamma_k^{(2)} \big)_{ab} (p,q) \equiv \frac{\overrightarrow{\delta}}{\delta \Phi_a^T(-p)} \Gamma_k \frac{\overleftarrow{\delta}}{\delta \Phi_b(q)}}\,,$
and the supertrace, $\mathrm{STr}$, integrates over momenta and sums over all fields, including a minus for the fermion sector. The arrow on top of the functional derivatives indicate the direction in which they are acting.

At finite temperature, the time-domain compactifies, and frequency integrals are replaced by Matsubara sums,
\begin{equation}
    q_0\to i\omega_n,\quad \int\frac{d^dq}{(2\pi)^d}\to   T\sum_{n\in\mathbb{Z}}\int\frac{d^{d-1}q}{(2\pi)^{d-1}}\,,
\end{equation}
where the Matsubara frequencies for the order-parameter fields are given by ${\omega_n \vert_\phi=2\pi n T}$  and for the fermionic fields by ${\omega_n \vert_\psi=2\pi (n+1/2) T}$  with $n\in\mathbb{Z}$.

\subsubsection{Effective action}

For practical computations with the Wetterich equation in Gross--Neveu--Yukawa-type models, we choose a leading-order derivative expansion for the flowing action ${\Gamma_k=\Gamma_k[\bar{\psi},\psi,\phi]}$. 
This approach is well established in the fRG literature and has been successfully applied to a wide range of systems, see the review~\cite{Dupuis:2020fhh} and references therein.
Within this approximation, our ansatz takes the form
\begin{align}
\Gamma_k =& \int_0^{1/T}d \tau \int d^{d-1}x \Big[  Z_{\psi,k} \, \bar{\psi} D_{\psi,k}  \psi  \notag\\[5pt]
		&- \frac{1}{2} Z_{\phi,k} \phi D_{\phi,k} \phi +  h_k  \phi \bar{\psi} \psi + U_k(\rho)\Big]\,,
		\label{eq:EffAction}
\end{align}
where we consider scale-dependent and uniform wave-function renormalizations $Z_{\phi,k}$ and $Z_{\psi,k}$ for bosons and fermions, respectively, and a scale-dependent Yukawa coupling~$h_k$. 
Furthermore, we include a scale-dependent effective potential $U_k$, which depends on the field invariant $\rho=\frac{1}{2}\phi^2$ and includes bosonic scatterings to all orders.
Also, we have introduced the kinetic differential operators
\begin{align}\label{eq:KineticOperatorsb}
	D_{\phi,k} &= \partial_0^2 +  v_{\phi,k}^2 \, \vec{\partial}^{\,2} \,, \\ 
	D_{\psi,k} &=   \gamma_0 \partial_0 +v_{\psi,k} \, \vec{\gamma} \cdot \vec{\partial}  \,,
    \label{eq:KineticOperatorsf}
\end{align}
with scale-dependent velocities $v_{\phi,k}$ and $v_{\psi,k}$ for the order-parameter and fermionic fields, respectively.
%

\subsubsection{Regulators}\label{sec:regulators}

In this work, we explore (1)~emergent relativistic symmetry and (2)~the finite-temperature phase diagram of the model.
To that end, we employ two different types of regulators. 
In principle, physical observables should be independent of the explicit choice of the regulator. 
However, the fact that we need to employ a truncation, as in \Eqref{eq:EffAction}, introduces an artificial regulator dependence, which we aim to minimize by appropriate choices.
Generally, it is advantageous to use a regulator scheme that respects the global symmetries of the theory.
In particular, for the study of emergent relativistic symmetry, we use a regulator that does not explicitly break relativistic symmetry at intermediate RG scales.

A class of regulators that fulfills this can be parametrized in terms of bosonic and fermionic shape functions, $r_B$ and $r_F$, respectively,
\begin{align}\label{eq:cov_regs}
   R_k^\phi(p,q) &= Z_\phi p_B^2 r_B(p_B^2/k^2) \delta(p - q), \\
   R_k^\psi(p,q) &= iZ_\psi \slashed{p}_F r_F(p_F^2/k^2) \delta(p -q).
\end{align}
Here, we have defined the rescaled momenta as ${p_{B/F} = (p_0, v_{\phi/\psi,k}\, \vec{p}^{\,} )}$ and ${\slashed{p}_F=\gamma_0p_0 + v_{\psi,k} \vec\gamma\cdot\vec{p}}$. For the explicit evaluation of the appearing momentum integrals, we choose the shape functions
\begin{align}\label{eq:litim_shape}
    r_B(x) &= \Big( \frac{1}{x} - 1 \Big) \theta(1-x),\\
    r_F(x) &= \Big( \frac{1}{\sqrt{x}} - 1 \Big) \theta(1-x).\label{eq:litim_shape2}
\end{align}
While the above regulator choice is suitable to address emergent relativistic symmetry, it is disadvantageous in other aspects. 
In particular, at finite temperature, where the relativistic symmetry is explicitly broken, we generally expect $v_{\phi,k} \neq v_{\psi,k}$. 
Then, the analytic evaluation of mixed boson-fermion loops becomes difficult due to the presence of different momentum-shells for bosons and fermions. Additionally, the temperature dependence of the couplings is non-analytical, and the determination of thermodynamic properties becomes challenging.

To circumvent these issues in the computation of flows at finite temperatures, we use spatial regulators
\begin{align}
\label{eq:Regulatorb}
	R_k^\phi(p,q) &=Z_\phi v_{\phi}^2 \, \vec{p}^{\,2}  \,  r_B (\vec{p}^{\,2}/k^2) \delta(p-q)\,, \\
	  R_k^\psi(p,q) &= i Z_\psi  v_{\psi}  \, \vec{\gamma}\cdot\vec{p}  \,  r_F (\vec{p}^{\,2}/k^2) \delta(p-q)\,.\label{eq:Regulatorf}
\end{align}
They allow us to calculate all Matsubara sums analytically before having to specify the shape functions $r_B$ and $r_F$. The remaining spatial momentum integrals can then be conveniently calculated by the use of the shape functions in Eqs.~\eqref{eq:litim_shape} and \eqref{eq:litim_shape2}. 
This choice is also advantageous for analytic continuation to complex frequencies and the computation of spectral functions.

\subsection{Flow of Fermi and order-parameter velocities}

The presence of scale-dependent velocities in the kinetic operators in Eqs.~\eqref{eq:KineticOperatorsb} and \eqref{eq:KineticOperatorsf}, requires a distinction of temporal and spatial directions in the projection of the flow equations. 
Consequently, the bosonic and fermionic two-point functions are given by ${\big(\Gamma_k^{(2)} \big)_{ab} (p,q)\!=\!\delta (p -q)\Gamma_{ab}^{(2)}(p)}$ with  $p\!=\!(p_0, \vec{p}^{\,})$ reading
\begin{align}\label{eq:two-pointFuncs}
	\Gamma_{\phi \phi}^{(2)}(p) &= Z_{\phi, \parallel} \, p_0^2 + Z_{\phi, \perp}  v_{\phi,\Lambda}^2 \,\vec{p}^{\,2} + \partial_\rho U_k(\rho) + 2 \rho\, \partial_\rho^2 U_k(\rho)\,, \notag \\
	\Gamma_{\bar \psi \psi}^{(2)}(p) &=  i Z_{\psi, \parallel} \, \gamma_0 p_0  + i   Z_{\psi, \perp}  v_{\psi,\Lambda} \, \vec{\gamma}\cdot \vec{p}  + \sqrt{2 \rho } h_k \,.
\end{align}
The wave-function renormalizations $Z_{\phi/\psi, \parallel} $ and $ Z_{\phi/\psi, \perp}$ follow directly from the projection onto frequency and spatial momenta, respectively.
The velocities of the model at the initial cutoff scale $\Lambda$ are denoted as $ v_{\phi/\psi,\Lambda}$. 

The order-parameter expectation value $\phi_{0,k} = \sqrt{2 \rho_{0,k}}$ is determined from the minimum of the effective potential,
\begin{align}\label{eq:eom}
    \partial_\rho U (\rho)|_{\rho = \rho_{0,k}}  =  0 \,.
\end{align}
In the symmetric regime the minimum sits at the origin, $\rho_{0,k}=0$, while in the SSB regime it shifts to a finite value, $\rho_{0,k}>0$.
The scale-dependent bosonic and fermionic masses are obtained from the two-point functions in \Eqref{eq:two-pointFuncs} evaluated at vanishing external momentum and at the running minimum
\begin{align}
\label{eq:BosonicMass}
    m_\phi^2=& (\partial_{\rho} U_k(\rho) + 2 \rho \, \partial^2_{ \rho} U_k(  \rho) )|_{\rho = \rho_{0,k}} \,, \\
    m_\psi=& \sqrt{2\rho_{0,k}} h_k  \,. \label{eq:FermionicMass}
\end{align}

The scale-dependent velocities in Eqs.~\eqref{eq:KineticOperatorsb} and~\eqref{eq:KineticOperatorsf}~are
\begin{align}
	 v_{\phi,k}  = v_{\phi,\Lambda} \left(\frac{Z_{\phi, \perp}}{Z_{\phi, \parallel} }\right)^{1/2}  \,,
	 \quad 	v_{\psi, k} = v_{\psi,\Lambda}\frac{Z_{\psi, \perp}}{Z_{\psi, \parallel} } \,,\label{eq:velocities}
\end{align}
and their fRG flow can be written in terms of the anomalous dimensions associated with the wave-function renormalization of the spatial and frequency directions
\begin{align}
    &\partial_t v_{\phi,k} = \frac{v_{\phi,k}}{2}\left( \eta_{\phi,\parallel} -\eta_{\phi, \perp}  \right)\,,\nonumber\\
    &\partial_t v_{\psi,k} = v_{\psi,k} \left( \eta_{\psi,\parallel} -\eta_{\psi, \perp}  \right) \,,
\end{align}
where $\eta_{i,j}\! =\! -\partial_t \ln Z_{i,j}$ with $i\!\in\!\{\psi,\phi\}$ and $ j\!\in\!\{\parallel,\perp\}$.
Details on the projection prescriptions and the full expressions of the flow equations are provided in~\Cref{app:projections,app:Regulators}. 
In the remaining text, we drop the index~$k$ to simplify the notation.

Near a QCP, the dynamical critical exponent~$z$ governs the relative scaling of characteristic time and length scales of zero-temperature fluctuations, and also controls the temperature scaling of observables at finite temperature. 
In relativistic quantum field theories, time and space appear symmetrically, implying $z=1$. 
In the presence of different $Z_{\phi/\psi, \parallel} $ and $ Z_{\phi/\psi, \perp}$, however, this symmetry is broken, and $z$ can in principle differ from one.

Consider our ansatz for the effective action in \Eqref{eq:EffAction} at zero temperature. In the presence of a general $z$, the dimensions of space and time are given by ${[\vec{x}^{\,}] = -1}$ and ${[x_0] = -z}$. 
Hence, the engineering dimensions of $\phi$ and $\psi$ are $\Delta_\phi = (d-3+z)/2$ and $\Delta_\psi = (d-2+z)/2$, respectively. 
For $z \neq 1$, the velocities become dimensionful with ${[v_{\phi}] = [v_{\psi}] = z - 1}$.
At a QCP, the flowing action is at a fixed point, implying that all dimensionless and renormalized parameters of the theory are scale-invariant. In particular, the dimensionless renormalized velocities, $k^{1 - z} v_{\phi}$ and $k^{1 - z} v_{\psi}$, become scale-invariant, implying 
\begin{align}\label{eq:DefZ}
    z &= 1 + \frac{\eta_{\phi,\parallel} - \eta_{\phi,\perp}}{2}  = 1 + \eta_{\psi,\parallel} - \eta_{\psi,\perp} \,.
\end{align}
The scaling dimensions of the fields can be read-off the two-point function at the fixed point, and are given by the standard expressions
\begin{align}
    \Delta_\phi &= \frac{d-3+z+\eta_{\phi,\parallel}}{2} \,, \quad \Delta_\psi = \frac{d-2+z+\eta_{\psi,\parallel}}{2}  \,.
\end{align}
%


\subsection{Flow of effective potential and Yukawa coupling}\label{sec:Potentialflow}
The flow of the effective potential is obtained by evaluating the Wetterich equation~\eqref{eq:WetterichEq} for a constant scalar field $\phi(x) = \phi$ and vanishing fermionic fields $\bar \psi = \psi = 0$.
For the spatial regulator in \Eqref{eq:Regulatorb} and \Eqref{eq:Regulatorf}, this yields
\begin{align}\label{eq:PotFlow}
    \partial_t U (\rho)=& A_d k^{d+1} \left[ \left(1 - \frac{\eta_{\phi,\perp}}{d+1}  \right) v_\phi^2 \,  \frac{\coth\left(\frac{ \epsilon_\phi}{2 T }\right)}{2\epsilon_\phi}\right.  \notag\\[1ex]
    &\hspace{4mm}\left.-4 N_\mathrm{f} \left(1 - \frac{\eta_{\psi,\perp}}{d}  \right) v_\psi^2 \,  \frac{\tanh\left(\frac{\epsilon_\psi}{2 T }\right)}{2 \epsilon_\psi}\right]\,,
\end{align}
with the prefactor $A_{d+1}= 2 \pi^{d/2} /(d \, \Gamma(d/2) (2 \pi)^d)$.
The dispersion relations are given by
\begin{align}\label{eq:masses}
    \epsilon_\phi &= \left( k^2 v_\phi^2 +  \partial_{\bar \rho} U(\bar \rho) + 2  \bar \rho \, \partial^2_{\bar \rho} U( \bar \rho)\right)^{1/2} \,,\notag\\[1ex]
    \epsilon_\psi &= \left( k^2 v_{\psi}^2 + 2 \bar h^2 \bar \rho \right)^{1/2} \,,
\end{align}
where the renormalized fields are defined by
\begin{align}
    \bar \phi = Z_\phi^{1/2} \phi\,, \qquad  \bar \rho = Z_\phi \rho\,.
\end{align}
Comparing the flowing action in \Eqref{eq:EffAction} with the corresponding two-point function in \Eqref{eq:two-pointFuncs}, it is evident that the global wave-function renormalization is defined by the temporal wave-function renormalization, i.e.,~${Z_{\phi}=Z_{\phi,\parallel}}$, and analogously for the fermions, i.e.,~$Z_{\psi}=Z_{\psi,\parallel}$. Throughout the manuscript, the bar on top of a parameter indicates that it is rescaled by the temporal  wave-function renormalizations $Z_{\phi/\psi,\parallel}$.

The renormalized Yukawa coupling reads
\begin{align}
	\bar h = Z_{\psi}^{-1} Z_{\phi}^{-1/2} h \,, 
\end{align}
and the renormalized scalar and fermionic masses are given by
\begin{align}\label{eq:renBosFermMasses}
\bar m_{\phi}^2:= m_{\phi, \parallel}^2 = \frac{ m_\phi^2}{Z_{\phi,\parallel}}\,, \quad\bar m_{\psi}^2:= m_{\psi,\parallel}^2 = \frac{ m_\psi^2}{Z_{\psi,\parallel}^2}   \,,
\end{align}
respectively. 

At finite temperature, critical fluctuations leading to scaling behavior are driven by fluctuations at vanishing frequencies, i.e., static fluctuations, in the bosonic sector. Therefore, we will also be interested in quantities renormalized with respect to the spatial wave-function renormalization $Z_{\phi/\psi,\perp}$. 

In the following, we refer to masses renormalized by $Z_{\phi/\psi,\perp}$ as screening masses, as they effectively describe how correlation functions decay (are screened) in space. Moreover, they are the quantities of interest when assessing critical exponents at finite temperature. We define the screening masses as
\begin{align}\label{eq:ScreeningMasses}
m_{\phi, \perp}^2 = \frac{ m_\phi^2}{Z_{\phi,\perp}v_{\phi,\Lambda}^2}\,, \quad
m_{\psi,\perp}^2 = \frac{ m_\psi^2}{Z_{\psi,\perp}^2v_{\psi,\Lambda}^2}\,.
\end{align}
For the flow of the Yukawa coupling~$h$ and the remaining flow equations, see~\Cref{app:projections}.

\subsubsection{Initial conditions} \label{sec:InitialConditions}

We need to choose a set of initial conditions at the UV scale $\Lambda$, which we identify to be close to the pertinent QCP.
For this purpose, we choose the initial Yukawa coupling $h_\Lambda$ as tuning parameter and fix the remaining quantities as follows:
The fully field-dependent order-parameter potential is chosen to be non-interacting, i.e.,
\begin{align}\label{eq:InitialPotential}
    U_\Lambda(\rho) = m_\Lambda^2 \rho, \qquad m^2_\Lambda = 1\,,
\end{align}
where the initial mass also corresponds to the UV cutoff scale $\Lambda = 1$.
In the UV limit, the wave-function renormalizations are set to recover the classical action in \Eqref{eq:lagrangian}, i.e.,~$Z_{\psi,\Lambda} = Z_{\phi,\Lambda}=1$.

We choose distinct initial bosonic and Fermi velocities $v_{\phi,\Lambda}=1$ and $v_{\psi,\Lambda} = 1.4$, respectively.
While this choice is somewhat arbitrary in the context of the present chiral Ising model, it is guided by previous studies of the chiral Heisenberg model relevant to twisted tetralayer tungsten diselenide near the quantum critical transition. 
In particular, in Ref.~\cite{lr2x-nnks}, it has been estimated that the velocity ratio at the moiré lattice scale is roughly given by $v_{\psi,\Lambda}/v_{\phi,\Lambda}\sim 1.35$.
We then use the Yukawa coupling to iteratively fine-tune the fRG flow to the quantum critical point~$h_c$.

\subsubsection{Numerical implementation of the potential}
\label{app:NumericsOfFullPotential}
%
\begin{figure}[t!]
    \includegraphics[width=\columnwidth]{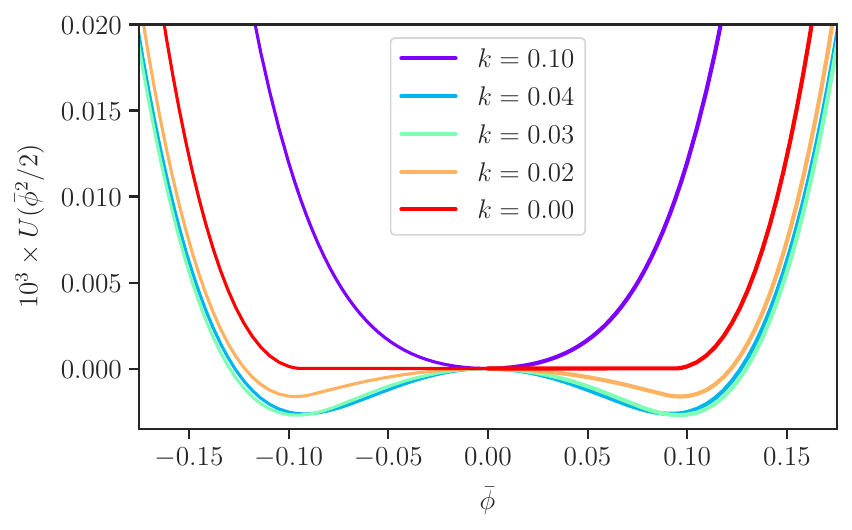}
    \includegraphics[width=\columnwidth]{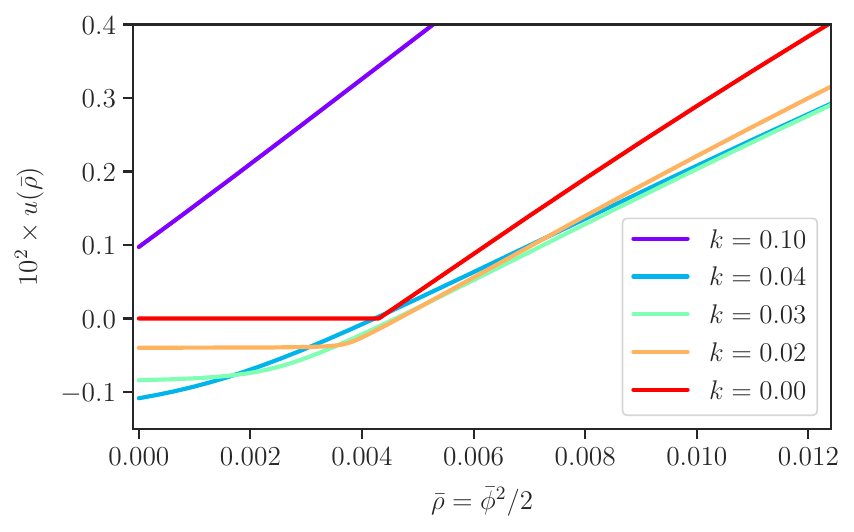}
    \caption{\textbf{RG evolution of the effective potential. Top}: Scale- and field-dependence of the potential $U_k$. \textbf{Bottom}: 
    Field derivative of the effective potential $u=\partial_{\bar{\rho}}U$ as a function of $\bar\rho$. Different RG scales~$k$ are indicated by the color scheme, at temperature $T=0$ and Yukawa coupling $h_\Lambda=2.55$. }
    \label{fig:Potential}
\end{figure}

While the RG~scale evolution of the field-independent couplings $h$, $Z_{\phi, \perp}$, $Z_{\phi, \parallel}$, $Z_{\psi, \perp}$, and $Z_{\psi, \parallel}$ corresponds to solving a set of algebraic and simple coupled ordinary differential equations, the resolution of the fully field-dependent potential requires solving a highly nonlinear partial differential equation.
In the present work, we solve this entire system of differential equations using the \texttt{DiFfRG} framework \cite{Sattler:2024ozv}, which builds on the use of fluid dynamical methods in the fRG. 
To that end, the flow of the effective potential is rewritten as a closed partial differential equation
\begin{align}\label{eq:defU}
    \partial_t u = \partial_{\bar \rho} \left( f(u(\bar \rho),u'(\bar \rho); \bar \rho, t) + \eta_{\phi, \parallel} \, \bar \rho u \right) \,,
\end{align}
where $u = \partial_{\bar \rho}U$  and the flux $f$ is defined by the flow of the potential, see the right hand side of \Eqref{eq:PotFlow}. 
From this point of view, the RG~scale evolution of $u$ can be understood as a convection-diffusion process with a highly non-linear wave-speed $c_f=\partial_u f(u,u'; \bar \rho, t)$, which measures how fast structures travel through the solution $u$ of the differential equation. 

Notably, $c_f$ becomes very large in the flat region of the potential, i.e., for field values $\bar \rho<\bar \rho_{0}$, where $\rho_{0}$ is the solution to the equation of motion~\Eqref{eq:eom}. 
This region is clearly discernible in \Cref{fig:Potential} for $k\to 0$ and a clear flattening of the potential for small field values starts around $k \approx0.03$. 
Resolving this gradual flattening requires the use of implicit time-stepping algorithms~\cite{Ihssen:2023qaq} and a fully field-dependent potential, since a simple Taylor expansion can not resolve the increasingly sharp kink indicating the end of the flat region, see e.g.~\cite{Ihssen:2024miv}, where a comparison to Taylor expansions was performed in the context of the phase structure of quantum chromodynamics. 
For further numerical developments towards the resolution of non-analytic structures in RG flows see \cite{Grossi:2019urj, Grossi:2021ksl, Stoll:2021ori, Ihssen:2022xkr,Zorbach:2024rre,Koenigstein:2025sse}. 
The occurrence of the flat region manifests the convexity of the effective action $\Gamma_k + \Delta S_k$, which is required by its definition as a Legendre transform. Physically, the convexity of $\Gamma_k + \Delta S_k$ implies a positive second derivative, i.e., positive dispersion relations \Eqref{eq:masses} and, in particular, real masses as $k\to 0$. 
In terms of the fully field-dependent potential, convexity implies that $u + 2 \bar \rho u' > -k^2 v_\phi^2$ at all $k$, which also explains the gradual flattening as the regulator insertion is successively removed by the RG time integration.

The resolution of field space is done using continuous Galerkin methods. 
For more details, we refer to Ref.~\cite{Sattler:2024ozv}.
An example of the RG scale-dependent potential in the ordered phase ($\bar \rho_0 >0$) is shown in~\Cref{fig:Potential}.

\section{Results}
\label{sec:results}

\subsection{Emergence of relativistic symmetry at $T=0$}\label{sec:emLS}

%
In Refs.~\cite{PhysRevB.89.205403,Tolosa-Simeon:2025fot}, is has been shown that in the case of $v_\psi / v_\phi = 1$, the fRG flow features a stable fixed point at zero temperature, which we identify as the underlying QCP of our model. 
Here, we establish that at zero temperature, the theory space with relativistic symmetry, in which this fixed point resides, is closed under the fRG.
Moreover, we show that small perturbations out of this subspace are irrelevant.

As explicitly discussed in \Cref{sec:model}, if $v_\psi / v_\phi = 1$, we recover a fully Lorentz-symmetric microscopic action. 
If the regulator scheme in \Eqref{eq:deltaS} respects this symmetry, then the RG trajectory generated by the flowing action~$\Gamma_k$ resides in a Lorentz-symmetric subspace for all scales~$k$, provided that $v_\psi / v_\phi = 1$ already at the ultraviolet scale. 
Hence, $v_\psi / v_\phi = 1$ must be a fixed point of the flows of $v_\phi$ and $v_\psi$.

This rather general argument can be explicitly verified within our truncation. 
In \Cref{app:der_flows_vels}, we derive the beta functions of $v_\phi$ and $v_\psi$ for the covariant regulator scheme and show that $v_\psi / v_\phi = 1$ is indeed a fixed point. 
We explicitly evaluate the appearing loop integrals for the shape functions defined in Eqs.~\eqref{eq:litim_shape} and~\eqref{eq:litim_shape2}, and show that in $d$ spacetime dimensions
\begin{align}
    \partial_t\left(\frac{v_\phi^2}{v_\psi^2}\right) = C\left( \frac{v_\phi^2}{v_\psi^2} - 1 \right) + \mathcal{O}\left(\left(\frac{v_\phi^2}{v_\psi^2} - 1\right)^2\right),
\end{align}
with constant
\begin{align}
\label{eq:ProportionalityConstant}
    C&= \eta_{\phi,\parallel}^{F}\!+\! \frac{4 A_{d+1}\tilde{h}^2v_\psi^{1-d}}{(d\!+\!1)(d\!+\!2)}\frac{2(d\!+\!1) \!-\! (1\!+\!\tilde{m}_\psi^2)\eta_{\phi,\parallel}}{(1+\tilde{m}_\phi^2)^2(1+\tilde{m}_\psi^2)^2}\,,
\end{align}
where $\tilde{m}_{\phi/\psi}^2\geq 0$ 
and $\eta_{\phi,\parallel}^{F}$ denotes only the fermionic contribution to $\eta_{\phi,\parallel}$. The expression for $\eta_{\phi,\parallel}^F$ reads
\begin{align}
    \eta_{\phi,\parallel}^F = 2 A_{d+1} N_\mathrm{f} d_\gamma \frac{\tilde{h}^2}{v_\psi^{d-1}}&\left[m_4^{(F)d}(\tilde{m}_\psi^2;\eta_{\psi,\parallel}) \right. \notag \\
    &\left.- 2 \tilde{\rho} \tilde{h}^2 m_2^{(F)d}(\tilde{m}_\psi^2;\eta_{\psi,\parallel})\right],
\end{align}
which, up to the additional factor of $v_\psi^{1-d}$, coincides with the results of previous literature, cf.  Ref.~\cite{Tolosa-Simeon:2025fot}. 
The appearing threshold functions ${m_n^{(F)d}}$ with $n=\{2,4\}$ can be found in \Cref{app:der_flows_vels}. 
We defined the dimensionless, rescaled quantities ${\tilde{m}_{\phi/\psi} = \bar{m}_{\phi/\psi} /k}$, and $\tilde{h}^2 = \bar{h}^2/k^{d-4}$, which are denoted by tildes throughout the manuscript.

At the QCP, $\tilde{m}_\psi^2 = 0$ and $\eta_{\phi,\parallel}^F = \eta_{\phi,\parallel}$ at all scales~\cite{PhysRevB.89.205403,Tolosa-Simeon:2025fot}, and positivity of $C$ is ensured, as we find $\eta_{\phi,\parallel} < 1 <  2(d+1)$. This is in agreement with the results of previous literature on closely related models~\cite{Roy:2015zna, lr2x-nnks}.
On the symmetric side of the QCP, $\bar{m}_\psi^2 = 0$ below a finite scale $k$, and hence $\tilde{m}_\psi^2 = 0$ and $C > 0$ towards the IR. In the symmetry-broken regime, $\bar{m}_\psi^2 \sim \bar{\rho}$ becomes finite in the IR, and hence, $\tilde{m}_\psi = \bar{m}_\psi / k$ grows towards the IR. At leading order in $1/\tilde{m}_\psi^2$, the proportionality constant reduces to $C = \eta_{\phi,\parallel}^F$.

In summary, this establishes the existence of a Lorentz-symmetric subspace that is closed under RG and that linear perturbations out of this subspace are irrelevant close to the QCP, i.e., relativistic symmetry is emergent. In particular, we also provide evidence for this on the side of the transition where the discrete chiral $\mathbbm{Z}_2$~symmetry is spontaneously broken, i.e., where $\bar{m}_\psi>0$.
More details on the derivation can be found in \Cref{app:der_flows_vels}. 

\subsection{Quantum critical transition}
\label{sec:QCbehavior}

We continue to investigate our model at $T=0$, but now we employ the spatial regulator in \Eqref{eq:Regulatorb} to establish the connection to  the finite-temperature study below and to benchmark with the well-understood quantum criticality.
As noted in \Cref{sec:regulators}, the spatial regulator explicitly breaks relativistic symmetry for all $k > 0$. 
This implies that the corresponding RG flow cannot have a fixed point with relativistic symmetry.
Instead, the QCP with $z = 1$ can only be reached asymptotically when $k \to 0$. 
Hence, we will explore quantum criticality in the following by solving fRG flows numerically as described in \Cref{app:NumericsOfFullPotential}.
More specifically, we fine-tune the value of the ultraviolet Yukawa coupling~$h_\Lambda$ through the continuous quantum phase transition and determine the quantum critical scaling as well as the flow of the velocities near the QCP from numerical integration of the functional RG flow equations, i.e., the flow of the potential in \Eqref{eq:PotFlow} and the equations listed in \Cref{app:projections}.

\begin{figure}[t!]
    \includegraphics[height=3.2cm]{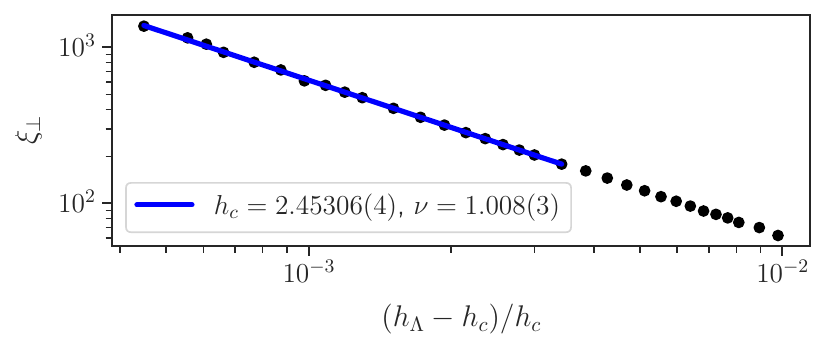}
    \includegraphics[height=3.2cm]{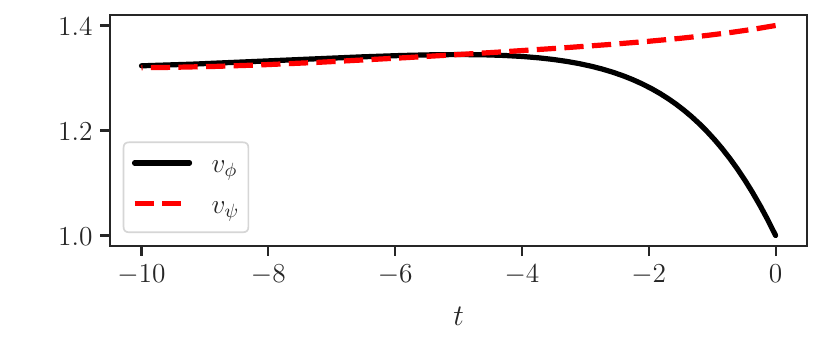}
    \caption{\textbf{Correlation length and velocity flows} from numerical integration of the functional RG flow equations. 
    \textbf{Top:}~Static correlation length as a function of the tuning parameter $\delta h = (h_\Lambda - h_c)/h_c$, used to extract the correlation length exponent via $\xi_\perp\propto \delta h^{-\nu}$.
    \textbf{Bottom:}~RG flow of the bosonic (black solid line)  and fermionic (red dashed line) velocities at the QCP.
    }
    \label{fig:qcpflow}
\end{figure}
%
\begin{table}[t!]
\centering
\begin{tabular}{l|ccc}
 \hline \hline
chiral Ising& $1/\nu$ & $\eta_\phi$ &$\eta_\psi$\\ 
 \hline
  \textit{this work} & 0.992(9)  & 0.748 & 0.031 \\
  LPA${}'$4~\cite{Tolosa-Simeon:2025fot} & 1.00  & 0.76 & 0.032 \\
  fRG (NLO) \cite{Knorr2016}  & 0.994(2) & 0.7765 & 0.0276 \\
  \hline
  $\epsilon$-exp w/ DREG3 \cite{PhysRevB.98.125109}  & 0.993(27) & 0.704(15) & 0.043(12) \\
     conformal bootstrap \cite{Erramilli2023}  & 0.998(12) & 0.7329(27) & 0.04238(11) \\
     QMC \cite{PhysRevB.108.L121112}  & 1.07(12) & 0.72(6) & 0.04(2)\\
     \hline\hline
\end{tabular}
\caption{\textbf{Chiral Ising universality} in $2+1$ spacetime dimensions for $N_{\mathrm{f}}=2$: correlation length exponent $\nu$ and anomalous dimensions $\eta_\phi$ and $\eta_\psi$ for bosons and fermions, respectively.
We compare our results with previous fRG studies using different truncation schemes, including improved local potential approximation~(LPA${}'$) and next-to-leading-order (NLO), as well as with estimates obtained from perturbative RG, conformal bootstrap, and quantum Monte Carlo (QMC) simulations. 
The results of \textit{this work} are obtained using the spatial regulator in Eqs.~\eqref{eq:Regulatorb} and \eqref{eq:Regulatorf}, and the numerical integration  techniques presented in  \Cref{app:NumericsOfFullPotential}. For the fermion anomalous dimension, we display the value obtained by the projection onto the spatial direction, $\eta_{\psi,\perp}$, as discussed in the text.}
\label{table:CriticalExponents}
\end{table}

We first determine the correlation length exponent by approaching the QCP at $h_c \approx2.45306$ from the symmetry-broken side of the transition.
We obtain the static correlation length from the curvature of the effective potential at its minimum, i.e.,
\begin{align}
    \xi_\perp = \left.\frac{1}{m_{\phi,\perp}} \right|_{\rho=\rho_0} \,,
\end{align}
with the scalar mass defined in \Eqref{eq:ScreeningMasses}. 
We then fit the divergence of the correlation length with a power law,  $\xi_\perp\propto \delta h^{-\nu}$, to the tuning parameter $\delta h=(h_\Lambda-h_c)/h_c$ measuring the distance from the QCP, which provides an estimate for the correlation length exponent $\nu$, see \Cref{fig:qcpflow}.
Approaching the QCP from the symmetry-broken side, we find $\nu_{\mathrm{QCP},\,\mathrm{SSB}}\approx 1.008(9)$, which agrees well with the estimates from previous fRG studies employing an algebraic approach solving fixed-point equations, cf. \Cref{table:CriticalExponents} and references therein.

We note that we cannot obtain a prediction for the correlation length exponent employing the numerical integration coming from the symmetric side of the QCP, because here, the scalar mass does not freeze out towards the infrared.
This is due to the leading fermionic loop diagram in the scalar anomalous dimension, where the internal fermionic lines at zero temperature do not acquire a finite fermionic mass gap and therefore continue to contribute to the flow in the infrared, see \Cref{app:SYMmassflow} for more details.
The reason for this technical shortcoming can be traced back to an insufficient momentum dependence of our approximations and will be addressed in future work.
At finite temperatures, which is the main focus of the remainder of this work, the flow of the mass freezes out and we can directly extract the correlation length, see below.

We also show the $T=0$ flow of the fermion and order-parameter velocities in \Cref{fig:qcpflow} for a parameter choice very close to the QCP.
The flows show that the velocities approach each other towards the infrared and saturate at some terminal velocity where relativistic symmetry with $v_\psi|_{t\to-\infty}=v_\phi|_{t\to-\infty}$ emerges.
The final velocity is determined by the initial conditions and this behavior is in agreement with the results reported in Ref.~\cite{Roy:2015zna, lr2x-nnks}.
\begin{figure}[t!]
    \includegraphics[width=\linewidth]{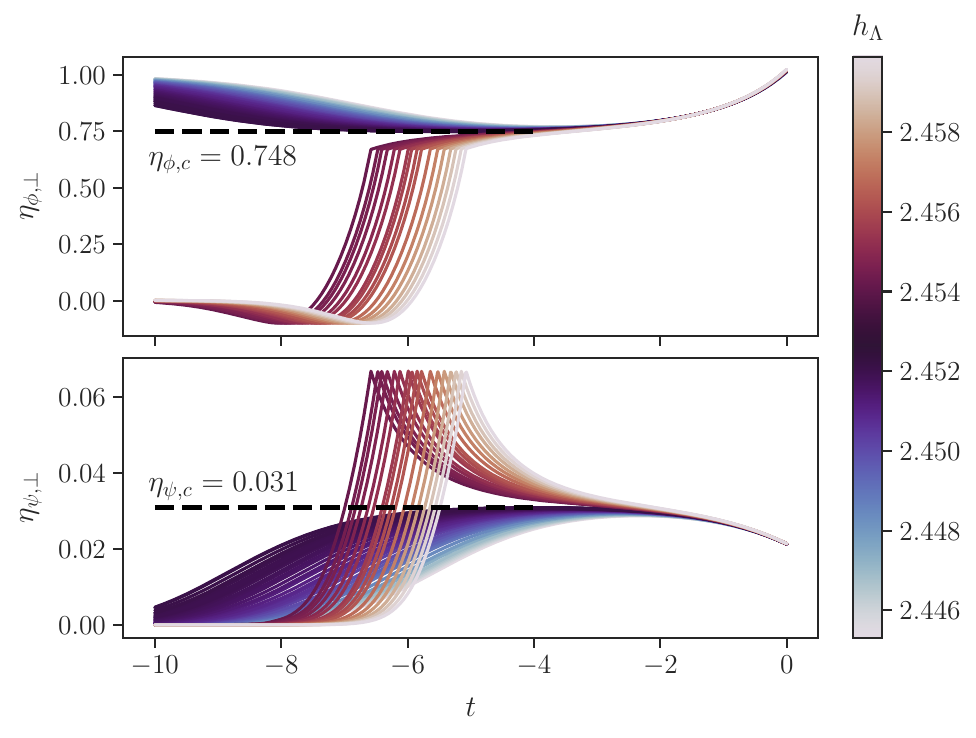}
    \caption{\textbf{Anomalous dimensions at the QCP}
    from the numerical integration of the functional RG flow equations. 
    The anomalous dimensions are extracted by projecting the wave-function renormalizations onto the spatial (perpendicular) direction.
    We show the RG time dependence of the anomalous dimensions for different initial values of the Yukawa coupling $h_\Lambda$ close to the QCP. 
    The extracted critical values of the anomalous dimensions are indicated by the dashed black lines and shown in \Cref{table:CriticalExponents}.}
    \label{fig:qcpflow2}
\end{figure}

We also track the behavior of the flows of fermion and order-parameter anomalous dimensions, $\eta_{\psi,\perp}$ and $\eta_{\phi,\perp}$, respectively, see \Cref{fig:qcpflow2}.
Approaching the QCP, the flows of the anomalous dimensions exhibit a plateau over several orders of magnitude, which indicates their value at the QCP.
We use the plateaus to numerically extract the quantum critical values shown in \Cref{table:CriticalExponents}, which are consistent with previous fRG studies.

We note that on the symmetric side of the transition, $\eta_{\phi,\perp}$ flows to a value of unity in the deep infrared, which indicates another scaling regime.
In fact, within our truncation, we find a fully infrared-attractive fixed point that is always approached on the symmetric side of the QCP, see \Cref{app:SYMmassflow} for details.
This behavior has the same origin as the missing freeze-out of the scalar mass and can be considered as an artifact of our truncation. 
At any finite~$T$, however, the flow freezes out and $\eta_{\phi,\perp}$ eventually drops to zero.

Finally, the breaking of relativistic symmetry induced by the spatial regulator results in different values of the fermion anomalous dimensions $\eta_{\psi,j},\ j\in\{\parallel,\perp\}$ at the QCP.
While $\eta_{\psi,\perp}=0.031$, see  \Cref{table:CriticalExponents}, the corresponding value for the projection onto frequencies \Eqref{eq:fermionZparallelProjection} yields $\eta_{\psi,\parallel}=0.037$. The extraction of $\eta_{\psi,\parallel}$ can be found in \Cref{app:AdditionalData}.
The discrepancy between these values is related to the absence of a fixed point and, in consequence, we find that the relation in \Eqref{eq:DefZ} is slightly violated.
In contrast, we find $\eta_{\phi,\parallel}=\eta_{\phi,\perp}$, indicating that the bosonic anomalous dimension is independent of the projection.

We conclude that overall,  our truncation and spatial regulator choice faithfully capture the well-established quantum critical behavior of the chiral Ising model upon numerical evaluation even at $T=0$.

\subsection{Finite-temperature transition}
\label{sec:classicalIsing}

In the ordered phase of the (2+1)-dimensional chiral Ising model, the $\mathbbm{Z}_2$ symmetry is spontaneously broken.
As a discrete symmetry, its spontaneous breaking is not excluded by any no-go theorem. 
In fact, we find an ordered phase up to a critical temperature $T_c$, which will generally depend on microscopic details of the system, e.g., the microscopic choices for the Yukawa coupling and the order-parameter mass~\cite{Tolosa-Simeon:2025fot}.
At finite~$T$, the fermions are always gapped due to the odd Matsubara frequencies and therefore do not contribute to the singular behavior at the phase transition.
Hence, the $\mathbbm{Z}_2$~transition at finite~$T$ can be expected to belong to the two-dimensional Ising universality class, where the critical exponents are exactly known, i.e., the  correlation length exponent $\nu_{2\mathrm{D},\,\mathrm{Ising}}=1$ and the  anomalous dimension $\eta_{2\mathrm{D},\,\mathrm{Ising}}=1/4$~\cite{PhysRev.65.117}.
\begin{figure}[t!]
    \includegraphics[width=0.96\columnwidth]{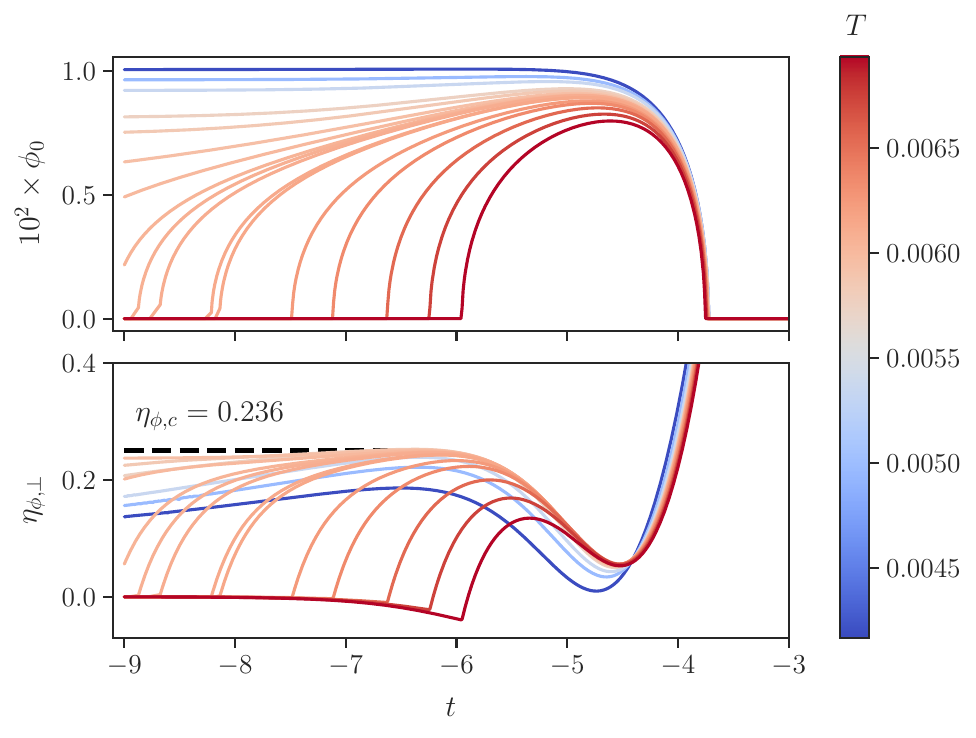}
    \includegraphics[width=\columnwidth]{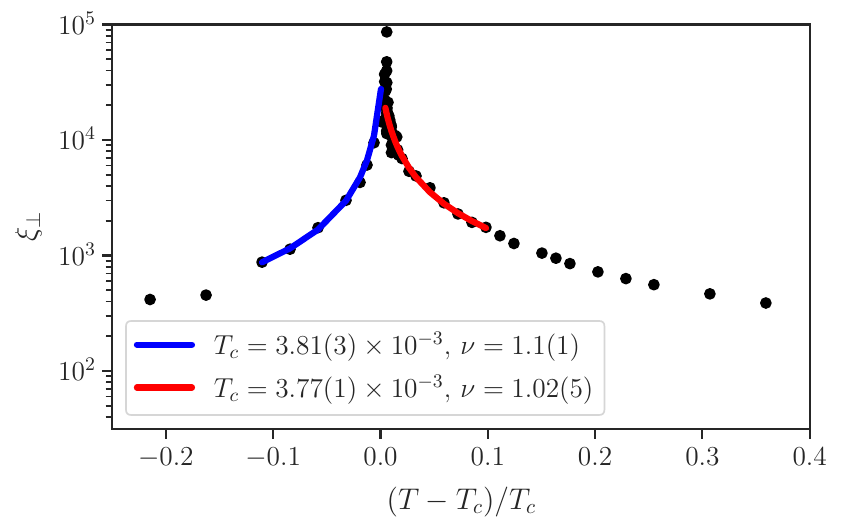}
    \caption{\textbf{Critical exponents at finite-$T$ transition.}
    \textbf{Top}: Expectation value of the order parameter $\phi_0$, c.f.~\Eqref{eq:eom}, and its corresponding anomalous dimension in the vicinity of the finite-temperature phase transition for $h_\Lambda = 2.48$ at various temperatures indicated by the color shading. The dashed line indicates the exact solution $\eta_{2\mathrm{D},\,\text{Ising}}=1/4$. \textbf{Bottom}: Fit of the critical exponent $\nu$ from the scaling of the correlation length given in~\Eqref{eq:corrLen}. The correlation length is extracted from the inverse screening mass in~\Eqref{eq:ScreeningMasses} at $t=-9$.}
    \label{fig:OPflow}
\end{figure}
\begin{figure*}[t!]
    \includegraphics[height=5.2cm]{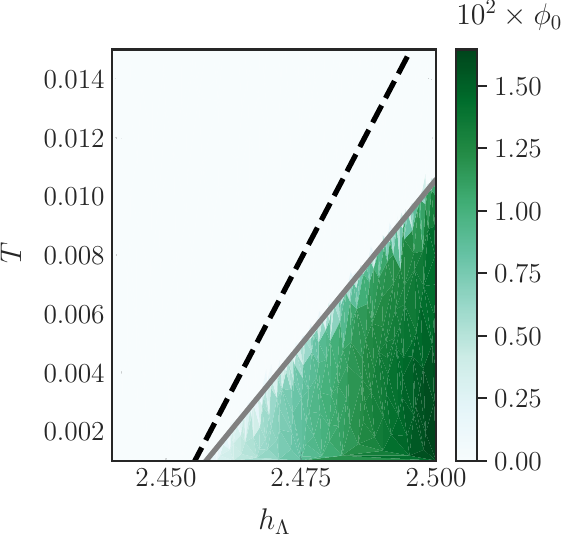}
    \hspace{3mm}
    \includegraphics[height=5.2cm]{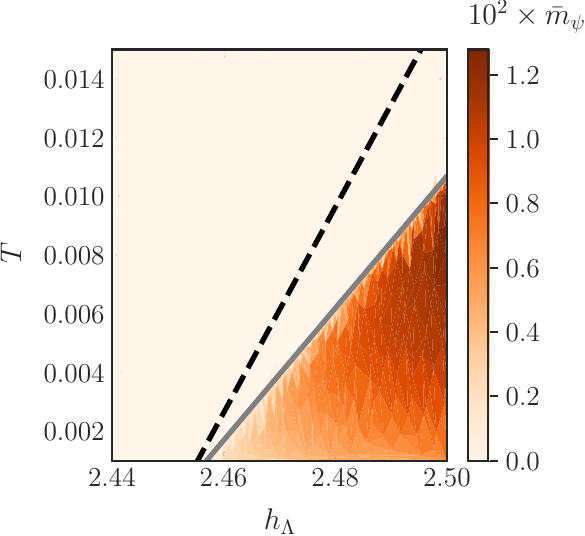}
    \hspace{3mm}
    \includegraphics[height=5.2cm]{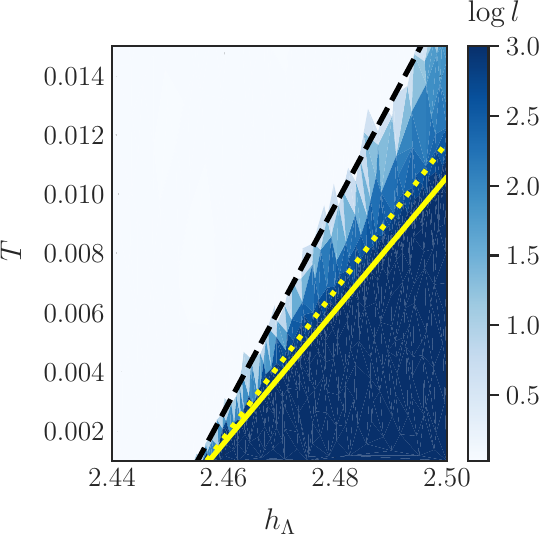}
    \caption{\textbf{Finite-temperature transition.} 
    The condensation and precondensation lines are indicated consistently across all panels by solid and dashed lines, respectively. \textbf{Left panel}: Order parameter $\phi_0$, c.f.~\Eqref{eq:eom}, as a function of the Yukawa coupling $h_\Lambda$ and temperature $T$ in the vicinity of the QCP.  
    \textbf{Middle panel}: Renormalized fermionic mass $\bar m_\psi$, c.f.~\Eqref{eq:masses},  as a function of the Yukawa coupling $h_\Lambda$ and temperature $T$ in the vicinity of the QCP. 
    The fermionic mass increases with temperature, reflecting the temperature dependence of the fermionic velocity, c.f.~\Cref{fig:ZPsi}.
    \textbf{Right panel}: 
    Domain size $l$ as a function of the Yukawa coupling $h_\Lambda$ and temperature $T$ in the vicinity of the QCP. The simulated RG time stops at $t=-9$, thus we consider all domains of size $\log(l)>3$ as fully ordered. 
    The dotted line indicates the boundary beyond which precondensation persists over more than three orders of magnitude in length scale.}
    \label{fig:OPphasediagram}
\end{figure*}

In two spatial dimensions and at finite temperature, the quantitative extraction of critical exponents is challenging, due to a breakdown of the convergence of local expansions in the effective potential.
This is hinted at by the canonical dimensionality of the operators appearing in a local expansion, i.e., the terms $\sim \phi^{2n}$ with $n\in \mathbbm{N}$. 
Indeed, in that case, the canonical dimension $[\cdot]$ of the order-parameter field reads $[\phi]=(d-2)/2$, i.e., all couplings of the operators $\sim \phi^{2n}$ are relevant with a canonical dimension of two.
To extract critical exponents in the present renormalization group setup, this observation therefore suggests that we need to take into account the renormalization of the full potential and not only a finitely truncated series expansion in $\phi^{2n}$.
Here, we can exploit the powerful numerical framework, \cite{Sattler:2024ozv} as described in \Cref{app:NumericsOfFullPotential}, which resolves the RG time evolution of the fully field-dependent potential, cf.~\Eqref{eq:PotFlow}.

To obtain the critical exponents for the finite-$T$ transition of our model, we first follow the flow of the order-parameter expectation value as a function of~$T$. 
We use the initial conditions specified below \Eqref{eq:InitialPotential} and fix the Yukawa coupling to $h_\Lambda = 2.48$, which is on the symmetry-broken side of the  transition.
We show fRG flows of the order-parameter expectation value in \Cref{fig:OPflow} for different temperatures across the phase transition.
They show that, as a function of RG time~$t$, a finite expectation value develops at some intermediate $t_{\mathrm{i}}\sim -4$ and then, if $T>T_c$, it vanishes again, before all scales are integrated out at $t \to -\infty$.
Upon approaching $T \to T_c$ from above, the expectation value remains finite for longer and longer RG times, and eventually persists for $t \to -\infty$.

This is known as precondensation~\cite{Boettcher_2012,khan2015phasediagramqc2dfunctional,Tolosa-Simeon:2025fot,Pastor-Gutierrez:2026rsy} and will be discussed in more detail below.
For numerical reasons, we stop the flow at $t_{\mathrm{IR}}=-9$. 
In the middle panel of \Cref{fig:OPflow}, we follow the fRG flow trajectories of the order-parameter anomalous dimension, approaching $T_c$. 
The value of $\eta_{\phi,c}\approx 0.236$  is approached asymptotically when $T\to T_c$ from both sides of the transition.
The correlation length, shown in the lower panel of \Cref{fig:OPflow}, develops a singularity as expected and we can fit the power law
\begin{align}\label{eq:corrLen}
    m_{\phi,\perp}\vert_{\rho = \rho_0}=\frac{1}{\xi_\perp }\propto | T - T_c |^{-\nu} \,,
\end{align}
with an exponent $\nu_{T>T_c}\approx 1.02(5)$ and $\nu_{T<T_c}\approx 1.1(1)$.  

Both our results for the bosonic anomalous dimension and the correlation length exponent lie remarkably close to the exact  exponents of the 2D Ising universality class. 
Deviations can be rationalized by referring to the fact that we employed a truncation in terms of a derivative expansion.
Together with the results at zero temperature, we have therefore established that our approximation faithfully reproduces the quantum critical behavior as well as the classical critical behavior, of the chiral Ising model with good quantitative precision.

\subsubsection{Phase diagram and precondensation}
\label{sec:phasediagram}

To gain further insight on the phase diagram, we scan a range of Yukawa interactions and identify the order-parameter expectation value as well as the fermion mass gap, \Eqref{eq:renBosFermMasses}, see the left and middle panels of \Cref{fig:OPphasediagram}, respectively.
Temperatures for which a finite expectation value vanishes at a finite scale $k_{\mathrm{precond.}}>0$ are in the symmetric regime.
As noted in the previous section, sufficiently close to the transition temperature, the system exhibits the phenomenon of precondensation, see also \Cref{fig:OPflow}.
In this regime, a finite order parameter is generated at intermediate RG scales but disappears again upon integrating out longer-wavelength fluctuations.
In that sense, precondensation precedes the formation of order in the thermodynamic limit and it signals the formation of locally ordered regions of characteristic size $\sim k_{\mathrm{precond.}}^{-d}$, analogous to the magnetic domains forming slightly above the Curie temperature in a ferromagnet. 

The inverse scale $1/k_{\mathrm{precond.}}$ can be used to define a length scale which can be interpreted as a system size beyond which longer-ranged fluctuations are cut off.
We measure this length scale in units of the ultraviolet cutoff~$\Lambda$, defining the linear domain size ${l=\Lambda/k_\mathrm{precond.}}$.
Recall that in moir\'e materials the microscopic scale roughly corresponds to ${a_\mathrm{M}\sim \mathcal{O}(10\,\mathrm{nm})}$ and the IR scale is set by the sample size ${L\sim \mathcal{O}(10\,\mu\mathrm{m})}$, i.e., it ranges over three orders of magnitude.
In the right panel of \Cref{fig:OPphasediagram}, we introduced a dotted line to show where the precondensation regime still persists after three orders of magnitude have been integrated out.
We loosely interpret this as a regime where the ordered domains exceed the size of the sample and therefore are indistinguishable from true long-range order, i.e., $L/a_\mathrm{M}\to\infty$.

\subsubsection{Quantum critical fan}
\label{sec:QCF}

Quantum critical points induce a zero-temperature power-law scaling of the correlation length with critical exponent $\nu$. Accordingly, the characteristic energy scale~$\Delta$ associated with the ground state, which, for example, can be the gap to excited states, scales as~\cite{Sachdev2011}
\begin{align}\label{eq:Gap}
    \Delta \propto |h_\Lambda-h_c|^{z\nu}\,,
\end{align}
with the dynamical critical exponent $z$ of the QCP, i.e., $z=1$ for our model.
At finite temperature, the energy scale~$\Delta$ competes with the thermal energy scale $k_B T$. Importantly, the impact of the QCP extends to the finite-$T$ region of the phase diagram, where ${k_B T \gg \Delta}$, exhibiting a QCF above the QCP. 
Therein, the thermal correlation length~\cite{Vojta_2003,Sachdev2011} also scales with~$T$ as a power law, obeying
\begin{align}
    \xi_T\propto T^{-1/z}\,.
\end{align}
To identify the QCF region, we calculate $\xi_T$ at fixed Yukawa coupling~$h_\Lambda$ as a function of temperature and determine its scaling behavior using the ansatz
\begin{align}
\label{eq:ExponentA}
    \xi_T\propto T^{-1/a} \,,
\end{align}
with exponent~$a$. 
The numerical determination of~$a$ is shown in \Cref{fig:Fan_nonUniversal}, exhibiting an extended region above the QCP with $a\approx z=1$.
In contrast to the previous fRG study~\cite{Tolosa-Simeon:2025fot} this QCF exponent is not fixed \textit{a priori} by the relativistic symmetry of the model. 
Instead, it naturally emerges above the QCP in the presence of independently renormalized Fermi and order-parameter velocities.

General scaling arguments predict that the boundaries of the quantum critical regime follow~\cite{Sachdev2011,Vojta_2003}
\begin{align}
    T_\mathrm{QCF} = A_\pm|h_\Lambda -h_c|^{z\nu}\,,
    \label{eq:QCF}
\end{align}
with the correlation length exponent $\nu$.
In \Cref{fig:Fan_nonUniversal}, we indeed observe a cusp-like QCF, whose boundary is consistent with a linear dependence, as~$z\nu\approx 1$, cf. \Cref{table:CriticalExponents}.
The transition away from this quantum critical scaling region is a smooth crossover. 
Thus, to limit the quantum critical region, we determine the non-universal prefactors $A_\pm$ by fitting to a $\pm 10 \%$ deviation from the critical value of $z = 1$. 
The numerical values for $A_\pm$ are given by $A_+ = 1.96$ and $A_- = -2.17$.

\begin{figure*}[ht!]
    \includegraphics[width=0.325\linewidth]{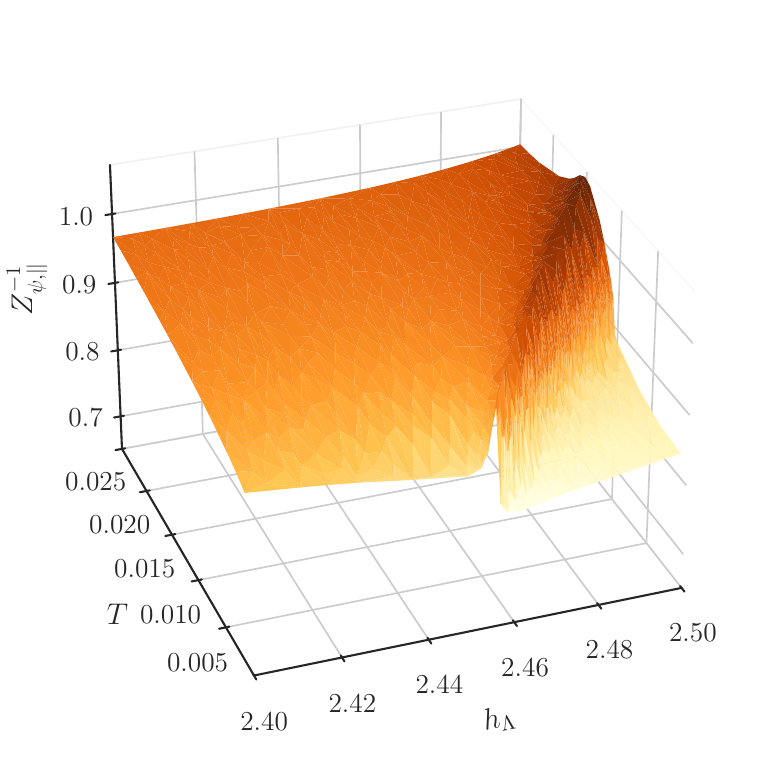}
\includegraphics[width=0.325\linewidth]{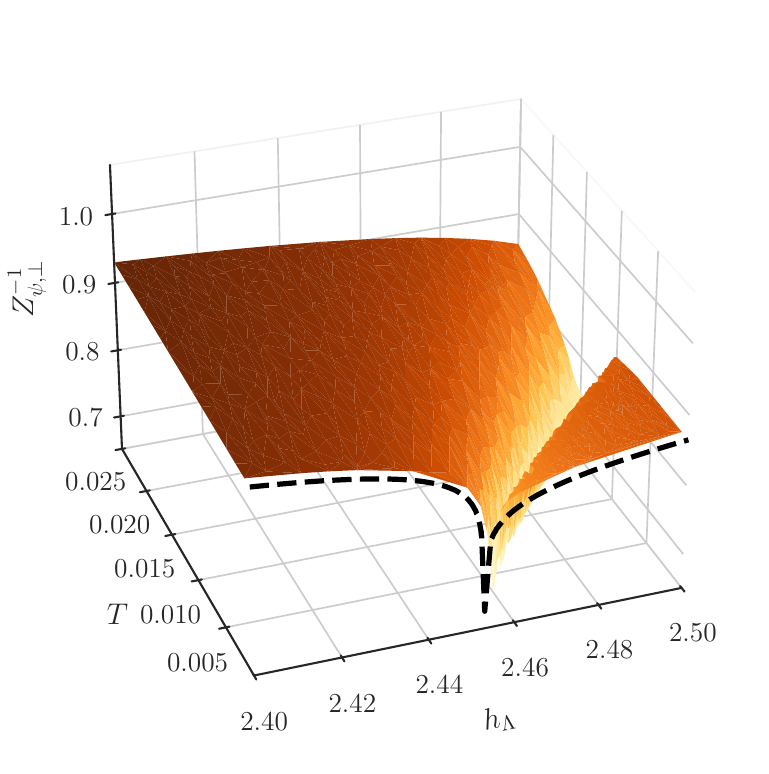}
\includegraphics[width=0.325\linewidth]{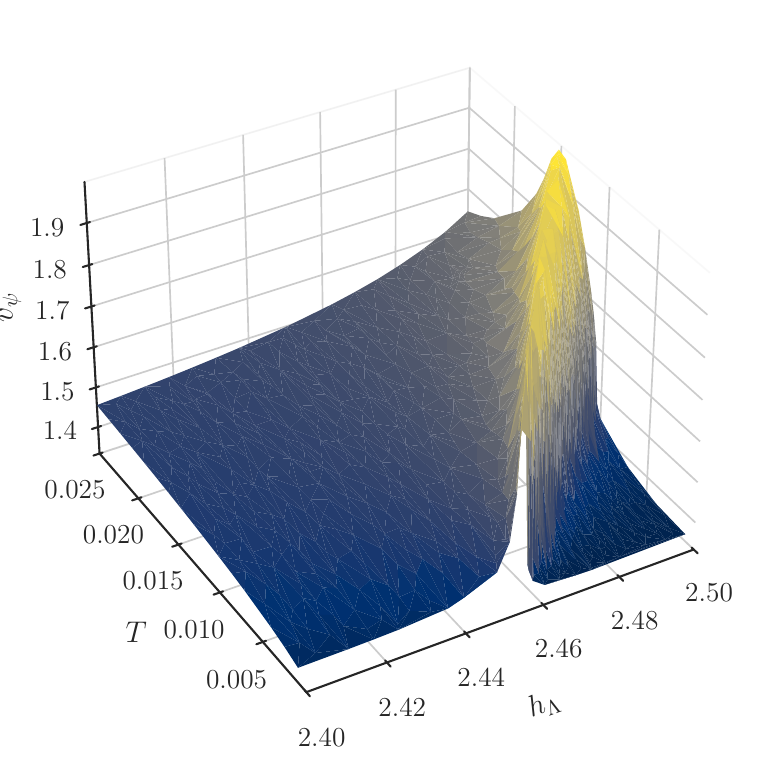}
    \caption{\textbf{Fermionic quasiparticle weight and Fermi velocity} as a function of the Yukawa coupling $h_\Lambda$ and temperature~$T$. \textbf{Left:}~Projection onto the frequency direction of the inverse of the fermionic wave-function renormalization $Z_{\psi,\parallel}^{-1}$.
    \textbf{Middle:}~Projection onto the spatial direction of the inverse of the fermionic wave-function renormalization  $Z_{\psi,\perp}^{-1}$. The corresponding $T=0$ result is indicated with a black dashed line. \textbf{Right:}~Fermi velocity $v_\psi$ as $k \to 0$.
    }
    \label{fig:ZPsi}
\end{figure*}

The present choice of spatial regulators allows for an analytical evaluation all Matsubara sums, which results in smooth physical temperature dependencies~\footnote{In previous work, cf.~Ref.~\cite{Tolosa-Simeon:2025fot}, a covariant regulator scheme was used which, at higher temperatures, exhibits discontinuities in the fRG flow as an artifact of the regulator choice.}.
We can hence continue our study towards higher temperatures, approaching the ultraviolet cutoff scale $\Lambda=1$.
In the phase diagram of \Cref{fig:Fan_nonUniversal}, we observe that the characteristic QCF scaling of the correlation length stops for $T\sim 0.2\Lambda$ and enters a non-universal regime.
In our model, this can be clearly traced back to the vicinity of the temperature to the ultraviolet cutoff.
Indeed, the scaling theory of quantum critical phenomena implies that universal behavior is cut off at high temperatures, exceeding the characteristic microscopic energy scales of the system~\cite{Vojta_2003,Sachdev2011}.
Our study therefore provides a full  account of the extent of the QCF in the direction of the tuning parameter $h_\Lambda$ as well as in temperature.

\subsubsection{Fermionic quasiparticle weight}
\label{sec:quasiparticle}

The present setup allows us to study the fermionic wave-function renormalization $Z_{\psi}$ which acts as a proxy for the quasiparticle weight. 
Its anisotropic dependence on the temporal and spatial direction is resolved minimally by projecting onto $Z_{\psi,\parallel}$ and  $Z_{\psi,\perp}$, respectively, and evaluating at $p = (\pi T, \vec{0}^{\,})$, cf.~also \Eqref{eq:two-pointFuncs}.
The infrared values of both quantities are shown in \Cref{fig:ZPsi} as a function of temperature and the tuning parameter~$h_\Lambda$.

In addition, we indicate the behavior of the fermionic wave-function renormalization~$Z_{\psi,\perp}$ at $T=0$ as a dashed line in \Cref{fig:ZPsi}. 
At $T=0$, the anisotropy, i.e., the difference between $Z_{\psi,\parallel}$ and $Z_{\psi,\perp}$, is relatively small
\begin{align}
\frac{1-Z_{\psi,\parallel}/Z_{\psi,\perp}}{Z_{\psi,\parallel}/Z_{\psi,\perp}} \lesssim 0.07 \,,
\end{align}
and can be considered an artifact of the spatial regulator.
Approaching the QCP, the fermionic quasiparticle weight vanishes, i.e., $Z_{\psi}^{-1} \to 0$ as $k\to 0$, for both projections, indicating the absence of coherent quasiparticles.

At finite temperatures, in contrast, we find that the temporal and spatial fermionic wave-function renormalizations  behave very differently.
This is because the finite temperature introduces a clear distinction of both directions: 
the spatial wave-function renormalization $Z_{\psi,\perp}$ mirrors the behavior in Ref.~\cite{Tolosa-Simeon:2025fot}, where we find a pronounced anti-symmetric suppression on both sides of the QCP and its minimum closely follows the precondensation line.
This behavior derives from the interactions with the bosonic sector in the precondensation regime that is near the finite-temperature transition. 
Here, the bosonic fluctuations predominantly transfer spatial momentum and are effectively governed by the lowest Matsubara mode, i.e., they are static in frequency, leading to the dimensional reduction and two-dimensional Ising criticality across the transition.

In comparison, the static frequency direction is slightly suppressed, which shows in the slight peak of $Z_{\psi,\parallel}^{-1}$ in \Cref{fig:ZPsi}, where $Z_{\psi,\perp}^{-1}$ clearly dips along the precondensation line. 
Together, these observations lead to an increase of the fermionic velocity $v_\psi \propto Z_{\psi,\perp}/Z_{\psi,\parallel}$ along the precondensation line.
Such behavior of the Fermi velocity in the precondensation region of a pertinent finite-temperature ordering transition could be experimentally scrutinized with measurements of the Dirac fermion's cyclotron mass~$m^\ast=\hbar\sqrt{\pi n}/v_\psi$~\cite{Novoselov:2005kj,Zhang_2005,ma2024relativisticmotttransitionstrongly}.

We note that the present setup does not show any signatures of emergent Lorentz symmetry at finite~$T$ and~${k \to 0}$. The corresponding figures are shown in \Cref{app:AdditionalData}. 
However, we observe that the ratio $v_\phi/v_\psi$ approaches unity at finite~$k$ for large parts of the phase structure, hinting at remnants of this feature in the full momentum dependence. A more thorough analysis of this is deferred to future work.

\section{Conclusions and Outlook}
\label{sec:outlook}

In this contribution, we employed a non-perturbative fRG approach to explore emergent relativistic symmetry and the phase diagram in strongly-correlated Dirac semimetals, employing the chiral Ising model with different bare Fermi and order-parameter velocities.
We carefully benchmarked with well-established results on quantum critical exponents, emergent relativistic symmetry near the quantum critical point, and classical two-dimensional Ising criticality near the finite-temperature transition.
To that end, we made use of the state-of-the-art numerical framework provided by \texttt{DiFfRG}~\cite{Sattler:2024ozv}, which allows us to resolve the fully field-dependent effective potential with fluid dynamical methods.

For the first time, we have calculated the scaling behavior and extent of the quantum critical fan and mass gaps in the symmetry-broken regime in the presence of different bare bosonic and fermionic velocities.
In total, we provide a complete unified framework that quantitatively describes the universal quantum and finite-temperature critical behavior of Gross--Neveu--Yukawa-type models as well as the non-universal regions in the phase diagram near the quantum critical point.

Our developments prepare the calculation of thermodynamic observables and spectral functions for correlated Dirac semimetals at zero and finite temperatures in the fRG approach, see, e.g., Refs.~\cite{Helmboldt:2014iya,Braun:2022mgx,Horak:2023hkp}, which we aim for in future work.
Perspectively, this will allow us to attach realistic scales, e.g., millikelvin or electronvolt, to our physical quantities and provide material-specific descriptions of phase diagrams of Dirac semimetals near a quantum critical point, e.g., for the case of the strongly-correlated Dirac material twisted tetralayer tungsten diselenide.

Moreover, we plan to investigate the impact of possible secondary many-body instabilities above the QCP through the inclusion of additional fluctuations by means of dynamical bosonization.
This will allow us to explore the possible appearance of superconductivity from incoherent quasiparticles~\cite{sgnp-ywsh,x7lc-ztqn}.

\section*{Acknowledgments}

We thank L.~Classen and 
J.~M.~Pawlowski for discussions.
F.I.~acknowledges funding by the DFG within Project-ID 277146847, SFB 1238 (project C02),
M.M.S. was funded by DFG within Project-ID 277146847, SFB 1238 (project C02), and the DFG Heisenberg programme (Project-ID 452976698).

\begingroup
\appendix
\allowdisplaybreaks

\section{Projection of flow equations}\label{app:projections}

We employ the Fourier transforms
\begin{align} \label{eq:FourierTransforms}
    \phi(x) =& \int_p \phi(p) e^{ipx}\,,\\  \psi(x) =& \int_p \psi(p) e^{ipx}\,,\qquad   \bar\psi(x) = \int_p \bar\psi(p) e^{ipx}\,.
\end{align}
With these conventions, the effective average action in \Eqref{eq:lagrangian} takes the form 
\begin{align}
        \Gamma_k =& i Z_\psi \int_{p} \bar{\psi}(-p) \slashed{p}_F \psi(p)  + \frac{Z_\phi}{2} \int_p \phi(-p) p_B^2 \phi(p) \notag \\[1ex] 
    &  + h \int_{p,q} \phi(p-q) \bar{\psi}(-p)\psi(q) + \int_x U[\phi] \,.
    \label{eq:EffectiveActionMomentum}
\end{align}
With the momentum-space action written in this form, the projection prescriptions for the flow equations of the running couplings can be introduced straightforwardly.

The general projection strategy in the derivative expansion is to consider the Wetterich equation~\eqref{eq:WetterichEq} of the $n$-point function of interest evaluated at a constant background field configuration with constant order-parameter field and vanishing fermionic fields, denoted by $\Phi(x) = (\phi(x), \psi(x), \bar \psi^T(x)) \to \Phi_0 = (\phi ,0,0)$ in the super-field notation. In a second step, the momentum dependence of the correlation functions is obtained by considering momentum derivatives at this specific field configuration and the lowest momentum configuration.

We have derived the flow of the fully field-dependent potential $U(\rho)$ in \Cref{sec:Potentialflow}. In the present Section, we state the remaining flow equations for completeness. The functional derivatives of the Wetterich equation~\eqref{eq:WetterichEq} have been derived using \texttt{FunKit} \cite{Sattler:2026csm}.

\subsection{Flow of the Yukawa coupling}
    
The flow of the Yukawa coupling can be evaluated from the projection of the Wetterich flow onto the three point vertex $\Gamma_{\phi \bar \psi  \psi }$ at the minimal external fermionic momentum $ p_{\mathrm{ex}}=( p_{\mathrm{ex},0}, \vec{0}^{\,})$ with $p_{\mathrm{ex},0}=\pi T$ and vanishing bosonic external momentum
\begin{subequations}
\begin{align}\label{eq:YukawaProjection2}
		\partial_t h =\left. \frac{1}{d_\gamma N_\mathrm{f} }\Tr\left[
		 \frac{\delta}{\delta  \phi(p-q)} \frac{ \overrightarrow{\delta}}{\delta \bar\psi(-p)} \partial_t\Gamma_k \frac{ \overleftarrow{\delta}}{\delta \psi(q)}\right]\right\rvert_{\tiny\begin{array}{l}
p=q=p_\mathrm{ex} \\
\Phi(x)= \Phi_0
\end{array}}
\end{align}
where $d_\gamma = 4$ and $N_\mathrm{f}=2$. 

Alternatively, one can evaluate the Yukawa coupling from the fermionic two-point function as
\begin{align}\label{eq:YukawaProjection}
	\partial_t h =\frac{1}{d_\gamma N_\mathrm{f} \phi}\left. 
	 \Tr \left[ \frac{ \overrightarrow{\delta}}{\delta \bar\psi(-p)} \partial_t\Gamma_k \frac{ \overleftarrow{\delta}}{\delta \psi(p)}\right]\right\rvert_{\scriptsize\begin{array}{l}
p=p_\mathrm{ex} \\
\Phi(x)= \Phi_0
\end{array}}\,.
\end{align}
\end{subequations}
This choice corresponds to a projection onto a Goldstone mode and the subsequent limit $N \to 1$, see also~\cite{Pawlowski:2014zaa}. We opt for this projection in the present work to remain consistent with the projection of the anomalous dimension $\eta_\phi$ in \Cref{sec:anomalousdim}. The latter is usually projected in this way as it produces quantitatively better results \cite{Codello:2012sc}. Note that numerically \Eqref{eq:YukawaProjection} may contain a non-vanishing imaginary part from the threshold function \Eqref{eq:FB11}, which is remedied by considering the real part of the flow.
	
With \Eqref{eq:YukawaProjection}, we find the flow of the renormalized Yukawa coupling 
\begin{align}\label{eq:YukawaFlow}
	\partial_t \bar h =& \left(\frac{\eta_{\phi,\parallel}}{2} + \eta_{\psi,\parallel} \right)\bar h  +  2 A_d k^{d+1} \bar h^3 \notag \\[1ex] 
    & \times\left\{v_{\psi}^2  \left(1-\frac{\eta_{\psi,\perp}}{d} \right)  	\mathcal{FB}_{21}(\epsilon_\psi, \epsilon_\phi ; p_{\mathrm{ex},0} ) \right.\notag \\[1ex] 
	& \left.+  v_{\phi}^2  \left(1-\frac{\eta_{\phi,\perp}}{d+1} \right)  	\mathcal{FB}_{12}(\epsilon_\psi, \epsilon_\phi  ; p_{\mathrm{ex},0})\right\} \,,
\end{align}
with the threshold functions $\mathcal{FB}_{21}$ and $\mathcal{FB}_{12}$ indicated in \Cref{app:Regulators}. As a general feature, the renormalized quantities are proportional to the direction parallel to the heat bath, i.e., the projection onto frequencies, whereas the loops are corrected with the spatial projection of the wave-function $\eta_\perp$. This is due to the current (spatial) regulator choice.

\subsection{Flow of the fermionic anomalous dimensions}

The spatial momentum and frequency dependence of the fermionic two-point function is described by the wave-functions parallel and perpendicular to the heat-bath as defined in \Eqref{eq:two-pointFuncs}. 
Thus in the projection procedure, we take derivatives with respect to the corresponding external momentum $p=(p_0,\vec{p}^{\,})$. 

Then the spatial fermionic anomalous dimension is obtained from
\begin{align}\label{eq:fermionAnomDimProjection}
	&\eta_{\psi,\perp}  =\frac{i}{2 d_\gamma N_\mathrm{f} Z_{\psi, \perp} v_{\psi,\Lambda}} \notag \\[1ex]
	&\hspace{2mm}\times \left. \frac{\partial^2}{\partial |\vec{p}\,|^2} \Tr\left[ \vec{\gamma}\cdot\vec{p}  \frac{ \overrightarrow{\delta}}{\delta \bar\psi(-p)} \partial_t\Gamma_k \frac{ \overleftarrow{\delta}}{\delta \psi(p)}\right]\right|_{\tiny
		\begin{array}{l}
			\Phi(x)= \Phi_0 \\ 
			p = p_{\mathrm{ex}}
		\end{array}}
	\,,
\end{align}
yielding
	\begin{align}
		\eta_{\psi,\perp} =& 2 A_d k^{d+1} \bar h^2 v_{\phi}^2 \left(1-\frac{\eta_{\phi,\perp}}{d} \right)  	\mathcal{FB}_{12}(\epsilon_\psi, \epsilon_\phi ; p_{\mathrm{ex},0} ) \,.
	\end{align}
	Similarly, we obtain the projection parallel to the heat bath from a frequency derivative of the fermionic two-point function
	\begin{align}\label{eq:fermionZparallelProjection}
	\eta_{\psi, \parallel } =&\frac{i}{ d_\gamma N_\mathrm{f}  Z_{\psi, \parallel } }  \notag \\[1ex]
		& \hspace{-5mm}\left. \times \frac{\partial}{\partial p_0} \Tr\left[  \gamma_0  \frac{ \overrightarrow{\delta}}{\delta \bar\psi(-p)} \partial_t\Gamma_k \frac{ \overleftarrow{\delta}}{\delta \psi(p)}\right]\right|_{\tiny
			\begin{array}{l}
				\Phi(x)= \Phi_0 \\ 
				p = p_{\mathrm{ex}}
		\end{array}}\,,
	\end{align}
	which evaluates to
	\begin{align}
		&\eta_{\psi, \parallel }= -2 A_d k^{d+1} \bar h^2  \left\{ \left(1-\frac{\eta_{\psi,\perp}}{d} \right)  v^2_{\psi} \right. \notag \\[1ex] 
		 &\hspace{2mm} \times\left[3	\mathcal{FB}_{21}(\epsilon_\psi, \epsilon_\phi ; p_{\mathrm{ex},0} ) -4 \epsilon_\psi^2 \mathcal{FB}_{31}(\epsilon_\psi, \epsilon_\phi ; p_{\mathrm{ex},0} ) \right]
		\notag \\[1ex] 
		& + \left(1-\frac{\eta_{\phi,\perp}}{d+1} \right)  v^2_{\phi} \notag \\[1ex] 
		&\hspace{3mm} \left.\times \left[ \mathcal{FB}_{12}(\epsilon_\psi, \epsilon_\phi ; p_{\mathrm{ex},0} )  -2 \epsilon_\psi^2	\mathcal{FB}_{22}(\epsilon_\psi, \epsilon_\phi ; p_{\mathrm{ex},0} ) \right]\right\}\
		 \,.
	\end{align}
Again the threshold functions $\mathcal{FB}_{nm}$ are derived in \Cref{app:Regulators}.

\subsection{Flow of the bosonic anomalous dimensions}
\label{sec:anomalousdim}

For the computation of the bosonic two-point function, we use a projection onto the Goldstone mode of a $\mathrm{O}(N)$ type theory and subsequently take the limit $N\to 1$. This trick is commonly applied in $\mathbb{Z}_2$-type theories, as it yields quantitatively better results for the scaling exponents: For example in $d=2$, the direct projection onto the scalar mode yields a critical anomalous dimension of $\eta_{\phi, c} = 0.436$ \cite{Codello:2012sc}, whereas the modified projection procedure captures the exact solution $\eta_\phi = 0.25$ much better, see e.g., \Cref{fig:OPflow} for the result of the present work.
On a technical level, the improved quantitative accuracy is explained by the fact, that a direct projection onto the bosonic two-point function contains contamination from higher-derivative terms $\propto (\partial_\mu \rho)^2$ which are not accurately accounted for in LPA'.

Note that the chiral Ising model does not contain Goldstone modes.
Therefore, to avoid the projection onto the radial mode, we promote the bosonic field to an $N$-component field and take the limit $N\to1$ after performing the projection~\cite{Classen:2016}.
Now, the flow of the anomalous dimension is obtained by projecting onto the momentum dependence of the Goldstone mode $\phi_G$ and evaluating at zero bosonic external momentum 
\begin{align}\label{eq:bosonAnomDimProjection}
	&\eta_{\phi,\perp}  =\lim_{N\to1}\left[-\frac{1}{2 Z_{\phi, \perp} v_{\phi, \Lambda}^2} \right.\notag\\
    &\quad\times\left.\left. \frac{\partial^2}{\partial |\vec{p}^{\,}|^2} \Tr\left[\frac{\delta^2 \partial_t \Gamma_k}{\delta \phi_G(-p) \delta \phi_G (p)}\right]\right|_{\tiny
		\begin{array}{l}
			\Phi(x)= \Phi_0 \\ 
				p = 0
	\end{array}}\right].
\end{align}
It is important to note that the fermionic loop which contributes to the flow \Eqref{eq:bosonAnomDimProjection} is plagued by artifacts linked to the parametrization of the fermionic regulator \Eqref{eq:Regulatorf}. The modifications to the computation of this specific diagram are outlined in \Cref{app:loopIntegrals}, where we also comment on the integration of the momentum loops in general. Importantly, this artifact is not linked to the effect discussed in \Cref{app:SYMmassflow}, which also occurs for different regulator schemes. 

The flow evaluates to
	\begin{align}
		\eta_{\phi,\perp} =&  A_d k^{d+1} \left\{4  \bar \rho \left(\partial_{\bar \rho} ^2 U \right)^2 v_\phi^2  	\mathcal{BB}_{22}(\epsilon_\phi, \epsilon_\phi'  )\right.
		\notag \\[1ex]
		& \left.
		 -8 \bar h^2 \frac{v_\psi^4}{v_\phi^2}  \left[ 
		\Big(\frac{3}{2} -\eta_{\psi,\perp} \Big)\frac{\mathcal{F}_{2}(\epsilon_\psi)}{k^2  v_\psi^2}  \right.\right.
		\notag \\[1ex]
		& \left.\left. -
		(4- 2\eta_{\psi,\perp}) \mathcal{F}_{3}(\epsilon_\psi )
		 +  4  \bar{m}_{\psi}^2 	\mathcal{F}_{4}(\epsilon_\psi )
		\right]
		\right\} 
		 \,,
	\end{align}
    where the dispersion relation of the fictitious Goldstone mode reads 
    \begin{align}
        \epsilon_\phi' = \left( k^2 v_\phi^2 + \partial_{\bar \rho} U(\bar \rho ) \right)^{1/2}\,,
    \end{align}
    and the mixed bosonic threshold function $\mathcal{BB}_{22}$ is derived in \Cref{app:Regulators}.

    Similarly, the parallel anomalous dimension can be derived by a projection onto the lowest bosonic frequency using the same projection onto $\phi_G$ to remain consistent with \Eqref{eq:bosonAnomDimProjection},
	\begin{align}\label{eq:bosonZparallelProjection}
		\eta_{\phi,\parallel}=&\lim_{N\to1}\left[- \frac{1}{2  Z_{\phi, \parallel } } \left. \frac{\partial^2}{\partial p_0^2} \Tr\left(  \frac{\delta^2 \partial_t \Gamma_k}{\delta \phi_G(-p) \delta \phi_G (p)}\right)\right|_{\tiny
			\begin{array}{l}
				\Phi(x)= \Phi_0 \\ 
				p = p_{\mathrm{ex}}
		\end{array}}\right]\,.
	\end{align}
	After tracing this expression evaluates to
	\begin{align}
		&\eta_{\phi,\parallel}= -16 A_d k^{d+1} \left(1-\frac{\eta_{\psi,\perp}}{d} \right) \bar h^2 v^2_\psi  \notag \\[1ex] 
		&  \times\left[  16 \bar m_\psi^2 \epsilon_\psi^2	\mathcal{F}_{5}(\epsilon_\psi )-12  \bar m_\psi^2 \mathcal{F}_{4}(\epsilon_\psi) -	\mathcal{F}_{3}(\epsilon_\psi )  \right]
		\notag \\[1ex] 
		& - 8 A_d k^{d+1}    \left(1-\frac{\eta_{\phi,\perp}}{d+1} \right)  \bar \rho \left(\partial_{\bar \rho} ^2 U\right)^2 v^2_\phi   \left[3	\mathcal{BB}_{22}( \epsilon_\phi) \right.
		\notag \\[1ex] 
		&  \left.  -2 \left(\epsilon_\phi^2 \mathcal{BB}_{32}( \epsilon_\phi,\epsilon_\phi' )+ \epsilon_\phi'^2 \mathcal{BB}_{23}( \epsilon_\phi,\epsilon_\phi' )\right)	 \right]
		\,.
	\end{align}

\section{Regulators and threshold functions} \label{app:Regulators}

This appendix gathers technical details on the evaluation of the loop-integrals and Matsubara sums of the Wetterich equation.

\subsection{Matsubara summation}\label{app:Matsubara}

The summation over all frequencies in \Eqref{eq:EffAction} can be done analytically when using a spatial regulator. To retain meaningful expressions it is useful to cast the flows in a standard form.
To this aim, we consider the scalar parts of the propagators 
\begin{align}\label{eq:propagators}
	G_{B,n}(M^2(\vec{p}^{\,}))&= \frac{1}{  \omega_{\phi, n}^2 + M^2(\vec{p}^{\,})} \,, \notag  \\[1ex]
	G_{F,n}(M^2(\vec{p}^{\,}))&= \frac{1}{ \omega_{\psi, n} ^2+M^2(\vec{p}^{\,})} \,,
\end{align}
with the respective bosonic and fermionic Matsubara frequencies $ \omega_{\phi, n} = 2 n \pi T$ and $ \omega_{\psi, n} =( 2 n + 1) \pi T$  with ${n \in \mathbb{Z}}$. The momentum dependent dispersion relations for the different particle species are given by
\begin{align}\label{eq:Dispersion}
	M^2_\psi(\vec{p}^{\,}) &= v_\psi^2 \left(|\vec{p}^{\,}| + R_k^\psi(\vec{p}^{\,})\right)^2  + \bar  m^2_\psi \,, \notag \\[1ex]
	M^2_\phi(\vec{p}^{\,}) &=  v_\phi^2 \left(\vec{p}^{\,2} + R_k^\phi(\vec{p}^{\,})\right)  +\bar  m^2_\phi  \,,
\end{align}
and contain the full regulator dependence.
When inserting the Litim regulator-shape function, see Eqs.~\eqref{eq:litim_shape} and \eqref{eq:litim_shape2}, the dispersion reduce to \Eqref{eq:masses} and we use the scale-dependent velocities defined in \Eqref{eq:velocities}.

Using \Eqref{eq:propagators}, the flow equations can be rewritten as sums of different products of these scalar propagators such that the only frequency dependence is contained in $G_{F,n}$ or $G_{B,n}$.  It remains to compute the sums for the single particle species loops
\begin{subequations}\label{eq:ThermalTHRS}
    \begin{align}
	s\mathcal{F}_{n}(M_\psi(\vec{p}^{\,})) &= T \sum_i (G_{F,i}(M_\psi(\vec{p}^{\,}))^n \,, \notag  \\[1ex]
	s\mathcal{B}_{n}(M_\phi(\vec{p}^{\,})) &=  T \sum_i (G_{B,i}(M_\phi(\vec{p}^{\,})))^n   \,, 
\end{align}
as well as the mixed particle loops,
\begin{align}
	s\mathcal{FB}_{nm}(M_\psi(\vec{p}^{\,}),M_\phi(\vec{p}^{\,})) &=  \notag  \\[1ex]
    &\hspace{-2.5cm}T \sum_i (G_{F,i}(M_\psi(\vec{p}^{\,})))^n  (G_{B,i}(M_\phi(\vec{p}^{\,})))^m  \,, \notag  \\[1ex]
    s\mathcal{BB}_{nm}(M_\phi(\vec{p}^{\,}),M_\phi'(\vec{p}^{\,})) &=  \notag  \\[1ex]
    &\hspace{-2.5cm}T \sum_i (G_{B,i}(M_\phi(\vec{p}^{\,})))^n (G_{B,i}(M_\phi'(\vec{p}^{\,})))^m  \,,
\end{align}
\end{subequations}
where the first expression corresponds to mixed fermionic/bosonic loops and the second one considers different bosonic particle species such as the massive and Goldstone modes.

The task of solving the Matsubara sums is simplified by using the identities
\begin{align}
	 s\mathcal{F}_{n}(\vec{p}^{\,})  =  -\frac{1}{(n-1)}  \partial_{\bar m^2_\psi}  (s\mathcal{F}_{n-1}(\vec{p}^{\,}) ) \,, \notag \\[1ex]
	 s\mathcal{B}_{n}(\vec{p}^{\,})  =  -\frac{1}{(n-1)}  \partial_{\bar m^2_\phi}  (s\mathcal{B}_{n-1}(\vec{p}^{\,}) ) \,, 
\end{align}
which are derived from the structure of the scalar propagators. A similar identity holds to increase the indices $n$ and $m$ in the mixed loops $s\mathcal{FB}_{nm}$ and $s\mathcal{BB}_{nm}$.

Finally, we indicate the expressions for the lowest sums for completeness
\begin{align}
	\mathcal{F}_{1}(M_\psi(\vec{p}^{\,}))  &= \frac{\tanh(\frac{  M_\psi(\vec{p}^{\,})}{2 T })}{2 M_\psi(\vec{p}^{\,})} \,, \notag\\[1ex]
	\mathcal{B}_{1}(M_\phi(\vec{p}^{\,}))  &= \frac{\coth(\frac{  M_\phi(\vec{p}^{\,})}{2 T })}{2 M_\phi(\vec{p}^{\,})} \,, \notag\\[1ex]
\end{align}
and 
\begin{widetext}
	\begin{align}\label{eq:FB11}
		\mathcal{FB}_{11}(M_\psi(\vec{p}^{\,}),M_\phi(\vec{p}^{\,}); p_{\mathrm{ex},0}) &= 
		-\frac{\coth(\frac{  M_\phi(\vec{p}^{\,})}{2 T }) }{2 M_\phi(\vec{p}^{\,})}  \frac{M_\phi(\vec{p}^{\,})^2 -M_\psi(\vec{p}^{\,})^2  - p_{\mathrm{ex},0}^2}{M_\phi(\vec{p}^{\,})^4 + \left[M_\psi(\vec{p}^{\,})^2 +p_{\mathrm{ex},0}^2\right]^2 - 2 M_\phi(\vec{p}^{\,})^2 \left[M_\psi(\vec{p}^{\,})^2- p_{\mathrm{ex},0}^2\right] } \notag \\[1ex]
		&\hspace{-35mm}
		+
		\frac{\coth(\frac{ M_\psi(\vec{p}^{\,})  - i  p_{\mathrm{ex},0}}{2 T })\left[ M_\phi(\vec{p}^{\,})^2 -( i p_{\mathrm{ex},0} +M_\psi(\vec{p}^{\,}) )^2\right]+\coth(\frac{ M_\psi(\vec{p}^{\,})  + i  p_{\mathrm{ex},0}}{2 T })\left[ M_\phi(\vec{p}^{\,})^2 -( i p_{\mathrm{ex},0} -M_\psi(\vec{p}^{\,}) )^2\right]}{4 M_\psi(\vec{p}^{\,}) \left( M_\phi(\vec{p}^{\,})^4 +\left[ p_{\mathrm{ex},0}^2 + M_\psi(\vec{p}^{\,})^2 \right]^2+2 M_\phi(\vec{p}^{\,})^2 \left[  p_{\mathrm{ex},0}^2 - M_\psi(\vec{p}^{\,})^2 \right]\right)}
		\,,
	\end{align}
as well as 
\begin{align}
    \mathcal{BB}_{11}(M_\phi(\vec{p}^{\,}),M_\phi'(\vec{p}^{\,})) &= \frac{1}{ (M_\phi^2(\vec{p}^{\,}) - M_\phi'^2(\vec{p}^{\,}))}\left[ \frac{\coth(\frac{M_\phi'(\vec{p}^{\,})}{2 T})}{2 M_\phi'(\vec{p}^{\,})}-\frac{\coth(\frac{M_\phi(\vec{p}^{\,})}{2 T})}{2 M_\phi(\vec{p}^{\,})}\right]\,.
\end{align}
\end{widetext}
Again, we reiterate that for the present choice of regulator all momentum dependencies vanish from these expressions. This is discussed in the following subsection.

\subsection{Loop integration}\label{app:loopIntegrals}

In this work, we use the spatial flat (or Litim) regulator~\cite{Litim:2001fd,Pawlowski:2015mlf}, defined in the main text by the shape functions in Eqs.~\eqref{eq:litim_shape} and \eqref{eq:litim_shape2}.
With this choice all momentum loops can be performed analytically as the momentum dependence in the dispersion relations \Eqref{eq:Dispersion} drops out. 
Consequently, the momentum dependence in the threshold functions is also removed and the loop integrals remain as purely polynomial integrals, with the exception of the fermionic loop contributing to the bosonic anomalous dimension \Eqref{eq:bosonAnomDimProjection} which we discuss in the following:

In specific dimensionalities the fermionic loop features a divergence when using the Litim shape functions defined in Eqs.~\eqref{eq:litim_shape} and \eqref{eq:litim_shape2}: It occurs in $d=2$ in the case of a covariant regularization scheme, see, e.g.~\Eqref{eq:m4m4}, or previous works \cite{Scherer:2013many,Tolosa-Simeon:2025fot}, and in $d=3$ for spatial regularization schemes.
This contribution is generated by hitting the fermionic regulator function with the momentum-derivative
\begin{align}\label{eq:naiveDiv}
    \eta_{\phi,\perp} \propto \int_{\mathbb{R}^2} d^2 \vec q \,   f(\vec q^{\,}) \frac{\partial^2}{\partial |\vec p^{\,}|^2}R_k^\psi ( \vec p + \vec q^{\,}) + \dots \,,
\end{align}
where $``\dots"$ denotes the remaining, integrable $\vec p$ dependent contributions to the flow and $f(0) \neq 0$ corresponds to the remaining contributions in the loop.
Consequently, after hitting the square-root in \Eqref{eq:litim_shape2} twice with the $|\vec p^{\,}|$-derivative and then setting $\vec p=0$, the integrand scales with $|q|^{-1}$, and thus induces a log-divergence in the integral.
At closer inspection, this divergence is caused by a non-analyticity in the momentum dependence of $\partial_t \Gamma_{\phi\phi}^{(2)}(p)$, which is in turn linked to the regulator insertion. 
For the calculation of the present work, this term -- which does not appear, e.g.,  in the covariant (or sharp) regularization scheme -- is identified as spurious and  dropped. 
The resulting flow coincides with \Eqref{eq:c36} up to the threshold functions.

\section{Derivation of velocity flows for covariant regulators}\label{app:der_flows_vels}

For convenience, we adopt index notation in this appendix, instead of the vector notation used throughout the rest of the manuscript.

\subsection{RG flow of the bosonic and Fermi velocities}

The RG flow of the bosonic and Fermi velocities are defined in terms of the scale-dependent wave-function renormalizations as
\begin{align}
    \partial_t v_{\phi}^2 &= \frac{1}{Z_{\phi}} \left[\partial_t (Z_{\phi} v_{\phi}^2) - v_\phi^2 \partial_t Z_{\phi} \right],  \notag\\[1ex]
    &= v_\phi^2 (\eta_{\phi, \perp} - \eta_{\phi, \parallel})  \\[1ex]
    \partial_t v_{\psi} &= \frac{1}{Z_{\psi}} \left[ \partial_t (Z_{\psi} v_{\psi}) - v_{\psi} \partial_t Z_{\psi} \right]\notag\\[1ex]
    &= v_\psi (\eta_{\psi, \perp} - \eta_{\psi, \parallel}) \,.
\end{align}
The projection prescriptions for $\eta_{\psi, \perp},\, \eta_{\psi, \parallel}, \,\eta_{\phi, \perp}$, and $ \eta_{\phi, \parallel}$ are given in Eqs.~\eqref{eq:fermionAnomDimProjection}, \eqref{eq:fermionZparallelProjection}, \eqref{eq:bosonAnomDimProjection}, and \eqref{eq:bosonZparallelProjection}, respectively,
evaluated at a constant background field configuration and vanishing external momenta indicated by $|_{0}=|_{\Phi=\Phi_0, p = q= 0}$ and $\Phi_0 = (\phi_0,0,0)^T$.
To evaluate these projections, we expand the Wetterich equation~\eqref{eq:WetterichEq} around the background field $\Phi = \Phi_0 + \delta \Phi$ with fluctuations $\delta \Phi = (\xi, \psi, \bar{\psi}^T)^T$. 
Since each projection only involves two-point functions, it is sufficient to expand $\partial_t \Gamma_k$ to second order in fluctuations,
\begin{align}
    \partial_t \Gamma_k &\supset -\frac{1}{4} \mathrm{STr}\tilde\partial_t\left[\left(\Gamma_{k,0}^{(2)} + R_k\right)^{-1} \Delta\Gamma_k^{(2)}\right]^2 .
\end{align}
At second order in fluctuations, the flow equation reduces to
\begin{samepage}
\begin{align}
    \partial_t &\Gamma_k \supset -\frac{1}{4} \int_p \left\{ (U''')^2 \tilde{\partial}_t \int_q  G_\phi(q_B) G_\phi(p_B+q_B) \right. \notag\\[1ex]
    & \left. - 2 h^2 \tilde{\partial}_t\int_q \Tr\left[G_\psi(q_F) G_\psi(p_F+q_F) \right] \right\} \xi(-p) \xi(p) \notag\\[1ex]
    &  - h^2 \int_p \bar{\psi}(-p)  \tilde{\partial}_t \int_q G_\phi(q_B) G_\psi(p_F+q_F)  \psi(p)\,,
\end{align}    
\end{samepage}
where $G_{\phi(\psi)}^{-1} := \Gamma_{\phi\phi(\bar{\psi}\psi)}^{(2)} + R_k^{\phi(\psi)}$ denotes the full regularized bosonic (fermionic) propagator.
It is convenient to introduce the loop integrals
\begin{align}
I^\phi(p) &= \tilde{\partial}_t \int_q G_\phi(q_B)G_\phi(p_B+q_B)\,, \notag\\[1ex]
I^\psi(p) &= \tilde{\partial}_t \int_q \Tr\left[G_\psi(q_F)G_\psi(p_F+q_F)\right]\,, \notag\\[1ex]
I^{\psi\phi}(p) &= -\tilde{\partial}_t \int_q G_\phi(q_B)G_\psi(p_F+q_F)\,,
\end{align}
so that the flow equation takes the form
\begin{align}
    \partial_t \Gamma_k =:& -\frac{1}{4} \int_p \left[(U''')^2 I^\phi(p) - 2 h^2 I^\psi(p) \right] \xi(-p) \xi(p) \notag\\[1ex]
    & + h^2 \int_p \bar{\psi}(-p) I^{\psi\phi}(p) \psi(p)\,.
\end{align}
Inserting this expression into the projection prescriptions, the flow of the bosonic velocity becomes
\begin{align}
    \partial_t v_{\phi}^2
    =& -\frac{1}{4 Z_\phi} \left( \frac{1}{d-1} \partial_{p_i^2} - v_\phi^2 \partial_{p_0^2} \right)\notag\\[1ex]
    & \left. \times \left[ (U''')^2 I^\phi(p) - 2 h^2 I^\psi(p) \right] \right\rvert_{p=0}\,,
\end{align}
and, similarly, the flow of the Fermi velocity reads
\begin{align}
    \partial_t &v_{\psi}
    = \frac{h^2}{N_\mathrm{f} d_\gamma Z_\psi} \notag\\[1ex]
    &\left.\times\Tr\left[\left(\frac{1}{d-1} \gamma_i \partial_{p_i} - v_\psi \gamma_{0} \partial_{p_0} \right) I^{\psi\phi}(p)\right]\right\rvert_{p=0}\,.
    \label{eq:FlowFermiVelocity}
\end{align}
It is useful to introduce the differential operators for the combinations of derivatives appearing above,
\begin{align}
    D^\phi_p &:= \left(\frac{\delta_{ij}}{2(d-1)} \partial_{p_i}\partial_{p_j} - v_\phi^2 \partial_{p_0^2} \right)\bigg|_{p=0} \,, \\[1ex]
    D^\psi_p &:= \left(\frac{1}{d-1} \gamma_i \partial_{p_i} - v_\psi \gamma_{0} \partial_{p_0}\right)\bigg|_{p=0}\,,
\end{align}
so that the flows $\partial_t v_\phi^2 \propto D^\phi_p\left[(U''')^2 I^\phi(p) - 2h^2 I^\psi(p)\right]$ and ${\partial_t v_\psi \propto \Tr\left[D^\psi_p I^{\psi\phi}(p)\right]}$.

\subsection{Covariant regulator scheme}

Before evaluating these expressions explicitly, it is worth noting a general structural feature.
For $v_\psi/v_\phi = 1$, the microscopic action features full rotational (Lorentz) invariance after rescaling of spacetime. 
If the regulator insertion respects this symmetry, the theory space defined by $v_\psi / v_\phi = 1$ is a symmetry-protected subspace, which the effective average action can not leave. 
This implies that for $v_\phi = v_\psi$, $\beta_{v_\phi} = \beta_{v_\psi} = 0$. 
In the presence of $v_{\psi} / v_{\phi} \neq 1$, however, the spacetime symmetry is reduced to $\mathbb{Z}_2 \times \mathrm{O}(d-1)$. 
To understand whether or not the symmetry-enhanced subspace is attractive or repulsive, we consider regulator schemes of the covariant form
\begin{align}
    R_k^\phi(p,q) &= R_\phi(p_B^2) \delta(p-q)\,, \\[1ex]
    R_k^\psi(p,q) &= R_\psi(\slashed{p}_F) \delta(p-q)\,,
\end{align}
for which the bosonic and fermionic propagators are then functions of $p_B^2$ and $\slashed{p}_F$, respectively, and read
\begin{align}
    G_\phi^{-1}(p_B^2) &= Z_\phi p_B^2 + R_\phi(p_B^2) + m_\phi^2\,,  \\[1ex]
    G_\psi^{-1}(\slashed{p}_F) &= i Z_\psi \slashed{p}_F + R_\psi(\slashed{p}_F) + m_\psi\,,
\end{align}
with the bosonic and fermionic masses given in Eqs.~\eqref{eq:BosonicMass} and \eqref{eq:FermionicMass}, respectively.

Consider first the bosonic contribution to the flow of~$v_\phi$, 
\begin{align}
    D^\phi_{p} I^\phi(p) 
    =  \frac{1}{v_\phi^{d-3}}\tilde{\partial}_t  & \int_{q_B} \left(\frac{\delta_{ij}}{2(d-1)} \partial_{p_{B,i}}\partial_{p_{B,j}} - \partial_{p_{B,0}^2} \right) \notag \\[1ex]
    & \hspace{-5mm}\left. \times G_\phi(q_B^2) G_\phi((q_B+p_B)^2) \right\rvert_{p=0}\,,
\end{align}
where we have made the substitution $q \to q_B$. 
Note that after this rescaling, the integral measure as well as the integrand have full rotational symmetry in the space of rescaled momenta $q_B$ due to the the chosen regulator scheme, and hence, the integral vanishes. 
As a consequence, the flow of $v_\phi$ reduces to
\begin{align}
    \partial_t v_\phi^2 =& \frac{h^2}{2Z_\phi} D^\phi_{p} I^\psi(p)\,, 
\end{align}
i.e., the flow of the bosonic velocity is generated entirely by the fermion loop.
To extract the scale-dependence of the bosonic velocity, we expand $D^\phi_{p} I^\psi(p)$ to second order in external momentum $p$
\begin{samepage}
\begin{align}
        D^\phi_{p} I^\psi(p) &= \tilde{\partial}_t \int_q  D^\phi_{p} \Tr\left[G_\psi(\slashed{q}_F)  G_\psi(\slashed{p}_F+\slashed{q}_F) \right]\notag\\[1ex]
        &\hspace{-14mm}= \tilde{\partial}_t \int_q \left(\frac{\delta_{ij}}{2(d-1)} \partial_{p_i}\partial_{p_j} - v_\phi^2 \partial_{p_0^2} \right) \notag\\[1ex]
        &\left.\times\Tr \left[G_\psi(\slashed{q}_F) \frac{\partial^2}{\partial q_{F,\mu} \partial q_{F,\nu}} G_\psi(\slashed{q}_F) p_{F,\mu} p_{F,\nu}\right] \right\rvert_0\notag\\[1ex]
        &\hspace{-14mm}= -\tilde{\partial}_t \int_q \left(\frac{\delta_{i\mu}\delta_{i\nu}}{d-1}v_\psi^2-\delta_{0\mu}\delta_{0\nu}v_\phi^2\right)\notag\\[1ex]
        &\left.\times\Tr \left[ \frac{\partial G_\psi(\slashed{q}_F)}{\partial q_{F,\mu}} \frac{\partial G_\psi(\slashed{q}_F)}{\partial q_{F,\nu}} \right] \right\rvert_0 \,.
\end{align}
\end{samepage}
For the flow of the Fermi velocity $v_\psi$ in \Eqref{eq:FlowFermiVelocity}, we have to calculate
\begin{equation}
    D^\psi_p I^{\psi\phi}(p) = \tilde{\partial}_t \int_q  D^\psi_p G_\phi(q_B)G_\psi(\slashed{p}_F+\slashed{q}_F)\,,
\end{equation}
i.e., the flow of the Fermi velocity is generated by the mixed boson-fermion loop.

\subsection{Evaluation of the loop integrals}

To evaluate the loop integrals for the velocity flows, we introduce the bosonic and fermionic shape functions $r_B$ and $r_F$ via
\begin{align}
    R_k^\phi(p,q) &= Z_\phi p_B^2 r_B(p_B^2/k^2) \delta(p-q)\,, \\[1ex]
    R_k^\psi(p,q) &= i Z_\psi \slashed{p}_F r_F(p_F^2/k^2) \delta(p-q)\,,
\end{align}
and introduce the dressed propagator functions  
\begin{align}
        \mathcal{G}_\phi^{-1}(p^2) &= Z_\psi p^2\left(1+r_B(p^2/k^2)\right) + m_\phi^2\,,\\[1ex]
        \mathcal{G}_\psi^{-1}(p^2) &= Z_\psi^2 p^2\left(1+r_F(p^2/k^2)\right)^2 + m_\psi^2\,.
    \end{align}

We find the bosonic velocity flow
\begin{align}\label{eq:vel_flows}
    &\partial_t v_\phi =  \frac{d_\gamma N_\mathrm{f}}{ d Z_\phi} h^2 \frac{v_\phi}{v_\psi^{d-1}} \left(1 - \frac{v_\psi^2}{v_\phi^2}\right)\notag\\[1ex]
    &\times\bigg\{ \tilde{\partial}_t \int_{p_F}  p_F^4 \left(\partial_{p_F^2} \left[(1+r_F(p_F^2/k^2))\mathcal{G}_\psi(p_F^2)\right] \right)^2\notag\\[1ex]
    & \hspace{1.4cm}- Z_\psi^2 h^2 \phi_0^2  \, \tilde{\partial}_t\int_{p_F} p_F^2 \left(\partial_{p_F^2} \mathcal{G}_\psi(p_F^2)\right)^2 \bigg\}\,, 
\end{align}
which can be written in the compact form
\begin{samepage}
\begin{align}
        \partial_t v_\phi=& -\frac{4 v_d}{d}  d_\gamma N_\mathrm{f} \tilde{h}^2 \frac{1}{v_\psi^{d-1}} \left(v_\phi^2-v_\psi^2\right)\notag\\
    &\times\left[ m_4^{(F)d}(\tilde{m}_\psi;\eta_\psi) - 2\tilde{\rho}\tilde{h}^2 m_2^{(F)d}(\tilde{m}_\psi;\eta_\psi) \right]\notag \\
    =& (v_\phi^2 - v_\psi^2) \eta_{\phi, \parallel}^F\,,
\end{align}
\end{samepage}
after introducing the dimensionless threshold functions
\begin{align}
        m_2^{(F)d} &(\omega;\eta_\psi) = - \frac{1}{4v_d} k^{6-d} \notag\\
        &\times\tilde{\partial}_t \int_q q^2 \left[\partial_{q^2} \frac{1}{q^2(1+r_F)^2 + k^2\omega}\right]^2,\\
        m_4^{(F)d} &(\omega;\eta_\psi) = - \frac{1}{4v_d} k^{4-d}  \notag\\
        &\times\tilde{\partial}_t \int_q q^4 \left[\partial_{q^2} \frac{1+r_F}{q^2(1+r_F)^2 + k^2\omega}\right]^2,
\end{align}
with the rescaled, dimensionless parameters ${\tilde{m}_\psi=2\tilde{h}^2\tilde{\rho}}$, ${\tilde h^2 = Z_{\phi}^{-1} Z_{\psi}^{-2} k^{d-4} h^2}$, and ${\tilde\rho= \frac{1}{2} Z_{\phi} k^{2-d} \phi^2}$, and the volume factor ${v_d^{-1}= 2^{d+1} \pi^{d/2} \Gamma(d/2)}$.

The corresponding flow of the Fermi velocity is
\begin{samepage}
\begin{align}
    \partial_t v_\psi =& 2 h^2 v_\psi  \tilde{\partial}_t  \int_{p} \left(\frac{v_\psi^2}{d-1}  p_{i}^2 - p_{0}^2\right) \mathcal{G}_\phi(p_B^2)\notag \\
    & \times \partial_{p_F^2} [(1+r_F(p_F^2/k^2))\mathcal{G}_\psi(p_F^2)]\,.\label{eq:vel_flows2}
\end{align}
\end{samepage}
Note that the flow for the bosonic velocity is fully rotational invariant, while the integrand for the Fermi velocity is not. Due to the fact that this introduces different momentum shells for bosons and fermions, the exact evaluation of this expression is challenging.

From the expressions above, we can see that the flow of $v_\phi$ is driven purely by fermionic fluctuations, while the flow of $v_\psi$ is driven by a mixed boson-fermion loop. 
This is a manifestation that a single velocity can be always scaled out of the theory by a redefinition of spacetime. 
From \Eqref{eq:vel_flows}, it is evident that $v_\psi / v_\phi = 1$ is a fixed point of the flows of $v_\phi$. That this is also the case for $v_\psi$ can be seen as follows: if $v_\psi = v_\phi$, we can substitute $p_i = p_{B,i}/v_\phi$. Then, the flow of $v_\psi$ is of the form
\begin{align}
    \partial_t v_\psi \propto \int_{p_B} \left( \frac{1}{d-1} p_{B,i}^2 - p_{B,0}^2 \right) f(p_B^2) = 0\,,  
\end{align}
which vanishes due to rotational invariance of measure and $f$. 

In order to simplify the calculation of the flow of $v_\psi$ and for our purposes, we restrict ourselves to linear perturbations out of the Lorentz-symmetric subspace, which we parametrize by $r = v_\phi^2/v_\psi^2 - 1$. As noted above, the zeroth order in $r$ vanishes, and hence, at leading non-trivial order in $r$, we find
\begin{samepage}
\begin{align}
    &\partial_t  v_\psi = 2 h^2 v_\psi \left(\frac{v_\phi^2}{v_\psi^2} - 1 \right) \tilde{\partial}_t  \int_{p} \left(\frac{v_\psi^2}{d-1}  p_{i}^2 - p_{0}^2\right)p_{i}^2  \notag \\
    & \times \mathcal{G}'_\phi(p_F^2) \partial_{p_F^2} [(1+r_F(p_F^2/k^2))\mathcal{G}_\psi(p_F^2)] + \mathcal{O}(r^2) \,.\label{eq:vel_flows2}
\end{align}
\end{samepage}
Next, we substitute $p \to p_F$, and subsequently rewrite 
\begin{align}
    \int d^d p_F \, f(p_F^2)= \frac{1}{2} \int_{S^{d-1}} \frac{d\Omega_{d-1}}{(2\pi)^d} \int_0^\infty dx \, f(x).
\end{align}
This yields
\begin{samepage}
\begin{align}
    \partial v_\psi = &\frac{h^2}{v_\psi^{d-2}} \left(\frac{v_\phi^2}{v_\psi^2} - 1 \right) \int_{S^{d-1}} \frac{d\Omega_{d-1}}{(2\pi)^d} \left(\frac{1}{d-1}  \hat{p}_{i}^2 - \hat{p}_{0}^2\right)\hat{p}_{i}^2 \notag \\
    &\times \tilde{\partial}_t \int_0^\infty dx \, x^{\frac{d}{2} + 1} G'_\phi(x) \partial_{x} [(1+r_F(x/k^2))\mathcal{G}_\psi(x)], \label{eq:dtvf_radial}
\end{align}    
\end{samepage}
where we have defined the normalized components ${\hat{p}_\mu = p_\mu/p}$. The angular integral can be evaluated with $d$-dimensional spherical coordinates; we find
\begin{align}
    \int_{S^{d-1}} \frac{d\Omega_{d-1}}{(2\pi)^d} \left(\frac{1}{d-1}  \hat{p}_{i}^2 - \hat{p}_{0}^2\right)\hat{p}_{i}^2 = \left[2^{d} \pi^{d/2} \Gamma\left(2+\frac{d}{2}\right)\right]^{-1}.
\end{align}

\begin{figure*}[ht!]
    \includegraphics[width=0.32\linewidth]{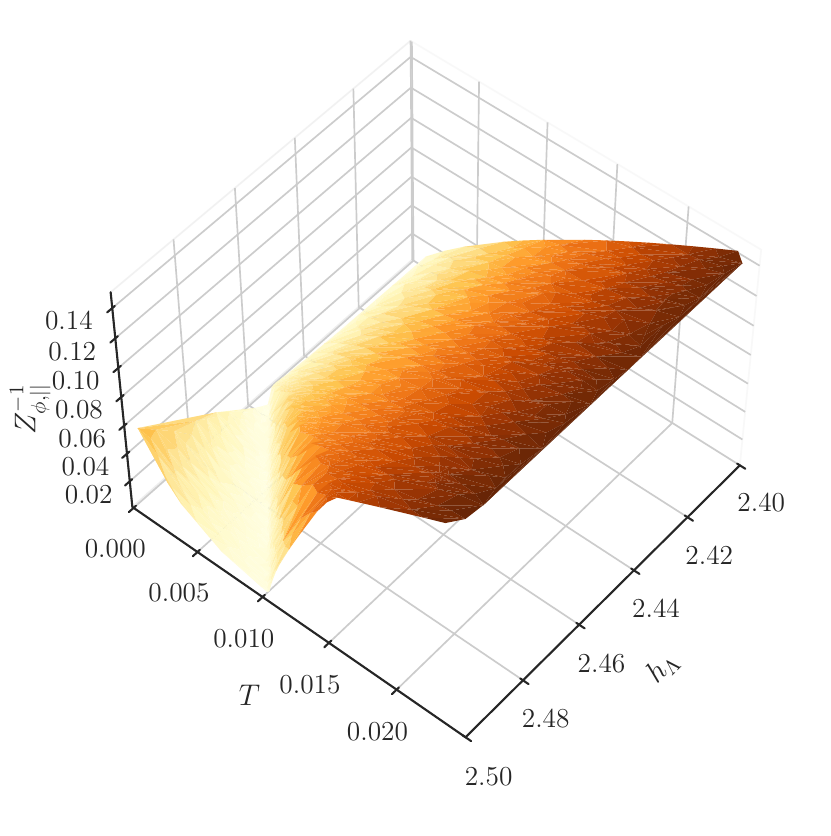}
\includegraphics[width=0.32\linewidth]{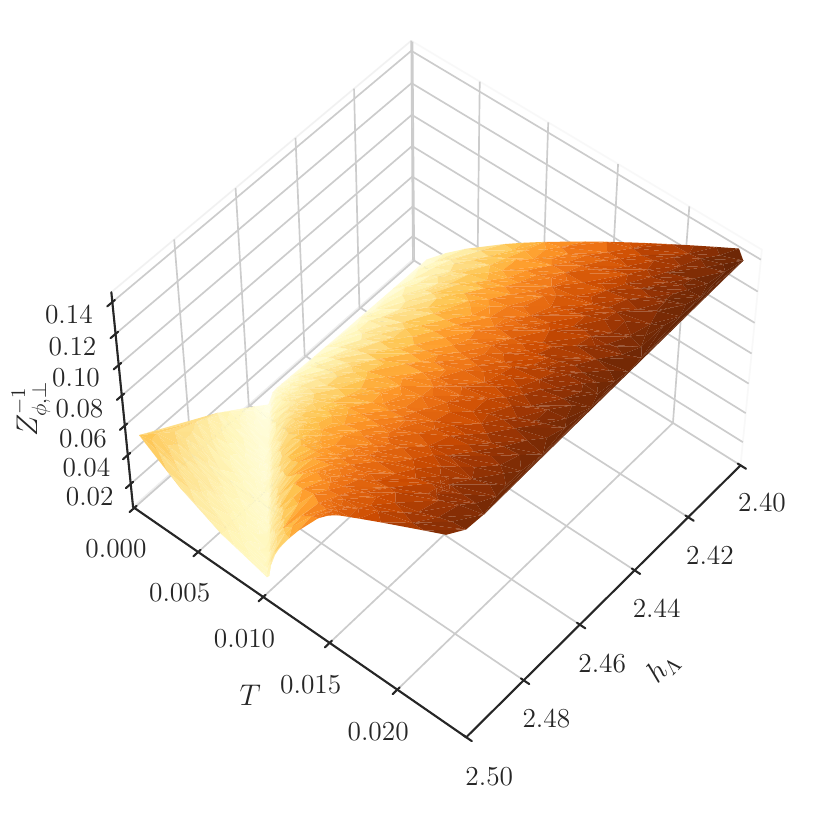}
\includegraphics[width=0.32\linewidth]{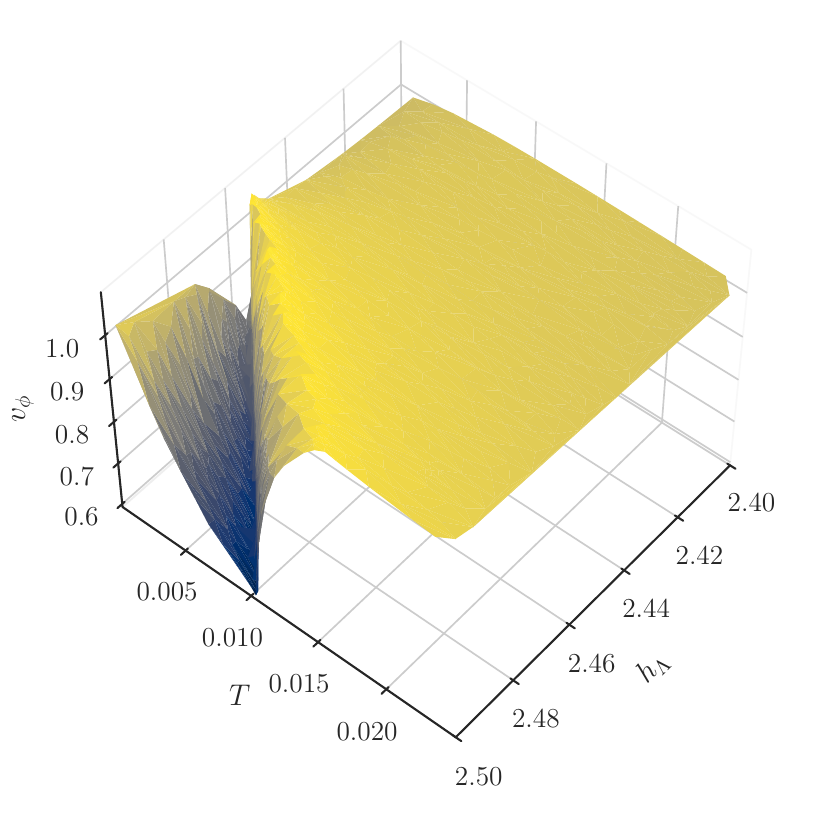}
    \caption{\textbf{Bosonic wave-function renormalization and bosonic velocity} as a function of the Yukawa coupling $h_\Lambda$ and temperature~$T$ at $t=-9$. \textbf{Left:} Projection onto the frequency direction of the inverse of the bosonic wave-function renormalization~$Z_{\phi,\parallel}^{-1}$. \textbf{Middle:} Projection onto the spatial direction of the inverse of the bosonic wave-function renormalization~$Z_{\phi,\perp}^{-1}$. \textbf{Right:} Bosonic velocity $v_\phi$.}
    \label{fig:Zphi}
\end{figure*}

\subsection{Covariant Litim regulator}

To explicitly evaluate the radial integral in \Eqref{eq:dtvf_radial}, we choose the covariant Litim regulator, defined in the main text by the shape functions in Eqs.~\eqref{eq:litim_shape} and \eqref{eq:litim_shape2}.
We find 
\begin{align}
    \tilde{\partial}_t& \int_0^\infty dx \, x^{\frac{d}{2} + 1} G'_\phi(x) \partial_{x} [(1+r_F(x/k^2))\mathcal{G}_\psi(x)] \notag \\
    &= \frac{k^{d-4}}{(d+1)Z_\phi Z_\psi^2} \frac{-2(d+1) + (1+m_\psi^2) \eta_{\phi, \parallel}}{(1+m_\phi^2)^2(1+m_\psi^2)^2}.
\end{align}
Hence, we find for the flow of $v_\phi^2$ and $v_\psi^2$
\begin{align}
    \partial v_\phi^2 &= \left(\frac{v_\phi^2}{v_\psi^2} - 1 \right) v_\psi^2 \eta_{\phi, \parallel}^F\, , \\
    \partial_t v_\psi &= \left(1 -  \frac{v_\phi^2}{v_\psi^2}\right) \frac{8 v_d}{d(d+1)(d+2)} \notag\\
    &\quad \times\frac{\tilde{h}^2}{v_\psi^{d-2}} \frac{2(d+1) - (1+\tilde{m}_\psi^2) \eta_{\phi,\parallel}}{(1+\tilde{m}_\phi^2)^2(1+\tilde{m}_\psi^2)^2} + \mathcal{O}(r^2).
\end{align}
The fermionic contribution to the bosonic anomalous dimension is given by
\begin{align}\label{eq:c36}
    \eta_{\phi,\parallel}^F = \frac{8 v_d}{d} N_\mathrm{f} d_\gamma \frac{\tilde{h}^2}{v_\psi^{d-1}}&\left[m_4^{(F)d}(\tilde{m}_\psi^2;\eta_{\psi,\parallel}) \right. \notag \\
    &\left.- 2 \tilde{\rho} \tilde{h}^2 m_2^{(F)d}(\tilde{m}_\psi^2;\eta_{\psi,\parallel})\right],
\end{align}
with threshold functions evaluated for the covariant Litim regulator,
\begin{align}\label{eq:m2m4}
    m_2^{(F)d}(\omega;\eta_\psi) =& \frac{1}{\left(1+\omega\right)^{4}}  \, ,\\
    m_4^{(F)d}(\omega;\eta_\psi) =& \frac{1}{\left(1+\omega\right)^{4}} + \frac{1-\eta_\psi}{d-2} \frac{1}{\left(1+\omega\right)^{3}} \notag\\
   &- \left(\frac{1-\eta_\psi}{2d-4} + \frac{1}{4}\right) \frac{1}{\left(1+\omega\right)^{2}}\label{eq:m4m4}
    \, .
\end{align}
With these equations at hand, we obtain the result cited in the main text,
\begin{align}
    \partial_t\left(\frac{v_\phi^2}{v_\psi^2}\right) = C\left( \frac{v_\phi^2}{v_\psi^2} - 1 \right) + \mathcal{O}\left(\left(\frac{v_\phi^2}{v_\psi^2} - 1\right)^2\right),
\end{align}
with constant
\begin{align}
    C&= \eta_{\phi,\parallel}^F + \frac{16 v_d}{d(d+1)(d+2)} \frac{\tilde{h}^2}{v_\psi^{d-1}} \frac{2(d+1) - (1+\tilde{m}_\psi^2)\eta_{\phi,\parallel}}{(1+\tilde{m}_\phi^2)^2(1+\tilde{m}_\psi^2)^2}\,.
\end{align}

\section{Flow of the boson mass in the symmetric regime}
\label{app:SYMmassflow}

%
\begin{figure*}[bt]
    \includegraphics[width=0.4\linewidth]{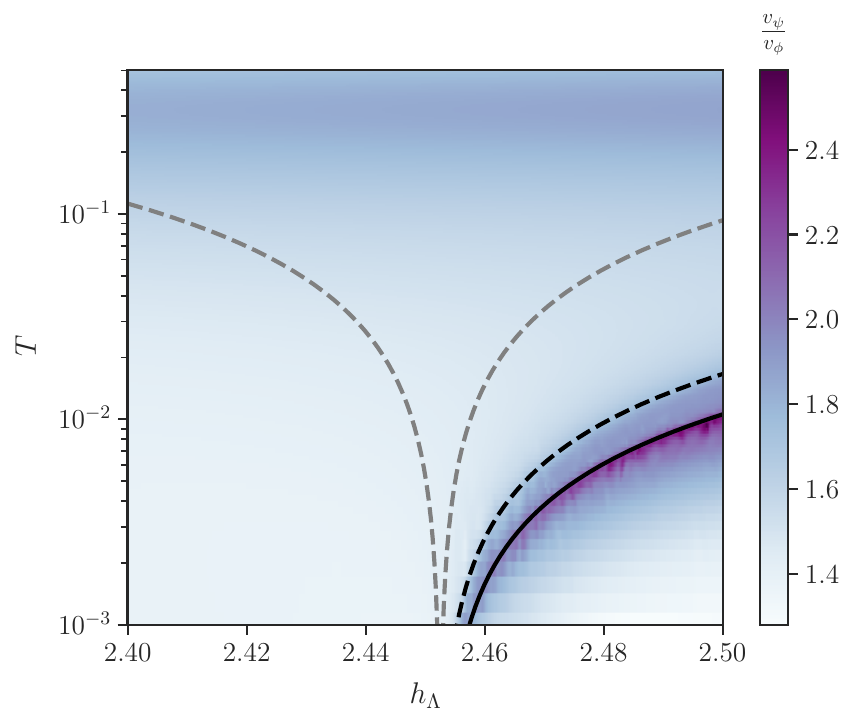}
    \hspace{3mm}
    \includegraphics[width=0.4\linewidth]{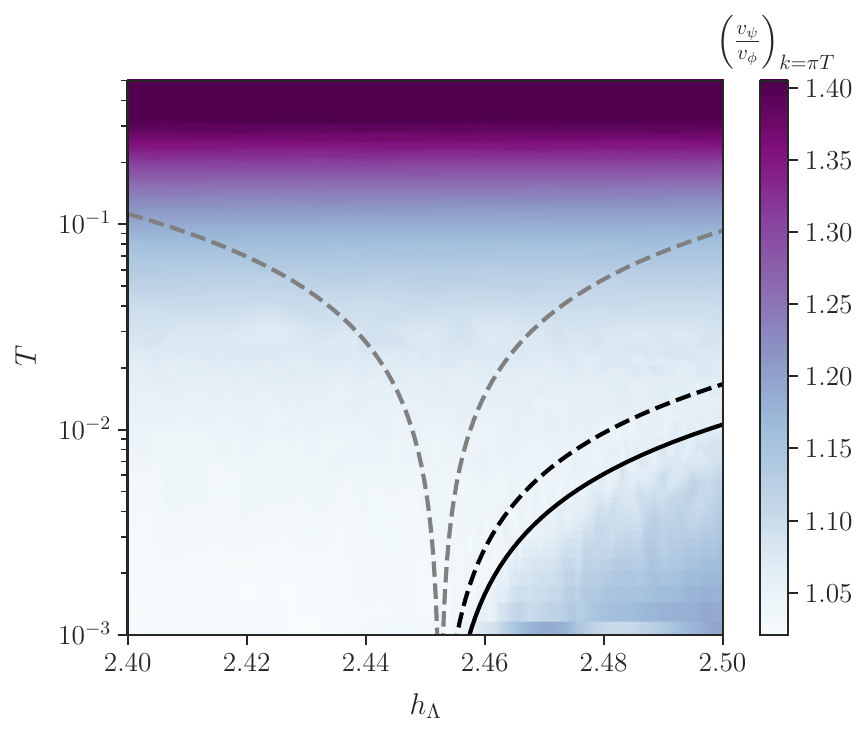}
    \caption{\textbf{Velocity ratio}: We depict the ratio of the fermionic and bosonic velocity at different RG scales in the vicinity of the QCP. The condensation line (solid black), precondensation line (dashed black) as well as the QCF (dashed gray) are indicated for better orientation. \textbf{Left:} $k \to 0$ , \textbf{Right:} $k = \pi T$.}
    \label{fig:velocities}
\end{figure*}

In the main text, we have extracted the correlation length by the inverse mass of the order parameter field. We have, however, noted that the bosonic mass does not freeze out towards the IR on the symmetric side of the QCP, attributed to the fact that the fermions are gapless. 

To see this explicitly, we consider a truncated effective action that includes only operators up to fourth order in the fields, i.e., 
\begin{align}
    U(\rho) = m_\phi^2 \rho + \frac{\lambda_\phi}{2} (\rho - \rho_0)^2.
\end{align}
In the symmetric phase, $m_\phi^2 > 0$ and $\rho_0 = 0$, while in the symmetry-broken phase, $m_\phi^2 = 0$ and $\rho_0 > 0$. The beta function of the renormalized mass, $\bar{m}_\phi^2$, is then given by
\begin{samepage}
\begin{align}
    \partial_t \bar{m}_\phi^2&= - \eta_{\phi, \parallel}\bar{m}_\phi^2 + (\partial_{\bar \rho} \partial_t U)( \bar \rho = 0)/Z_{\phi,\parallel} \notag  \\[1ex]
    &=
    \begin{cases}
        \frac{3 \bar{h}^2 \bar{m}_\phi^2}{4 \pi k} + \mathcal{O}(k)\,, & \text{SYM}\,, \\
        \frac{N_\mathrm{f} \bar{h}^2}{2 \pi} k + \mathcal{O}(k^2)\,, & \text{SSB}\,.
    \end{cases}
\end{align}    
\end{samepage}
Evidently, assuming that $h^2$ freezes out towards the IR, the bosonic mass freezes out as well in the symmetry-broken phase. 
In the symmetric phase, however, the mass keeps flowing due to the massless fermionic loop appearing in $\eta_{\phi, \parallel}\propto \tilde h^2$  which itself increases after flowing out of the QCP regime until it reaches a saturated value at $\eta_{\phi, \parallel}=1$, see \Cref{fig:qcpflow2}.
The saturated value is reached due as a result of the dimensional scaling of the Yukawa coupling $\partial_t \tilde h^2 = (d-4+\eta_{\phi,\parallel}-2\eta_{\psi,\parallel})\tilde h^2+\mathrm{loops}$ which in $d=3$ and away from the QCP in the symmetric regime reduces to $\partial_t \tilde h^2 \approx (-1+\eta_{\phi,\parallel})\tilde h^2$, i.e.~the $\eta_{\phi, \parallel}$ saturates at unity.
We numerically observe that this induces a growth of the dimensionless renormalized mass $\tilde{m}_\phi^2 = \bar{m}_\phi^2 / k^2$ towards the IR. At leading order in $1/\tilde{m}_\phi^2$, the beta functions of the remaining dimensionless couplings feature another attractive fixed point, given by
\begin{align}
    (\tilde{h}^2 {}^{*}, \tilde{\lambda}^*) = \left(\frac{3 \pi^2}{10}, \frac{24 \pi^2}{25}\right).
\end{align}
We further note that $1/\tilde{m}_\phi^2 = 0$ is a fixed point of $\partial_t (1/\tilde{m}_\phi^2)$, and hence, the theory does, in fact, feature a fully IR attractive fixed point that is present only in the symmetric parametrization. We attribute the existence of this fixed point to an insufficient resolution of momentum dependence in the LPA'. We will further address this issue with a more elaborate truncation in future work.

\section{Additional data}  \label{app:AdditionalData}

In this Appendix, we provide additional plots of the spatial and frequency wave-function renormalizations. This data was moved to the Appendix, due to the limited momentum resolution in the present setup and we defer a more in depth analysis of these observations to future work.

\subsection{Velocity flow at finite temperature}

We show the different components of the bosonic wave-function renormalization in \Cref{fig:Zphi}. 
In comparison to both fermionic wave-function renormalizations, the bosonic ones grow large, highlighting the change from a system that is dominated by fermions at $k=\Lambda$ to a system that is mainly determined by bosons at $k \to 0$.
\begin{figure}[t]
    \includegraphics[width=0.93\linewidth]{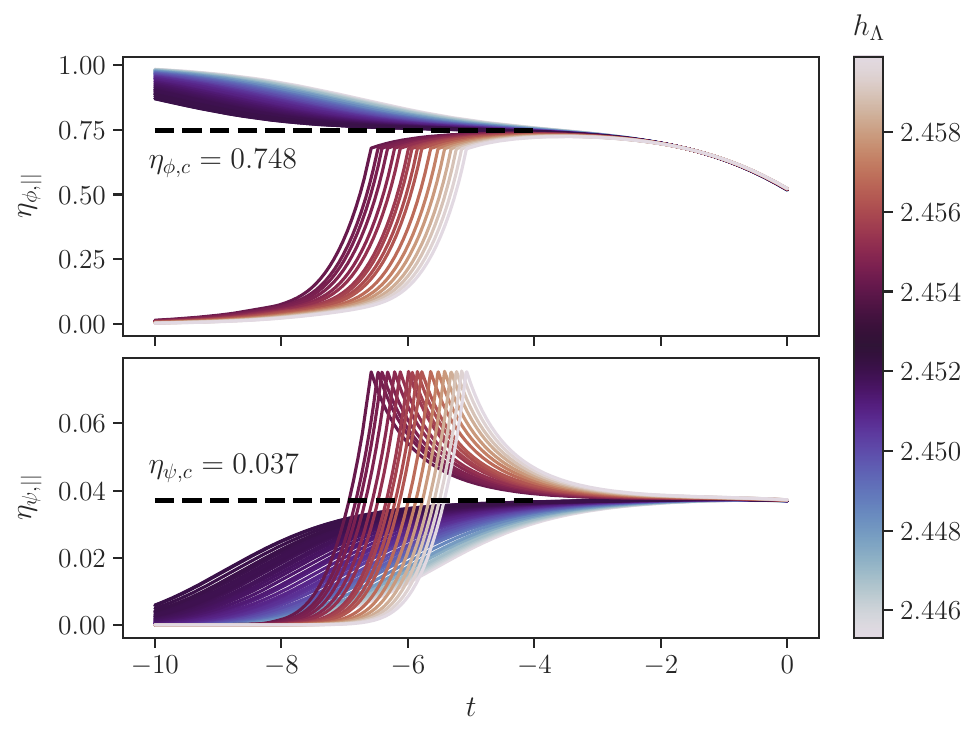}
    \caption{
    \textbf{Anomalous dimensions at the QCP}
    from the numerical integration of the functional RG flow equations. 
    The anomalous dimensions are extracted by projecting the wave-function renormalizations onto the frequency (parallel) direction.
    We show the RG time dependence of the anomalous dimensions for different initial values of the Yukawa coupling~$h_\Lambda$ close to the QCP. 
    The extracted critical values of the anomalous dimensions are indicated by the dashed black lines.} 
    \label{fig:criticalParEtas}
\end{figure}

We find that the bosonic velocity remains very close to its initial value $v_{\phi,\Lambda} = 1$ in the symmetric region of the phase diagram and only severely dips around the finite-temperature phase transition, where $\eta_{\phi, \perp}$ and $\eta_{\phi, \parallel}$ remain finite but distinct at the critical point.
This behavior is expected, as the introduction of a finite temperature explicitly breaks Lorentz symmetry and allows the temporal and spatial components of the wave-function renormalization to evolve independently.

Finally, we depict the ratio of fermionic and bosonic velocities in the vicinity of the QCP in \Cref{fig:velocities}. We find that the velocity ratio away from the QCP assumes its original value  $v_{\psi,\Lambda}/v_{\phi,\Lambda} = 1.4$ throughout most of the symmetric phase at $k \to 0$. This is expected, as the different projections of the wave-function renormalizations, cf.~\Cref{fig:ZPsi} and \Cref{fig:Zphi}, remain very close to unity as well. It is, however, a good indication for the consistency of our spatial regulator scheme, which introduces a breaking of Lorentz symmetry at finite~$k$. 
For larger temperatures $T>0.2$, we find a slight increase of the initial velocity ratio, which we attribute to regulator artifacts of the scheme, i.e., the initial cutoff scale is to close to the temperature scale.

The velocity ratio is slightly larger in the precondensation regime, which is related to the increase of the fermionic velocity as discussed in \Cref{sec:quasiparticle}. Furthermore, we observe a pronounced peak on the phase transition line, which is associated to the decrease of bosonic velocity. 
However, these results should be interpreted with caution due to the limited momentum resolution of the present calculation, as the bosonic wave-function renormalizations are computed from a projection onto $p = 0$, cf.~\Cref{app:projections}.

The emergence of relativistic symmetry in the present system is naturally a $T=0$ phenomenon.
However, even at finite $T$, a precursor for this zero-$T$ emergence of relativistic symmetry at finite momenta $ \vec p$ can be found in the behavior of the velocity ratio at finite $k$ as a heuristic connection $|k| \approx |\vec p^{\,}|$ can be made for the present cutoff choice. We find $v_{\psi,k}/v_{\phi,k}  \approx 1$ at $k = \pi T$ in a very large region of the phase diagram, cf.~the right panel of \Cref{fig:velocities}. After the fermions are gapped out by the lowest fermionic Matsubara mode, this effect is washed out.

\subsection{Fermionic anomalous dimension at the QCP}

As stated in the main text and below \Cref{table:CriticalExponents}, we find a small deviation of the fermionic quantum critical exponents $\eta_{\psi,\parallel}$ and $\eta_{\psi,\perp}$, whereas the bosonic $\eta_\phi$ coincides for both projections. 
The extraction of the exponents in the frequency direction is provided in \Cref{fig:criticalParEtas}. 
This artifact is connected to the regulator choice in Eqs.~\eqref{eq:Regulatorb} and \eqref{eq:Regulatorf}, which breaks relativistic symmetry explicitly.

\bibliography{references}

@article{Pawlowski:2015mlf,
    author = "Pawlowski, Jan M. and Scherer, Michael M. and Schmidt, Richard and Wetzel, Sebastian J.",
    title = "{Physics and the choice of regulators in functional renormalisation group flows}",
    eprint = "1512.03598",
    archivePrefix = "arXiv",
    primaryClass = "hep-th",
    doi = "10.1016/j.aop.2017.06.017",
    journal = "Annals Phys.",
    volume = "384",
    pages = "165--197",
    year = "2017"
}

@article{Helmboldt:2014iya,
    author = "Helmboldt, Alexander J. and Pawlowski, Jan M. and Strodthoff, Nils",
    title = "{Towards quantitative precision in the chiral crossover: masses and fluctuation scales}",
    eprint = "1409.8414",
    archivePrefix = "arXiv",
    primaryClass = "hep-ph",
    doi = "10.1103/PhysRevD.91.054010",
    journal = "Phys. Rev. D",
    volume = "91",
    number = "5",
    pages = "054010",
    year = "2015"
}

@article{Braun:2022mgx,
    title = {Renormalised spectral flows},
	pages = {061},
	author = {Braun, Jens and Chen, Yong-Rui and Fu, Wei-Jie and Geißel, Andreas and Horak, Jan and Huang, Chuang and Ihssen, Friederike and Pawlowski, Jan M. and Reichert, Manuel and Rennecke, Fabian and Tan, Yang-Yang and Töpfel, Sebastian and Wessely, Jonas and Wink, Nicolas},
	journal = {SciPost Phys. Core},
	volume = {6},
	year = {2023},
	publisher = {SciPost},
	doi = {10.21468/SciPostPhysCore.6.3.061},
	url = {https://scipost.org/10.21468/SciPostPhysCore.6.3.061}
}

@article{Horak:2023hkp,
    author = "Horak, Jan and Ihssen, Friederike and Pawlowski, Jan M. and Wessely, Jonas and Wink, Nicolas",
    title = "{Scalar spectral functions from the spectral functional renormalization group}",
    eprint = "2303.16719",
    archivePrefix = "arXiv",
    primaryClass = "hep-th",
    doi = "10.1103/PhysRevD.110.056009",
    journal = "Phys. Rev. D",
    volume = "110",
    number = "5",
    pages = "056009",
    year = "2024"
}

@article{lu2019superconductors,
  title={Superconductors, orbital magnets and correlated states in magic-angle bilayer graphene},
  author={Lu, Xiaobo and Stepanov, Petr and Yang, Wei and Xie, Ming and Aamir, Mohammed Ali and Das, Ipsita and Urgell, Carles and Watanabe, Kenji and Taniguchi, Takashi and Zhang, Guangyu and others},
  journal={Nature},
  volume={574},
  number={7780},
  pages={653--657},
  year={2019},
  doi= {doi.org/10.1038/s41586-019-1695-0},
  publisher={Nature Publishing Group UK London}
}

@misc{Koenigstein:2025sse,
    author = "Koenigstein, Adrian and Steil, Martin J. and Floerchinger, Stefan",
    title = "{Functional Renormalization Group flows as diffusive Hamilton-Jacobi-type equations}",
    eprint = "2512.05973",
    archivePrefix = "arXiv",
    primaryClass = "hep-th",
    month = "11",
    year = "2025"
}

@article{PhysRev.65.117,
  title = {Crystal Statistics. I. A Two-Dimensional Model with an Order-Disorder Transition},
  author = {Onsager, Lars},
  journal = {Phys. Rev.},
  volume = {65},
  issue = {3-4},
  pages = {117--149},
  numpages = {0},
  year = {1944},
  month = {Feb},
  publisher = {American Physical Society},
  doi = {10.1103/PhysRev.65.117},
  url = {https://link.aps.org/doi/10.1103/PhysRev.65.117}
}

@misc{Stoll:2021ori,
    author = "Stoll, Jonas and Zorbach, Niklas and Koenigstein, Adrian and Steil, Martin J. and Rechenberger, Stefan",
    title = "{Bosonic fluctuations in the $( 1 + 1 )$-dimensional Gross-Neveu(-Yukawa) model at varying $\mu$ and $T$ and finite $N$}",
    eprint = "2108.10616",
    archivePrefix = "arXiv",
    primaryClass = "hep-ph",
    month = "8",
    year = "2021"
}

@article{Zorbach:2024rre,
    author = "Zorbach, Niklas and Koenigstein, Adrian and Braun, Jens",
    title = "{Functional renormalization group meets computational fluid dynamics: RG flows in a multidimensional field space}",
    eprint = "2412.16053",
    archivePrefix = "arXiv",
    primaryClass = "cond-mat.stat-mech",
    doi = "10.1103/z954-46md",
    journal = "Phys. Rev. D",
    volume = "113",
    number = "3",
    pages = "036011",
    year = "2026"
}

@article{Ihssen:2022xkr,
    author = "Ihssen, Friederike and Pawlowski, Jan M. and Sattler, Franz R. and Wink, Nicolas",
    title = "{Local discontinuous Galerkin for the functional renormalisation group}",
    eprint = "2207.12266",
    archivePrefix = "arXiv",
    primaryClass = "hep-th",
    doi = "10.1016/j.cpc.2024.109182",
    journal = "Comput. Phys. Commun.",
    volume = "300",
    pages = "109182",
    year = "2024"
}

@article{Grossi:2021ksl,
    author = "Grossi, Eduardo and Ihssen, Friederike J. and Pawlowski, Jan M. and Wink, Nicolas",
    title = "{Shocks and quark-meson scatterings at large density}",
    eprint = "2102.01602",
    archivePrefix = "arXiv",
    primaryClass = "hep-ph",
    doi = "10.1103/PhysRevD.104.016028",
    journal = "Phys. Rev. D",
    volume = "104",
    number = "1",
    pages = "016028",
    year = "2021"
}

@article{Ihssen:2024miv,
    author = "Ihssen, Friederike and Pawlowski, Jan M. and Sattler, Franz R. and Wink, Nicolas",
    title = "{Toward quantitative precision in functional QCD}",
    eprint = "2408.08413",
    archivePrefix = "arXiv",
    primaryClass = "hep-ph",
    doi = "10.1103/lkwg-697n",
    journal = "Phys. Rev. D",
    volume = "113",
    number = "9",
    pages = "094038",
    year = "2026"
}

@misc{Sattler:2024ozv,
    author = "Sattler, Franz R. and Pawlowski, Jan M.",
    title = "{DiFfRG: A Discretisation Framework for functional Renormalisation Group flows}",
    eprint = "2412.13043",
    archivePrefix = "arXiv",
    primaryClass = "hep-ph",
    month = "12",
    year = "2024"
}

@article{Pawlowski:2014zaa,
    author = "Pawlowski, Jan M. and Rennecke, Fabian",
    title = "{Higher order quark-mesonic scattering processes and the phase structure of QCD}",
    eprint = "1403.1179",
    archivePrefix = "arXiv",
    primaryClass = "hep-ph",
    doi = "10.1103/PhysRevD.90.076002",
    journal = "Phys. Rev. D",
    volume = "90",
    number = "7",
    pages = "076002",
    year = "2014"
}

@misc{Sattler:2026csm,
    author = "Sattler, Franz R.",
    title = "{FunKit: A computer algebra toolkit for functional approaches}",
    eprint = "2605.28935",
    archivePrefix = "arXiv",
    primaryClass = "hep-ph",
    month = "5",
    year = "2026"
}

@article{Ihssen:2023qaq,
    author = "Ihssen, Friederike and Sattler, Franz R. and Wink, Nicolas",
    title = "{Numerical RG-time integration of the effective potential: Analysis and benchmark}",
    eprint = "2302.04736",
    archivePrefix = "arXiv",
    primaryClass = "hep-th",
    doi = "10.1103/PhysRevD.107.114009",
    journal = "Phys. Rev. D",
    volume = "107",
    number = "11",
    pages = "114009",
    year = "2023"
}

@article{Weeks2010,
  title = "{Interaction-driven instabilities of a Dirac semimetal}",
  author = {Weeks, C. and Franz, M.},
  journal = {Phys. Rev. B},
  volume = {81},
  issue = {8},
  pages = {085105},
  numpages = {8},
  year = {2010},
  month = {Feb},
  publisher = {American Physical Society},
  doi = {10.1103/PhysRevB.81.085105},
  url = {https://link.aps.org/doi/10.1103/PhysRevB.81.085105}
}

@article{xia2025superconductivity,
  author  = {Xia, Yiyu and Han, Zhongdong and Watanabe, Kenji and Taniguchi, Takashi and Shan, Jie and Mak, Kin Fai},
  title   = "{Superconductivity in twisted bilayer {WSe}$_2$}",
  journal = {Nature},
  year    = {2025},
  volume  = {637},
  number  = {8047},
  pages   = {833--838},
  doi     = {10.1038/s41586-024-08116-2},
  url     = {https://doi.org/10.1038/s41586-024-08116-2},
  issn    = {1476-4687}
}

@article{Tolosa-Simeon:2025fot,
    title = "{Relativistic Mott transitions and finite-temperature effects of quantum criticality in Dirac semimetals}",
  author = {Tolosa-Sime\'on, Mireia and Classen, Laura and Scherer, Michael M.},
  journal = {Phys. Rev. B},
  volume = {112},
  issue = {11},
  pages = {115133},
  numpages = {18},
  year = {2025},
  month = {Sep},
  publisher = {American Physical Society},
  doi = {10.1103/7kw4-8r3m},
  url = {https://link.aps.org/doi/10.1103/7kw4-8r3m}
}

@article{Codello:2012sc,
    author = "Codello, Alessandro",
    title = "{Scaling Solutions in Continuous Dimension}",
    eprint = "1204.3877",
    archivePrefix = "arXiv",
    primaryClass = "hep-th",
    doi = "10.1088/1751-8113/45/46/465006",
    journal = "J. Phys. A",
    volume = "45",
    pages = "465006",
    year = "2012"
}

@article{Grossi:2019urj,
    author = "Grossi, Eduardo and Wink, Nicolas",
    title = "{Resolving phase transitions with discontinuous Galerkin methods}",
    eprint = "1903.09503",
    archivePrefix = "arXiv",
    primaryClass = "hep-th",
    doi = "10.21468/SciPostPhysCore.6.4.071",
    journal = "SciPost Phys. Core",
    volume = "6",
    pages = "071",
    year = "2023"
}

@article{tchf-w8h7,
  title = "{Relativistic Mott transition and high-order van Hove singularity in twisted double bilayer ${\mathrm{WSe}}_{2}$: Mean-field and functional renormalization group study}",
  author = {Hawashin, Bilal and Kleeschulte, Julian and Kurz, David and Al-Eryani, Aiman and Scherer, Michael M.},
  journal = {Phys. Rev. B},
  volume = {113},
  issue = {7},
  pages = {075107},
  numpages = {15},
  year = {2026},
  month = {Feb},
  publisher = {American Physical Society},
  doi = {10.1103/tchf-w8h7},
  url = {https://link.aps.org/doi/10.1103/tchf-w8h7}
}

@article{Grushin2013,
  title = "{Charge instabilities and topological phases in the extended Hubbard model on the honeycomb lattice with enlarged unit cell}",
  author = {Grushin, Adolfo G. and Castro, Eduardo V. and Cortijo, Alberto and de Juan, Fernando and Vozmediano, Mar\'{\i}a A. H. and Valenzuela, Bel\'en},
  journal = {Phys. Rev. B},
  volume = {87},
  issue = {8},
  pages = {085136},
  numpages = {8},
  year = {2013},
  month = {Feb},
  publisher = {American Physical Society},
  doi = {10.1103/PhysRevB.87.085136},
  url = {https://link.aps.org/doi/10.1103/PhysRevB.87.085136}
}

@article{Raghu2008,
  title = "{Topological Mott Insulators}",
  author = {Raghu, S. and Qi, Xiao-Liang and Honerkamp, C. and Zhang, Shou-Cheng},
  journal = {Phys. Rev. Lett.},
  volume = {100},
  issue = {15},
  pages = {156401},
  numpages = {4},
  year = {2008},
  month = {Apr},
  publisher = {American Physical Society},
  doi = {10.1103/PhysRevLett.100.156401},
  url = {https://link.aps.org/doi/10.1103/PhysRevLett.100.156401}
}

@article{Vafek_2014,
   title="{Dirac Fermions in Solids: From High-Tc Cuprates and Graphene to Topological Insulators and Weyl Semimetals}",
   volume={5},
   ISSN={1947-5462},
   url={http://dx.doi.org/10.1146/annurev-conmatphys-031113-133841},
   DOI={10.1146/annurev-conmatphys-031113-133841},
   number={1},
   journal={Annual Review of Condensed Matter Physics},
   publisher={Annual Reviews},
   author={Vafek, Oskar and Vishwanath, Ashvin},
   year={2014},
   month=mar, pages={83–112} }

@article{Boyack:2020xpe,
    author = "Boyack, Rufus and Yerzhakov, Hennadii and Maciejko, Joseph",
    title = "{Quantum phase transitions in Dirac fermion systems}",
    doi = "10.1140/epjs/s11734-021-00069-1",
    journal = "Eur. Phys. J. ST",
    volume = "230",
    number = "4",
    pages = "979--992",
    year = "2021"
}

@misc{Pastor-Gutierrez:2026rsy,
    author = "Pastor-Guti{\'e}rrez, {\'A}lvaro and Pawlowski, Jan M. and Sattler, Franz R.",
    title = "{Thermal precondensation in gauge-fermion theories}",
    eprint = "2602.11265",
    archivePrefix = "arXiv",
    primaryClass = "hep-ph",
    month = "2",
    year = "2026"
}

@article{sorella1992semi,
   author = "Sorella, S. and Tosatti, E.",
    title = "{Semi-Metal-Insulator Transition of the Hubbard Model in the Honeycomb Lattice}",
    doi = "10.1209/0295-5075/19/8/007",
    journal = "EPL",
    volume = "19",
    number = "8",
    pages = "699",
    year = "1992"
}

@article{PhysRevLett.100.146404,
  title = "{Density Waves and Cooper Pairing on the Honeycomb Lattice}",
  author = {Honerkamp, Carsten},
  journal = {Phys. Rev. Lett.},
  volume = {100},
  issue = {14},
  pages = {146404},
  numpages = {4},
  year = {2008},
  month = {Apr},
  publisher = {American Physical Society},
  doi = {10.1103/PhysRevLett.100.146404},
  url = {https://link.aps.org/doi/10.1103/PhysRevLett.100.146404}
}

@article{Litim:2001fd,
    author = "Litim, Daniel F.",
    editor = "Arnone, S. and Kubyshin, Yu. A. and Morris, T. R. and Yoshida, K.",
    title = "{Mind the gap}",
    reportNumber = "CERN-TH-2001-013",
    doi = "10.1142/S0217751X01004748",
    journal = "Int. J. Mod. Phys. A",
    volume = "16",
    pages = "2081--2088",
    year = "2001"
}

@article{x7lc-ztqn,
  title = "{Strong-coupling superconductivity near Gross-Neveu quantum criticality in Dirac systems}",
  author = {Stangier, Veronika C. and Sheehy, Daniel E. and Schmalian, J\"org},
  journal = {Phys. Rev. B},
  volume = {113},
  issue = {8},
  pages = {085119},
  numpages = {15},
  year = {2026},
  month = {Feb},
  publisher = {American Physical Society},
  doi = {10.1103/x7lc-ztqn},
  url = {https://link.aps.org/doi/10.1103/x7lc-ztqn}
}

@article{sgnp-ywsh,
  title = "{Superconductivity of Incoherent Electrons near the Relativistic Mott Transition in Twisted Dirac Materials}",
  author = {Stangier, Veronika C. and Scheurer, Mathias S. and Sheehy, Daniel E. and Schmalian, J\"org},
  journal = {Phys. Rev. Lett.},
  volume = {136},
  issue = {17},
  pages = {176501},
  numpages = {7},
  year = {2026},
  month = {Apr},
  publisher = {American Physical Society},
  doi = {10.1103/sgnp-ywsh},
  url = {https://link.aps.org/doi/10.1103/sgnp-ywsh}
}

@article{Rosenstein:1993zf,
    author = "Rosenstein, B. and Yu, Hoi-Lai and Kovner, A.",
    title = "{Critical exponents of new universality classes}",
    reportNumber = "IP-ASTP-14-93, LA-UR-93-2562",
    doi = "10.1016/0370-2693(93)91253-J",
    journal = "Phys. Lett. B",
    volume = "314",
    pages = "381--386",
    year = "1993"
}

@article{Zhang_2005,
   title="{Experimental observation of the quantum Hall effect and Berry’s phase in graphene}",
   volume={438},
   ISSN={1476-4687},
   url={http://dx.doi.org/10.1038/nature04235},
   DOI={10.1038/nature04235},
   number={7065},
   journal={Nature},
   publisher={Springer Science and Business Media LLC},
   author={Zhang, Yuanbo and Tan, Yan-Wen and Stormer, Horst L. and Kim, Philip},
   year={2005},
   month=Nov, pages={201–204} }

@article{Novoselov:2005kj,
    author = "Novoselov, K. S. and Geim, A. K. and Morozov, S. V. and Jiang, D. and Katsnelson, M. I. and Grigorieva, I. V. and Dubonos, S. V. and Firsov, A. A.",
    title = "{Two-dimensional gas of massless Dirac fermions in graphene}",
    doi = "10.1038/nature04233",
    journal = "Nature",
    volume = "438",
    pages = "197",
    year = "2005"
}

@article{lr2x-nnks,
  title = {Dirac quantum criticality in twisted double bilayer transition metal dichalcogenides},
  author = {Biedermann, Jan and Janssen, Lukas},
  journal = {Phys. Rev. B},
  volume = {113},
  issue = {24},
  pages = {245123},
  numpages = {21},
  year = {2026},
  month = {Jun},
  publisher = {American Physical Society},
  doi = {10.1103/lr2x-nnks},
  url = {https://link.aps.org/doi/10.1103/lr2x-nnks}
}

@article{PhysRevB.103.155160,
  title = {Fractionalized quantum criticality in spin-orbital liquids from field theory beyond the leading order},
  author = {Ray, Shouryya and Ihrig, Bernhard and Kruti, Daniel and Gracey, John A. and Scherer, Michael M. and Janssen, Lukas},
  journal = {Phys. Rev. B},
  volume = {103},
  issue = {15},
  pages = {155160},
  numpages = {18},
  year = {2021},
  month = {Apr},
  publisher = {American Physical Society},
  doi = {10.1103/PhysRevB.103.155160},
  url = {https://link.aps.org/doi/10.1103/PhysRevB.103.155160}
}

@article{PhysRevB.97.041117,
  title = "{Compatible orders and fermion-induced emergent symmetry in Dirac systems}",
  author = {Janssen, Lukas and Herbut, Igor F. and Scherer, Michael M.},
  journal = {Phys. Rev. B},
  volume = {97},
  issue = {4},
  pages = {041117},
  numpages = {5},
  year = {2018},
  month = {Jan},
  publisher = {American Physical Society},
  doi = {10.1103/PhysRevB.97.041117},
  url = {https://link.aps.org/doi/10.1103/PhysRevB.97.041117}
}

@article{PhysRevX.11.011014,
  title = "{Correlation-Induced Insulating Topological Phases at Charge Neutrality in Twisted Bilayer Graphene}",
  author = {Da Liao, Yuan and Kang, Jian and Brei\o{}, Clara N. and Xu, Xiao Yan and Wu, Han-Qing and Andersen, Brian M. and Fernandes, Rafael M. and Meng, Zi Yang},
  journal = {Phys. Rev. X},
  volume = {11},
  issue = {1},
  pages = {011014},
  numpages = {17},
  year = {2021},
  month = {Jan},
  publisher = {American Physical Society},
  doi = {10.1103/PhysRevX.11.011014},
  url = {https://link.aps.org/doi/10.1103/PhysRevX.11.011014}
}

@article{PhysRevX.10.031034,
  title = "{Ground State and Hidden Symmetry of Magic-Angle Graphene at Even Integer Filling}",
  author = {Bultinck, Nick and Khalaf, Eslam and Liu, Shang and Chatterjee, Shubhayu and Vishwanath, Ashvin and Zaletel, Michael P.},
  journal = {Phys. Rev. X},
  volume = {10},
  issue = {3},
  pages = {031034},
  numpages = {13},
  year = {2020},
  month = {Aug},
  publisher = {American Physical Society},
  doi = {10.1103/PhysRevX.10.031034},
  url = {https://link.aps.org/doi/10.1103/PhysRevX.10.031034}
}

@article{PhysRevX.8.031089,
  title = "{Origin of Mott Insulating Behavior and Superconductivity in Twisted Bilayer Graphene}",
  author = {Po, Hoi Chun and Zou, Liujun and Vishwanath, Ashvin and Senthil, T.},
  journal = {Phys. Rev. X},
  volume = {8},
  issue = {3},
  pages = {031089},
  numpages = {31},
  year = {2018},
  month = {Sep},
  publisher = {American Physical Society},
  doi = {10.1103/PhysRevX.8.031089},
  url = {https://link.aps.org/doi/10.1103/PhysRevX.8.031089}
}

@article{PhysRevLett.98.186809,
  title = {Electron Fractionalization in Two-Dimensional Graphenelike Structures},
  author = {Hou, Chang-Yu and Chamon, Claudio and Mudry, Christopher},
  journal = {Phys. Rev. Lett.},
  volume = {98},
  issue = {18},
  pages = {186809},
  numpages = {4},
  year = {2007},
  month = {May},
  publisher = {American Physical Society},
  doi = {10.1103/PhysRevLett.98.186809},
  url = {https://link.aps.org/doi/10.1103/PhysRevLett.98.186809}
}

@article{PhysRevB.62.2806,
  title = {Solitons in carbon nanotubes},
  author = {Chamon, Claudio},
  journal = {Phys. Rev. B},
  volume = {62},
  issue = {4},
  pages = {2806--2812},
  numpages = {0},
  year = {2000},
  month = {Jul},
  publisher = {American Physical Society},
  doi = {10.1103/PhysRevB.62.2806},
  url = {https://link.aps.org/doi/10.1103/PhysRevB.62.2806}
}

@article{PhysRevResearch.5.043173,
  title = "{Realizing a tunable honeycomb lattice in ABBA-stacked twisted double bilayer ${\mathrm{WSe}}_{2}$}",
  author = {Pan, Haining and Kim, Eun-Ah and Jian, Chao-Ming},
  journal = {Phys. Rev. Res.},
  volume = {5},
  issue = {4},
  pages = {043173},
  numpages = {9},
  year = {2023},
  month = {Nov},
  publisher = {American Physical Society},
  doi = {10.1103/PhysRevResearch.5.043173},
  url = {https://link.aps.org/doi/10.1103/PhysRevResearch.5.043173}
}

@article{PhysRevResearch.2.013034,
  title = {Mass hierarchy in collective modes of pair-density-wave superconductors},
  author = {Jian, Shao-Kai and Scherer, Michael M. and Yao, Hong},
  journal = {Phys. Rev. Res.},
  volume = {2},
  issue = {1},
  pages = {013034},
  numpages = {8},
  year = {2020},
  month = {Jan},
  publisher = {American Physical Society},
  doi = {10.1103/PhysRevResearch.2.013034},
  url = {https://link.aps.org/doi/10.1103/PhysRevResearch.2.013034}
}

@article{PhysRevB.97.125137,
  title = "{Fermion-induced quantum criticality with two length scales in Dirac systems}",
  author = {Torres, Emilio and Classen, Laura and Herbut, Igor F. and Scherer, Michael M.},
  journal = {Phys. Rev. B},
  volume = {97},
  issue = {12},
  pages = {125137},
  numpages = {13},
  year = {2018},
  month = {Mar},
  publisher = {American Physical Society},
  doi = {10.1103/PhysRevB.97.125137},
  url = {https://link.aps.org/doi/10.1103/PhysRevB.97.125137}
}

@article{PhysRevB.96.115132,
  title = "{Fluctuation-induced continuous transition and quantum criticality in Dirac semimetals}",
  author = {Classen, Laura and Herbut, Igor F. and Scherer, Michael M.},
  journal = {Phys. Rev. B},
  volume = {96},
  issue = {11},
  pages = {115132},
  numpages = {15},
  year = {2017},
  month = {Sep},
  publisher = {American Physical Society},
  doi = {10.1103/PhysRevB.96.115132},
  url = {https://link.aps.org/doi/10.1103/PhysRevB.96.115132}
}

@article{PhysRevB.93.125119,
  title = "{Competition of density waves and quantum multicritical behavior in Dirac materials from functional renormalization}",
  author = {Classen, Laura and Herbut, Igor F. and Janssen, Lukas and Scherer, Michael M.},
  journal = {Phys. Rev. B},
  volume = {93},
  issue = {12},
  pages = {125119},
  numpages = {19},
  year = {2016},
  month = {Mar},
  publisher = {American Physical Society},
  doi = {10.1103/PhysRevB.93.125119},
  url = {https://link.aps.org/doi/10.1103/PhysRevB.93.125119}
}

@article{PhysRevB.89.205403,
  title = {Antiferromagnetic critical point on graphene's honeycomb lattice: A functional renormalization group approach},
  author = {Janssen, Lukas and Herbut, Igor F.},
  journal = {Phys. Rev. B},
  volume = {89},
  issue = {20},
  pages = {205403},
  numpages = {14},
  year = {2014},
  month = {May},
  publisher = {American Physical Society},
  doi = {10.1103/PhysRevB.89.205403},
  url = {https://link.aps.org/doi/10.1103/PhysRevB.89.205403}
}

@article{PhysRevB.94.245102,
  title = "{Ising and Gross-Neveu model in next-to-leading order}",
  author = {Knorr, Benjamin},
  journal = {Phys. Rev. B},
  volume = {94},
  issue = {24},
  pages = {245102},
  numpages = {11},
  year = {2016},
  month = {Dec},
  publisher = {American Physical Society},
  doi = {10.1103/PhysRevB.94.245102},
  url = {https://link.aps.org/doi/10.1103/PhysRevB.94.245102}
}

@article{PhysRevB.66.205111,
  title = "{Phase transition and critical behavior of the d=3 Gross-Neveu model}",
  author = {H\"ofling, F. and Nowak, C. and Wetterich, C.},
  journal = {Phys. Rev. B},
  volume = {66},
  issue = {20},
  pages = {205111},
  numpages = {5},
  year = {2002},
  month = {Nov},
  publisher = {American Physical Society},
  doi = {10.1103/PhysRevB.66.205111},
  url = {https://link.aps.org/doi/10.1103/PhysRevB.66.205111}
}

@article{PhysRevLett.86.958,
  title = "{Critical Exponents of the Gross-Neveu Model from the Effective Average Action}",
  author = {Rosa, L. and Vitale, P. and Wetterich, C.},
  journal = {Phys. Rev. Lett.},
  volume = {86},
  issue = {6},
  pages = {958--961},
  numpages = {0},
  year = {2001},
  month = {Feb},
  publisher = {American Physical Society},
  doi = {10.1103/PhysRevLett.86.958},
  url = {https://link.aps.org/doi/10.1103/PhysRevLett.86.958}
}

@article{PhysRevResearch.2.022005,
  title = "{Emergent symmetries and coexisting orders in Dirac fermion systems}",
  author = {Torres, Emilio and Weber, Lukas and Janssen, Lukas and Wessel, Stefan and Scherer, Michael M.},
  journal = {Phys. Rev. Res.},
  volume = {2},
  issue = {2},
  pages = {022005},
  numpages = {6},
  year = {2020},
  month = {Apr},
  publisher = {American Physical Society},
  doi = {10.1103/PhysRevResearch.2.022005},
  url = {https://link.aps.org/doi/10.1103/PhysRevResearch.2.022005}
}

@article{Herbut:2009qb,
    author = "Herbut, Igor F. and Juricic, Vladimir and Roy, Bitan",
    title = "{Theory of interacting electrons on the honeycomb lattice}",
    doi = "10.1103/PhysRevB.79.085116",
    journal = "Phys. Rev. B",
    volume = "79",
    pages = "085116",
    year = "2009"
}

@article{Herbut:2023xgz,
    author = "Herbut, Igor F.",
    title = "{Wilson-Fisher fixed points in the presence of Dirac fermions}",
    doi = "10.1142/S0217984924300060",
    journal = "Mod. Phys. Lett. B",
    volume = "38",
    number = "34",
    pages = "2430006",
    year = "2024"
}

@article{Wehling_2014,
   title={Dirac materials},
   volume={63},
   ISSN={1460-6976},
   url={http://dx.doi.org/10.1080/00018732.2014.927109},
   DOI={10.1080/00018732.2014.927109},
   number={1},
   journal={Advances in Physics},
   publisher={Informa UK Limited},
   author={Wehling, T.O. and Black-Schaffer, A.M. and Balatsky, A.V.},
   year={2014},
   month=jan, pages={1–76} }

@article{Vojta_2003,
doi = {10.1088/0034-4885/66/12/R01},
url = {https://dx.doi.org/10.1088/0034-4885/66/12/R01},
year = {2003},
month = {nov},
publisher = {},
volume = {66},
number = {12},
pages = {2069},
author = {Matthias Vojta},
title = {Quantum phase transitions},
journal = {Reports on Progress in Physics}
}

@article{Wetterich:1992yh,
    author = "Wetterich, Christof",
    title = "{Exact evolution equation for the effective potential}",
    reportNumber = "HD-THEP-92-61",
    doi = "10.1016/0370-2693(93)90726-X",
    journal = "Phys. Lett. B",
    volume = "301",
    pages = "90--94",
    year = "1993"
}

@article{PhysRevB.92.155137,
  title = {Correlated spinless fermions on the honeycomb lattice revisited},
  author = {Scherer, Daniel D. and Scherer, Michael M. and Honerkamp, Carsten},
  journal = {Phys. Rev. B},
  volume = {92},
  issue = {15},
  pages = {155137},
  numpages = {9},
  year = {2015},
  month = {Oct},
  publisher = {American Physical Society},
  doi = {10.1103/PhysRevB.92.155137},
  url = {https://link.aps.org/doi/10.1103/PhysRevB.92.155137}
}

@article{PhysRevB.87.094521,
  title = "{Correlated Dirac particles and superconductivity on the honeycomb lattice}",
  author = {Wu, Wei and Scherer, Michael M. and Honerkamp, Carsten and Le Hur, Karyn},
  journal = {Phys. Rev. B},
  volume = {87},
  issue = {9},
  pages = {094521},
  numpages = {12},
  year = {2013},
  month = {Mar},
  publisher = {American Physical Society},
  doi = {10.1103/PhysRevB.87.094521},
  url = {https://link.aps.org/doi/10.1103/PhysRevB.87.094521}
}

@book{Wipf2021,
    author = "Wipf, Andreas",
    title = "{Statistical Approach to Quantum Field Theory}",
    isbn = "978-3-030-83262-9, 978-3-030-83263-6",
    publisher = "Springer Cham",
    month = "10",
    year = "2021",
    doi = "https://doi.org/10.1007/978-3-030-83263-6"
}

@book{Sachdev2011,
    author = "Sachdev, Subir",
    title = "{Quantum Phase Transitions}",
    doi = "10.1017/cbo9780511973765",
    isbn = "978-0-511-97376-5",
    publisher = "Cambridge University Press",
    month = "4",
    year = "2011"
}

@article{biedermann2024twisttunedquantumcriticalitymoire,
  title = {Twist-tuned quantum criticality in moir\'e bilayer graphene},
  author = {Biedermann, Jan and Janssen, Lukas},
  journal = {Phys. Rev. B},
  volume = {112},
  issue = {4},
  pages = {L041109},
  numpages = {7},
  year = {2025},
  month = {Jul},
  publisher = {American Physical Society},
  doi = {10.1103/hj61-dw78},
  url = {https://link.aps.org/doi/10.1103/hj61-dw78}
}

@article{huang2024angletunedgrossneveuquantumcriticality,
	author = {Huang, Cheng and Parthenios, Nikolaos and Ulybyshev, Maksim and Zhang, Xu and Assaad, Fakher F. and Classen, Laura and Meng, Zi Yang},
	date = {2025/08/04},
	doi = {10.1038/s41467-025-62461-y},
	id = {Huang2025},
	isbn = {2041-1723},
	journal = {Nature Communications},
	number = {1},
	pages = {7176},
	title = {Angle-tuned Gross-Neveu quantum criticality in twisted bilayer graphene},
	url = {https://doi.org/10.1038/s41467-025-62461-y},
	volume = {16},
	year = {2025}}

@article{ma2024relativisticmotttransitionstrongly,
      author  = {Ma, Liguo and Chaturvedi, Raghav and Nguyen, Phuong X. and Watanabe, Kenji and Taniguchi, Takashi and Mak, Kin Fai and Shan, Jie},
  title   = {Relativistic {Mott} transition in twisted {WSe}$_2$ tetralayers},
  journal = {Nature Materials},
  year    = {2025},
  volume  = {24},
  number  = {12},
  pages   = {1935--1941},
  doi     = {10.1038/s41563-025-02359-8},
  url     = {https://doi.org/10.1038/s41563-025-02359-8},
  issn    = {1476-4660}
}

@article{Dupuis:2020fhh,
    author = "Dupuis, N. and Canet, L. and Eichhorn, A. and Metzner, W. and Pawlowski, J. M. and Tissier, M. and Wschebor, N.",
    title = "{The nonperturbative functional renormalization group and its applications}",
    doi = "10.1016/j.physrep.2021.01.001",
    journal = "Phys. Rept.",
    volume = "910",
    pages = "1--114",
    year = "2021"
}

@article{khan2015phasediagramqc2dfunctional,
      author = "Khan, Naseemuddin and Pawlowski, Jan M. and Rennecke, Fabian and Scherer, Michael M.",
    title = "{The Phase Diagram of QC2D from Functional Methods}",
    journal = "arXiv:1512.03673",
    month = "12",
    year = "2015" ,
url="https://arxiv.org/abs/1512.03673"
}

@article{Boettcher_2012,
   title={Ultracold atoms and the Functional Renormalization Group},
   volume={228},
   ISSN={0920-5632},
   url={http://dx.doi.org/10.1016/j.nuclphysbps.2012.06.004},
   DOI={10.1016/j.nuclphysbps.2012.06.004},
   journal={Nuclear Physics B - Proceedings Supplements},
   publisher={Elsevier BV},
   author={Boettcher, Igor and Pawlowski, Jan M. and Diehl, Sebastian},
   year={2012},
   month=jul, pages={63–135} }

@article{Scherer:2013many,
    author = "Scherer, Daniel D. and Braun, Jens and Gies, Holger",
    title = "{Many-flavor Phase Diagram of the (2+1)d Gross-Neveu Model at Finite Temperature}",
    doi = "10.1088/1751-8113/46/28/285002",
    journal = "J. Phys. A",
    volume = "46",
    pages = "285002",
    year = "2013"
}

@article{Berges:2002,
title = {Non-perturbative renormalization flow in quantum field theory and statistical physics},
journal = {Physics Reports},
volume = {363},
number = {4},
pages = {223-386},
year = {2002},
issn = {0370-1573},
doi = {https://doi.org/10.1016/S0370-1573(01)00098-9},
url = {https://www.sciencedirect.com/science/article/pii/S0370157301000989},
author = {Jürgen Berges and Nikolaos Tetradis and Christof Wetterich}
}

@article{Iliesiu:2015qra,
	author = {Iliesiu, Luca and Kos, Filip and Poland, David and Pufu, Silviu S. and Simmons-Duffin, David and Yacoby, Ran},
	date = {2016/03/17},
	doi = {10.1007/JHEP03(2016)120},
	id = {Iliesiu2016},
	isbn = {1029-8479},
	journal = {Journal of High Energy Physics},
	number = {3},
	pages = {120},
	title = "{Bootstrapping 3D fermions}",
	url = {https://doi.org/10.1007/JHEP03(2016)120},
	volume = {2016},
	year = {2016}}

@article{Iliesiu:2017nrv,
	author = {Iliesiu, Luca and Kos, Filip and Poland, David and Pufu, Silviu S. and Simmons-Duffin, David},
	date = {2018/01/09},
	doi = {10.1007/JHEP01(2018)036},
	id = {Iliesiu2018},
	isbn = {1029-8479},
	journal = {Journal of High Energy Physics},
	number = {1},
	pages = {36},
	title = {Bootstrapping 3D fermions with global symmetries},
	url = {https://doi.org/10.1007/JHEP01(2018)036},
	volume = {2018},
	year = {2018}}

@article{PhysRevB.101.064308,
  title     = "{Designer Monte Carlo simulation for the Gross-Neveu-Yukawa transition}",
  author    = {Liu, Yuzhi and Wang, Wei and Sun, Kai and Meng, Zi Yang},
  journal   = {Phys. Rev. B},
  volume    = {101},
  issue     = {6},
  pages     = {064308},
  numpages  = {12},
  year      = {2020},
  month     = {Feb},
  publisher = {American Physical Society},
  doi       = {10.1103/PhysRevB.101.064308},
  url       = {https://link.aps.org/doi/10.1103/PhysRevB.101.064308}
}

@article{PhysRevD.88.021701,
  title     = "{Quantum critical behavior in three dimensional lattice Gross-Neveu models}",
  author    = {Chandrasekharan, Shailesh and Li, Anyi},
  journal   = {Phys. Rev. D},
  volume    = {88},
  issue     = {2},
  pages     = {021701},
  numpages  = {5},
  year      = {2013},
  month     = {Jul},
  publisher = {American Physical Society},
  doi       = {10.1103/PhysRevD.88.021701},
  url       = {https://link.aps.org/doi/10.1103/PhysRevD.88.021701}
}

@article{Classen:2016,
  title = "{Competition of density waves and quantum multicritical behavior in Dirac materials from functional renormalization}",
  author = {Classen, Laura and Herbut, Igor F. and Janssen, Lukas and Scherer, Michael M.},
  journal = {Phys. Rev. B},
  volume = {93},
  issue = {12},
  pages = {125119},
  numpages = {19},
  year = {2016},
  month = {Mar},
  publisher = {American Physical Society},
  doi = {10.1103/PhysRevB.93.125119},
  url = {https://link.aps.org/doi/10.1103/PhysRevB.93.125119}
}

@article{Erramilli2023,
author = {Erramilli, Rajeev S. and Iliesiu, Luca V. and Kravchuk, Petr and Liu, Aike and Poland, David and Simmons-Duffin, David},
	date = {2023/02/03},
	doi = {10.1007/JHEP02(2023)036},
	id = {Erramilli2023},
	isbn = {1029-8479},
	journal = {Journal of High Energy Physics},
	number = {2},
	pages = {36},
	title = "{The Gross-Neveu-Yukawa archipelago}",
	url = {https://doi.org/10.1007/JHEP02(2023)036},
	volume = {2023},
	year = {2023}}

@article{Ihrig2018,
  title = "{Critical behavior of Dirac fermions from perturbative renormalization}",
  author = {Ihrig, Bernhard and Mihaila, Luminita N. and Scherer, Michael M.},
  journal = {Phys. Rev. B},
  volume = {98},
  issue = {12},
  pages = {125109},
  numpages = {20},
  year = {2018},
  month = {Sep},
  publisher = {American Physical Society},
  doi = {10.1103/PhysRevB.98.125109},
  url = {https://link.aps.org/doi/10.1103/PhysRevB.98.125109}
}

@article{Herbut2006,
  title = "{Interactions and Phase Transitions on Graphene's Honeycomb Lattice}",
  author = {Herbut, Igor F.},
  journal = {Phys. Rev. Lett.},
  volume = {97},
  issue = {14},
  pages = {146401},
  numpages = {4},
  year = {2006},
  month = {Oct},
  publisher = {American Physical Society},
  doi = {10.1103/PhysRevLett.97.146401},
  url = {https://link.aps.org/doi/10.1103/PhysRevLett.97.146401}
}

@article{Parthenios2023,
  title = "{Twisted bilayer graphene at charge neutrality: Competing orders of SU(4) Dirac fermions}",
  author = {Parthenios, Nikolaos and Classen, Laura},
  journal = {Phys. Rev. B},
  volume = {108},
  issue = {23},
  pages = {235120},
  numpages = {20},
  year = {2023},
  month = {Dec},
  publisher = {American Physical Society},
  doi = {10.1103/PhysRevB.108.235120},
  url = {https://link.aps.org/doi/10.1103/PhysRevB.108.235120}
}

@article{Zerf2017,
   author = "Zerf, Nikolai and Mihaila, Luminita N. and Marquard, Peter and Herbut, Igor F. and Scherer, Michael M.",
    title = "{Four-loop critical exponents for the Gross-Neveu-Yukawa models}",
    reportNumber = "DESY-17-133",
    doi = "10.1103/PhysRevD.96.096010",
    journal = "Phys. Rev. D",
    volume = "96",
    number = "9",
    pages = "096010",
    year = "2017"
}

@article{Knorr2016,
  title = "{Ising and Gross-Neveu model in next-to-leading order}",
  author = {Knorr, Benjamin},
  journal = {Phys. Rev. B},
  volume = {94},
  issue = {24},
  pages = {245102},
  numpages = {11},
  year = {2016},
  month = {Dec},
  publisher = {American Physical Society},
  doi = {10.1103/PhysRevB.94.245102},
  url = {https://link.aps.org/doi/10.1103/PhysRevB.94.245102}
}

@article{Knorr2018,
  title = "{Critical chiral Heisenberg model with the functional renormalization group}",
  author = {Knorr, Benjamin},
  journal = {Phys. Rev. B},
  volume = {97},
  issue = {7},
  pages = {075129},
  numpages = {9},
  year = {2018},
  month = {Feb},
  publisher = {American Physical Society},
  doi = {10.1103/PhysRevB.97.075129},
  url = {https://link.aps.org/doi/10.1103/PhysRevB.97.075129}
}

@article{PhysRevB.108.L121112,
  title = "{Quantum Monte Carlo calculation of critical exponents of the Gross-Neveu-Yukawa on a two-dimensional fermion lattice model}",
  author = {Wang, Ting-Tung and Meng, Zi Yang},
  journal = {Phys. Rev. B},
  volume = {108},
  issue = {12},
  pages = {L121112},
  numpages = {6},
  year = {2023},
  month = {Sep},
  publisher = {American Physical Society},
  doi = {10.1103/PhysRevB.108.L121112},
  url = {https://link.aps.org/doi/10.1103/PhysRevB.108.L121112}
}

@article{PhysRevB.98.125109,
  title = "{Critical behavior of Dirac fermions from perturbative renormalization}",
  author = {Ihrig, Bernhard and Mihaila, Luminita N. and Scherer, Michael M.},
  journal = {Phys. Rev. B},
  volume = {98},
  issue = {12},
  pages = {125109},
  numpages = {20},
  year = {2018},
  month = {Sep},
  publisher = {American Physical Society},
  doi = {10.1103/PhysRevB.98.125109},
  url = {https://link.aps.org/doi/10.1103/PhysRevB.98.125109}
}

@article{Erramilli:2022kgp,
    author = "Erramilli, Rajeev S. and Iliesiu, Luca V. and Kravchuk, Petr and Liu, Aike and Poland, David and Simmons-Duffin, David",
    title = "{The Gross-Neveu-Yukawa archipelago}",
    eprint = "2210.02492",
    archivePrefix = "arXiv",
    primaryClass = "hep-th",
    reportNumber = "CALT-TH 2022-027",
    doi = "10.1007/JHEP02(2023)036",
    journal = "JHEP",
    volume = "02",
    pages = "036",
    year = "2023"
}

@article{Roy:2015zna,
    author = "Roy, Bitan and Juricic, Vladimir and Herbut, Igor F.",
    title = "{Emergent Lorentz symmetry near fermionic quantum critical points in two and three dimensions}",
    doi = "10.1007/JHEP04(2016)018",
    journal = "Journal of High Energy Physics",
    volume = "04",
    pages = "018",
    year = "2016"
}

@article{RevModPhys.79.1015,
  title = {Fermi-liquid instabilities at magnetic quantum phase transitions},
  author = {L\"ohneysen, Hilbert v. and Rosch, Achim and Vojta, Matthias and W\"olfle, Peter},
  journal = {Rev. Mod. Phys.},
  volume = {79},
  issue = {3},
  pages = {1015--1075},
  numpages = {0},
  year = {2007},
  month = {Aug},
  publisher = {American Physical Society},
  doi = {10.1103/RevModPhys.79.1015},
  url = {https://link.aps.org/doi/10.1103/RevModPhys.79.1015}
}

@article{doi:10.1063/1.3554314,
author = {Sachdev,Subir  and Keimer,Bernhard },
title = {Quantum criticality},
journal = {Physics Today},
volume = {64},
number = {2},
pages = {29-35},
year = {2011},
doi = {10.1063/1.3554314},

URL = { 
        https://doi.org/10.1063/1.3554314
    
}
}

@article{doi:10.1146/annurev-conmatphys-031218-013339,
author = {Berg, Erez and Lederer, Samuel and Schattner, Yoni and Trebst, Simon},
title = "{Monte Carlo Studies of Quantum Critical Metals}",
journal = {Annual Review of Condensed Matter Physics},
volume = {10},
number = {1},
pages = {63-84},
year = {2019},
doi = {10.1146/annurev-conmatphys-031218-013339},

URL = { 
        https://doi.org/10.1146/annurev-conmatphys-031218-013339
    
}
}

@article{Xu_2019,
	author = {Xiao Yan Xu and Zi Hong Liu and Gaopei Pan and Yang Qi and Kai Sun and Zi Yang Meng},
	doi = {10.1088/1361-648x/ab3295},
	journal = {Journal of Physics: Condensed Matter},
	month = {aug},
	number = {46},
	pages = {463001},
	publisher = {{IOP} Publishing},
	title = "{Revealing fermionic quantum criticality from new Monte Carlo techniques}",
	url = {https://doi.org/10.1088/1361-648x/ab3295},
	volume = {31},
	year = 2019}

@article{RevModPhys.88.025006,
  title = {Metallic quantum ferromagnets},
  author = {Brando, M. and Belitz, D. and Grosche, F. M. and Kirkpatrick, T. R.},
  journal = {Rev. Mod. Phys.},
  volume = {88},
  issue = {2},
  pages = {025006},
  numpages = {71},
  year = {2016},
  month = {May},
  publisher = {American Physical Society},
  doi = {10.1103/RevModPhys.88.025006},
  url = {https://link.aps.org/doi/10.1103/RevModPhys.88.025006}
}

@article{RevModPhys.73.797,
  title = "{Non-Fermi-liquid behavior in $d$- and $f$-electron metals}",
  author = {Stewart, G. R.},
  journal = {Rev. Mod. Phys.},
  volume = {73},
  issue = {4},
  pages = {797--855},
  numpages = {0},
  year = {2001},
  month = {Oct},
  publisher = {American Physical Society},
  doi = {10.1103/RevModPhys.73.797},
  url = {https://link.aps.org/doi/10.1103/RevModPhys.73.797}
}

@article{doi:10.1146/annurev-conmatphys-031016-025531,
author = {Lee, Sung-Sik},
title = "{Recent Developments in Non-Fermi Liquid Theory}",
journal = {Annual Review of Condensed Matter Physics},
volume = {9},
number = {1},
pages = {227-244},
year = {2018},
doi = {10.1146/annurev-conmatphys-031016-025531},

URL = { 
        https://doi.org/10.1146/annurev-conmatphys-031016-025531
    
}
}

@article{RevModPhys.94.035004,
  title = "{Sachdev-Ye-Kitaev models and beyond: Window into non-Fermi liquids}",
  author = {Chowdhury, Debanjan and Georges, Antoine and Parcollet, Olivier and Sachdev, Subir},
  journal = {Rev. Mod. Phys.},
  volume = {94},
  issue = {3},
  pages = {035004},
  numpages = {78},
  year = {2022},
  month = {Sep},
  publisher = {American Physical Society},
  doi = {10.1103/RevModPhys.94.035004},
  url = {https://link.aps.org/doi/10.1103/RevModPhys.94.035004}
}

@article{doi:10.1080/001075199181602,
	author = {A. J. Schofield},
	doi = {10.1080/001075199181602},
	journal = {Contemporary Physics},
	number = {2},
	pages = {95-115},
	publisher = {Taylor & Francis},
	title = "{Non-Fermi liquids}",
	url = {https://doi.org/10.1080/001075199181602},
	volume = {40},
	year = {1999}}

@article{Classen_2022,
doi = {10.1088/2053-1583/ac6e71},
url = {https://dx.doi.org/10.1088/2053-1583/ac6e71},
year = {2022},
month = {jun},
publisher = {IOP Publishing},
volume = {9},
number = {3},
pages = {031001},
author = {Classen, Laura and Pixley, J H and König, Elio J},
title = {Interaction-induced velocity renormalization in magic-angle twisted multilayer graphene},
journal = {2D Materials}
}

@article{Braun_2012,
doi = {10.1088/0954-3899/39/3/033001},
url = {https://dx.doi.org/10.1088/0954-3899/39/3/033001},
year = {2012},
month = {jan},
publisher = {IOP Publishing},
volume = {39},
number = {3},
pages = {033001},
author = {Braun, Jens},
title = {Fermion interactions and universal behavior in strongly interacting theories},
journal = {Journal of Physics G: Nuclear and Particle Physics}
}

@article{PhysRevD.90.076002,
  title = "{Higher order quark-mesonic scattering processes and the phase structure of QCD}",
  author = {Pawlowski, Jan M. and Rennecke, Fabian},
  journal = {Phys. Rev. D},
  volume = {90},
  issue = {7},
  pages = {076002},
  numpages = {21},
  year = {2014},
  month = {Oct},
  publisher = {American Physical Society},
  doi = {10.1103/PhysRevD.90.076002},
  url = {https://link.aps.org/doi/10.1103/PhysRevD.90.076002}
}

@article{PhysRevX.12.011061,
  title = "{Fermionic Monte Carlo Study of a Realistic Model of Twisted Bilayer Graphene}",
  author = {Hofmann, Johannes S. and Khalaf, Eslam and Vishwanath, Ashvin and Berg, Erez and Lee, Jong Yeon},
  journal = {Phys. Rev. X},
  volume = {12},
  issue = {1},
  pages = {011061},
  numpages = {32},
  year = {2022},
  month = {Mar},
  publisher = {American Physical Society},
  doi = {10.1103/PhysRevX.12.011061},
  url = {https://link.aps.org/doi/10.1103/PhysRevX.12.011061}
}

\end{document}